\documentclass[12pt,a4paper,dvipsnames]{article}
\usepackage[T1]{fontenc}
\usepackage[latin9]{inputenc}
\usepackage{geometry}
\usepackage{xcolor}
\usepackage{longtable}
\usepackage{refstyle}
\usepackage{url}
\usepackage{amsmath}
\usepackage{amssymb}
\usepackage{graphicx}
\usepackage{setspace}

\makeatletter

\AtBeginDocument{\providecommand\subsecref[1]{\ref{subsec:#1}}}
\newcommand{\lyxmathsym}[1]{\ifmmode\begingroup\def\b@ld{bold}
  \text{\ifx\math@version\b@ld\bfseries\fi#1}\endgroup\else#1\fi}

\providecommand{\tabularnewline}{\\}
\RS@ifundefined{subsecref}
  {\newref{subsec}{name = \RSsectxt}}
  {}
\RS@ifundefined{thmref}
  {\def\RSthmtxt{theorem~}\newref{thm}{name = \RSthmtxt}}
  {}
\RS@ifundefined{lemref}
  {\def\RSlemtxt{lemma~}\newref{lem}{name = \RSlemtxt}}
  {}

\newcommand{\lyxrightaddress}[1]{
\par {\raggedleft \begin{tabular}{l}\ignorespaces
#1
\end{tabular}
\vspace{1.4em}
\par}
}

\usepackage[superscript]{cite}
\usepackage{notes2bib}
\usepackage{appendix}

\usepackage{tabularx} 				
\usepackage{helvet} 				
\usepackage[labelfont=bf,font=small]{caption} 		
\usepackage{microtype}				
\usepackage{collect}				
\usepackage[all]{nowidow}     			

\makeatother

\begin{document}

\title{A critical review of statistical\\
calibration/prediction models handling\\
data inconsistency and model inadequacy\\
}

\author{Pascal PERNOT \& Fabien CAILLIEZ}
\maketitle

\lyxrightaddress{CNRS, UMR8000, Laboratoire de Chimie Physique, F-91405 Orsay, France\\
Univ. Paris-Sud, UMR000, Laboratoire de Chimie Physique, F-91405 Orsay,
France\\
Contact: pascal.pernot@u-psud.fr}
\begin{abstract}
\noindent Inference of physical parameters from reference data is
a well studied problem with many intricacies (inconsistent sets of
data due to experimental systematic errors; approximate physical models...).
The complexity is further increased when the inferred parameters are
used to make predictions \textendash{} virtual measurements \textendash{}
because parameters uncertainty has to be estimated in addition to
parameters \emph{best value}. The literature is rich in statistical
models for the calibration/prediction problem, each having benefits
and limitations. 

We review and evaluate standard and state-of-the-art statistical models
in a common bayesian framework, and test them on synthetic and real
datasets of temperature-dependent viscosity for the calibration of
Lennard-Jones parameters of a Chapman-Enskog model.
\end{abstract}

\clearpage{}

\section{Introduction}

The past decades have seen a tremendous increase in the size and time-scales
of molecular systems accessible to computational chemistry. Estimation
of the prediction uncertainty of such simulations is the next challenging
step to reach virtual measurements \cite{Irikura2004}, \emph{i.e.},
to enable simulations or physical models\bibnote{Two types of models are referred to in this study: physical (or knowledge) models used to make predictions of physico-chemical properties and statistical models used to calibrate the knowledge models. Most of the times, the simple term model will be used when there is no ambiguity. The same holds for the parameters of both types of models.}
to replace experiments. This subject is taking momentum in the computational
chemistry community, and several studies have been reported recently
in the literature, mostly in the last 5 years, and notably for atomistic/molecular
simulation \cite{Cailliez2011,Angelikopoulos2012,Cailliez2013,Angelikopoulos2013,Chernatynskiy2013,Hadjidoukas2015,Wu2015},
density functional theory \cite{Mortensen2005,Wellendorff2012,Lejaeghere2013,Wellendorff2014,Pernot2015,Pandey2015,Lejaeghere2016},
quantum chemistry \cite{Edwards2014,Ruscic2014} and multiscale catalysis
studies \cite{Ulissi2011,Medford2014,Sutton2016}. Chemical engineering
is a companion field where uncertainty quantification (UQ) is becoming
crucial \cite{Vasquez2004,Androulakis2006,Frenklach2007,Sheen2009,Russi2010,Turanyi2012,Xing2015,Wang2015,Cheung2016,Frenklach2016,Kalyanaraman2016,Sutton2016}.

The estimation of model prediction uncertainty (MPU) is a complex
process, which requires a careful analysis of the main error sources:
(i) systematic errors due to the model formulation and approximations
(model inadequacy); (ii) numerical errors (notably for stochastic
models); and (iii) parameter uncertainty. Once the error sources are
well identified and quantified, uncertainty propagation to model predictions
is rather straightforward, even if often computationally challenging
\cite{GUM,GUMSupp1,deRocquigny2008}. 

Except for stochastic or chaotic models, numerical errors are expected
to be well controlled and kept to a negligible level \cite{Irikura2004,Williams2008,Feher2012}.
The remaining sources depend on comparisons with reference data: parameter
uncertainty is generally the result of a model calibration process
by which the parameters are identified; and quantification of a model's
systematic errors requires reference data. As reference data are generally
experimental, measurement errors, both systematic and random, contaminate
to some extent the estimation of model error sources \cite{Pernot2015}. 

In this study, we focus on the estimation of prediction uncertainty
in the typical calibration/prediction scenario, where one is faced
with inconsistent reference data \emph{and} an approximate or misspecified
model:
\begin{itemize}
\item \emph{Data inconsistency} occurs when the scatter of repeated measurements
is statistically inconsistent with the stated uncertainty of the individual
measurements \cite{Lira2007a,Toman2009}. This is an ubiquitous problem
in the comparison of data produced by different laboratories and/or
different measurement setups. It might result from incorrect quantification
of measurement uncertainty, but often has its origin in unidentified
systematic errors in the measurement process. If the data are abundant,
one has the option to reject data until one obtains a consistent dataset
\cite{Cox2007}. In this context, rejecting outliers is always dangerous,
as they might in fact be the nearest to the true value. Recently,
metrologists have developed statistical methods enabling to preserve
all data, notably through the use of bayesian hierarchical models
\cite{Toman2007,Toman2009,Yuan2015}.
\item \emph{Model inadequacy} results from approximations at various stages
of model development and is responsible for systematic errors in predictions
\cite{OHagan2013}. It should be identified and quantified by comparison
with reference data. Correction of model systematic errors can be
done by model improvement, or by a posteriori correction of model
predictions \cite{Pernot2015}. Model improvement is often impractical
or impossible, and the latter option is popular, for instance, in
the statistical correction of harmonic vibrational frequencies by
scaling \cite{ScoRad-96,Pernot2011}. If model inadequacy cannot be
corrected, it is essential that prediction uncertainty be large enough
to account for it. This leads to several options in prediction uncertainty
modeling reviewed recently by Sargsyan \emph{et al.},{\cite{Sargsyan2015}}
and that will be considered in the following.
\end{itemize}
The systematic errors linked to model inadequacy have a major impact
on the parameters recovered when calibrating the model, which is difficult
to unmix from the impact of data inconsistency \cite{Frenklach2016}.
In order to establish a reliable MPU, it is necessary to discriminate
correctly the error sources and their degeneracy in the model calibration
process, by the design of an adequate statistical calibration/prediction
process. 

Our aim in the present study is to review the main approaches used
in the computational chemistry literature to deal with data inconsistency
and/or model inadequacy. For instance, Wu \emph{et al. }\cite{Wu2015}
recently proposed a hierarchical model to calibrate the Lennard-Jones
parameters of an interatomic potential on inconsistent viscosity measurements.
Their use of physical parameters to represent systematic experimental
errors is intriguing and needs further consideration.

Assuming a valid calibration/prediction process, the question remains
of the transferability of model calibration to other observables \cite{Campbell2006,Oliver2015}.
All calibration methods based on the a posteriori correction of model
predictions are by nature not transferable. A solution to the transferability
problem is to affect to the model's parameters uncertainty the prediction
errors due to model inadequacy.{\cite{Sargsyan2015}}
This has been done in various ways: uncertainty scaling \cite{Mortensen2005,Wellendorff2012,Wellendorff2014},
embedded stochastic models \cite{Sargsyan2015,Cheung2016} and hierarchical
models \cite{Wu2015}. Parameters and their enlarged uncertainty are
then transferable, but this approach is not without drawbacks \cite{Kalyanaraman2016,Pernot2016}:
the transfer of parameter uncertainty to MPU is governed by the functional
shape (with respect to the control variables) of the model sensitivity
coefficients. This leads to confidence bands with a model-specific
shape, not necessarily representative of the actual model errors.
As demonstrated, for instance, in the case of scaling factors for
harmonic frequencies, or in the calibration of the mBEEF density functional,
the mean MPU is, by design, in good agreement with the error statistics,
but individual MPUs are unreliable, being over- or under-estimated
in different parts of the control space \cite{Pernot2016}. 

This paper attempts a critical review of classical and state-of-the-art
statistical models dealing with data inconsistency and model inadequacy.
For presentation consistency, the statistical models are cast in the
bayesian framework introduced in Section \ref{sec:Statistical-methods}.
The models themselves are presented in Section \ref{sec:Calibration-models}.
We then consider the advantages and limitations of each method (and
their combinations), illustrated on the calibration of the parameters
of a Lennard-Jones potential from synthetic and experimental viscosity
data (Section \ref{sec:Applications}). The discussion (Section \ref{sec:Discussion})
draws on these examples to propose guidelines for a successful calibration/prediction
process.

\section{Statistical methods\label{sec:Statistical-methods}}

Some of the simpler calibration/prediction methods are commonly presented
in a least-squares framework, whereas the most sophisticated ones
rely on a bayesian formulation. For homogeneity, the latter has been
used throughout this study, and we provide a short introduction below.
More detailed presentations of bayesian data analysis can be found
in textbooks \cite{Gregory2005,Gelman2013,McElreath2015}.

\subsection{Bayesian calibration and prediction}

One considers a model represented by the function $M(x;\boldsymbol{\vartheta})$\bibnote{Boldface type refers to vectors or matrices.},
depending on a (set of) control variable(s) $x$ (\emph{e.g.} temperature,
pressure...), and parameters $\boldsymbol{\vartheta}$ that have to
be identified, \emph{i.e.} characterized by their probability density
function (pdf) or, in the gaussian hypothesis, their ``best'' value
and covariance matrix \cite{OHagan2013}. Parameters inference is
done by calibration of the model on a set of reference data $\boldsymbol{D}$.

The \emph{posterior} probability density function $p(\boldsymbol{\vartheta}|\boldsymbol{D},M)$
for the parameters $\boldsymbol{\vartheta}$, conditional on $\boldsymbol{D}$
and $M$, is provided by Bayes formula
\begin{equation}
p(\boldsymbol{\vartheta}|\boldsymbol{D},M)=\frac{p(\boldsymbol{D}|\boldsymbol{\vartheta},M)\,p(\boldsymbol{\vartheta}|M)}{p(\boldsymbol{D}|M)},
\end{equation}
where 
\begin{itemize}
\item $p(\boldsymbol{D}|\boldsymbol{\vartheta},M)$ is the \emph{likelihood}
function, describing the distribution of the differences between model
and data, as detailed below; 
\item $p(\boldsymbol{\vartheta}|M)$ is the \emph{prior} pdf of the parameters,
to be defined later;
\item $p(\boldsymbol{D}|M)$ is the \emph{evidence}, a normalization constant
which we do not need to estimate in the following.
\end{itemize}
For simpler notations, $M$ will be kept implicit in the following.

The mode of the posterior pdf, $\hat{\boldsymbol{\vartheta}}$, or
\emph{maximum a posteriori} ($MAP$) solution, provides an estimate
of the ``best fit'' solution of the calibration problem. The mean
value of the parameters $\boldsymbol{\mu}_{\vartheta|D}$ and their
covariance matrix $\boldsymbol{V}_{\vartheta|D}$ are often used to
summarize the posterior pdf. 

The full \emph{posterior} pdf is necessary to account for parameter
uncertainty in prediction of the true value $\tilde{y}$ at a new
control point $\tilde{x}$ through the \emph{posterior predictive
distribution}
\begin{eqnarray}
p_{M}(\tilde{y}|\tilde{x},\boldsymbol{D}) & = & \int f(\tilde{y}|\tilde{x},\boldsymbol{\vartheta})\,p(\boldsymbol{\vartheta}|\boldsymbol{D})\,d\boldsymbol{\vartheta},\label{eq:conf}
\end{eqnarray}
where $f(\tilde{y}|\tilde{x},\boldsymbol{\vartheta})$ is a probability
density function which might be a Dirac function $f(\tilde{y}|\tilde{x},\boldsymbol{\vartheta})=\delta\left(\tilde{y}-M(\tilde{x};\boldsymbol{\vartheta})\right)$,
if there is no model uncertainty \cite{GUMSupp1}, or a pdf describing
$\tilde{y}$ as the output of a random process characterizing model
errors \cite{Gelman2013}. For deterministic models, the mean value
of a prediction at a new control value $\tilde{x}$ and its variance
can be approximated by linear uncertainty propagation\cite{GUM}
\begin{align}
\mu_{M|D}(\tilde{x}) & =M(\tilde{x};\boldsymbol{\mu}_{\vartheta|D})\label{eq:LUP-mean}\\
u_{M|D}^{2}(\tilde{x}) & =\boldsymbol{J}^{T}(\tilde{x};\boldsymbol{\mu}_{\vartheta|D})\boldsymbol{V}_{\vartheta|D}\boldsymbol{J}(\tilde{x};\boldsymbol{\mu}_{\vartheta|D}),\label{eq:LUP-var}
\end{align}
where $\boldsymbol{J}$ is a vector of \emph{sensitivity coefficients}
\begin{equation}
\boldsymbol{J}_{k}(x;\boldsymbol{\mu}_{\vartheta|D})=\left.\frac{\partial M(x;\boldsymbol{\vartheta})}{\partial\boldsymbol{\vartheta}_{k}}\right|_{\boldsymbol{\mu}_{\vartheta|D}}.\label{eq:Sensitivity}
\end{equation}

In order to predict a new experimental measurement, $p_{M}(\tilde{y}|\tilde{x},\boldsymbol{D})$
has to be convoluted with the pdf describing expected measurement
errors, $p(\tilde{y_{e}}|\tilde{y})$:
\begin{eqnarray}
p_{e}(\tilde{y_{e}}|\tilde{x},\boldsymbol{D}) & = & \int p(\tilde{y_{e}}|\tilde{y})\,p_{M}(\tilde{y}|\tilde{x},\boldsymbol{D})\,d\tilde{y}.\label{eq:pred}
\end{eqnarray}
In the linear uncertainty propagation framework, one would get the
variance of prediction $\tilde{y_{e}}$ as
\begin{equation}
u_{e|D}^{2}=u_{M|D}^{2}+u_{y_{e}}^{2},
\end{equation}
where $u_{y_{e}}^{2}$ is the measurement variance for $\tilde{y_{e}}$. 

\subsubsection{The likelihood}

For all models used in this study, we assume errors with normal distributions,
in which case the likelihood function can be expressed as a multivariate
normal distribution
\begin{equation}
p(\boldsymbol{D}|\boldsymbol{\vartheta}=\left\{ \boldsymbol{\vartheta}_{M},\boldsymbol{\vartheta}_{V}\right\} )=\frac{1}{\left(\sqrt{2\pi}\right)^{N}\left|\boldsymbol{V}\right|^{1/2}}\exp\left(-\frac{1}{2}\boldsymbol{R}^{T}\boldsymbol{V}^{-1}\boldsymbol{R}\right),\label{eq:likelihood}
\end{equation}
where 
\begin{itemize}
\item $\boldsymbol{R}$ is a $N$-vector of residuals $R_{i}=y_{i}-M(x_{i};\boldsymbol{\vartheta}_{M})$,
and 
\item $\boldsymbol{V}$ is a $N\times N$ covariance matrix, possibly parameterized
by $\boldsymbol{\vartheta}_{V}$. 
\end{itemize}
The latter has typically a contribution $\boldsymbol{V}_{D}$ built
from experimental measurement covariances, and, in some cases, a contribution
$\boldsymbol{V}_{M}$ characterizing model errors
\begin{equation}
\boldsymbol{V}=\boldsymbol{V}_{D}+\boldsymbol{V}_{M}.\label{eq:var}
\end{equation}

In the Approximate Bayesian Computing approach, this expression of
the likelihood might be different, as discussed in \subsecref{ABC}.

\subsubsection{Validation}

Noting $\boldsymbol{\hat{R}}$ the residuals at the $MAP$, $\hat{\boldsymbol{\vartheta}}$,
one defines the mean squared residuals ($MSR$), mean residuals ($MR$),
root mean squared deviation ($RMSD$): 
\begin{eqnarray}
MSR & = & \frac{1}{N}\sum_{i=1}^{N}\hat{R}_{i}^{2},\label{eq:mse}\\
MR & = & \frac{1}{N}\sum_{i=1}^{N}\hat{R}_{i},\label{eq:me}\\
RMSD & = & \sqrt{MSR-MR^{2}}.\label{eq:rmsd}
\end{eqnarray}
A convenient validation statistics accounting for the covariance structure
of the statistical model is the Birge ratio $R_{B}$,\cite{Birge1932,Kacker2008}
\begin{eqnarray}
R_{B} & = & \frac{1}{N-\nu}\boldsymbol{\hat{R}}^{T}\boldsymbol{V}^{-1}\boldsymbol{\hat{R}},\label{eq:birge}
\end{eqnarray}
where $\nu$ is the number of degrees of freedom of the fit. $R_{B}$
is a dimensionless quantity, which should be close to $1$ when calibration
is statistically sound. A limitation of $R_{B}$ is that for complex
statistical models, such as hierarchical models or gaussian processes,
$\nu$ is not easily estimated.

Various prediction statistics can be used also for validation \cite{Vehtari2012,Gelman2013,Oliver2015}.
Posterior predictive assessment compares model predictions with reference
data and/or validation data. 

Using the calibration dataset $\boldsymbol{D}=\{x_{i},y_{i},u_{i};\,i=1,\,N\}$
for validation, one defines the mean prediction variance as 
\begin{equation}
MPV=\frac{1}{N}\sum_{i=1}^{N}\left(u_{M}^{2}(x_{i},\boldsymbol{D})+u_{i}^{2}\right),
\end{equation}
where $u_{M}(x_{i},\boldsymbol{D})$ is the standard deviation of
$p_{M}(\tilde{y}|x_{i},\boldsymbol{D})$ and $u_{i}$ is the measurement
uncertainty on $y_{i}$. The mean prediction uncertainty is noted
\begin{equation}
u_{e|D}=\sqrt{MPV}.
\end{equation}
In absence, or after correction, of data inconsistency, the residuals
result from model inadequacy and data uncertainty. In such cases,
one should thus expect that 
\begin{equation}
u_{e|D}\gtrsim RMSD.\label{eq:valid}
\end{equation}

As evoked in the introduction, mean statistics can hide underlying
problems, and it is always important to perform a visual check of
the prediction bands, built at a series of points $\tilde{x}$ in
control space from prediction uncertainties $u_{M}(\tilde{x},\boldsymbol{D})$
and $u_{e}(\tilde{x},\boldsymbol{D})$, the standard deviation of
$p_{M}(\tilde{y}|\tilde{x},\boldsymbol{D})$ and $p_{e}(\tilde{y}|\tilde{x},\boldsymbol{D})$,
respectively.\textcolor{purple}{}

\subsection{Implementation}

The bayesian models are implemented in \texttt{Stan} \cite{Gelman2015},
using the \texttt{rstan} interface package for \texttt{R} \cite{RTeam2015}.
\texttt{Stan} is a very flexible and efficient probabilistic programming
language to implement bayesian statistical models, including advanced
ones, such as Hierarchical Models, Gaussian Processes or Approximate
Bayesian Computation (ABC). 

The main outputs are samples of the posterior pdf and posterior predictive
pdf, from which statistics and plots can be generated in \texttt{R}.
We used the No-U-Turn sampler \cite{Hoffman2014} throughout this
study. Convergence of the sampling was assessed by examining the parameters
samples and the \emph{split Rhat} statistics provided by \texttt{rstan}.

Uniform prior pdfs have been used for location parameters, and log-uniform
for scaling parameters, unless stated explicitly. All models were
run with 4 parallel Markov Chains of 5000 iterations each, 1000 of
which are used as warm-up for the No-U-Turn sampler. The convergence
criteria and parameters statistics are thus estimated on four samples
of 4000 points.

The data and codes necessary to reproduce the results of the present
study are provided as Supporting Information.

\section{Calibration models\label{sec:Calibration-models}}

One considers a calibration set of $N$ data points issued from $N_{s}$
experimental series, $\boldsymbol{D}=\{d_{i};\,i=1,\,N\}$ with $d_{i}=(x_{i},y_{i},u_{i},\kappa_{i})$,
where $x_{i}$ and $y_{i}$ are the values of the control and measured
variables, $u_{i}$ is the corresponding measurement uncertainty\bibnote{The uncertainty on the control variable is assumed to be negligible in all experiments.},
and $\kappa_{i}\in1:N_{s}$ is the series index of datum $i$. An
experimental measurement series is noted $D^{(j)}$, and its cardinal
is noted $n^{(j)}$.

We first consider the simplest statistical model as a basis for the
definition and discussion of more elaborate models accounting for
data inconsistency and model inadequacy.

\subsection{Std: standard model}

The most basic model (named Std in the following) treats all data
in a single series ($N_{s}=1$), with equal and unknown uncertainty.
The set of calibration parameters comprises the model parameters $\boldsymbol{\vartheta}_{M}$
and a parameter $\vartheta_{V}=\sigma$, accounting for the dispersion
of residuals. Assuming independent errors with normal distribution
\begin{align}
y_{i} & =M(x_{i};\boldsymbol{\vartheta}_{M})+\epsilon_{i},\\
\epsilon_{i} & \sim\mathcal{N}(0,\sigma^{2}),
\end{align}
the elements of $\boldsymbol{R}$ and $\boldsymbol{V}$ (Eq.\,\ref{eq:likelihood})
are
\begin{eqnarray}
R_{i} & = & y_{i}-M(x_{i};\boldsymbol{\vartheta}_{M}),\\
V_{ij} & = & \sigma^{2}\delta(i-j).\label{eq:stdVar}
\end{eqnarray}

In absence of prior information on the parameters, this method is
akin to Ordinary Least Squares Regression (OLSR) \cite{Hibbert2016},
and the optimal value of the dispersion parameter is $\sigma\simeq RMSD$,
ensuring \emph{de facto} the statistical validity of the fit ($R_{B}\simeq1$)
(Eq. \ref{eq:birge}). 

In the general case of data with non-negligible measurement uncertainties,
there is an ambiguity about the meaning of $\sigma$ and its attribution
to experimental or model uncertainty, which is a source of problems
at the prediction level (Eqns.\,\ref{eq:conf}-\ref{eq:pred}) \cite{Frenklach2016,Kalyanaraman2016}.
The use of a single dispersion parameter should therefore be reserved
to cases where a single uncertainty source is strongly dominant: 
\begin{itemize}
\item if measurement errors are dominant, $\sigma$ should be considered
as an estimate of a uniform measurement uncertainty and used for prediction
in Eq.\,\ref{eq:pred}; 
\item if model inadequacy is the main error source, $\sigma$ should be
linked to model prediction error and used in Eq.\,\ref{eq:conf}. 
\end{itemize}
An easy mistake in the second case is to ignore $\sigma$ at the prediction
level and consider only model parameters uncertainty. The latter is
a decreasing function of $N$ and becomes rapidly negligible for large
calibration sets. In this scenario, model prediction by Eq.\,\ref{eq:conf}
is unable to explain the dispersion of the residuals ($u_{e|D}\ll RMSD$),
and the calibration model should not be validated.

\subsection{Experimental uncertainty and data inconsistency}

We present now various options to deal with random and systematic
errors in the calibration data, by adapting either the model, or the
data covariance matrix $\boldsymbol{V}_{D}$.

\subsubsection{WLS: independent data, known uncertainty}

A first refinement from the standard model includes experimental uncertainty
in the covariance matrix
\begin{eqnarray}
V_{D,ij} & = & u_{i}^{2}\delta(i-j),\label{eq:std-unc-var}
\end{eqnarray}
Without prior information on the parameters, this calibration model
(WLS) is similar to the Weighted Least Squares Regression (WLSR) method
\cite{Hibbert2016}. 

However, ``experimental uncertainty'' is a rather vague expression,
and the interpretation of $u_{i}$ has to be carefully considered:
\begin{itemize}
\item if it represents only random errors, systematic errors are being ignored
and, if necessary, should be introduced by one of the approaches detailed
below (Section \ref{subsec:Serial-data});
\item if systematic errors have been aggregated in the experimental uncertainty
of calibration data (not an uncommon practice, notably in chemical
kinetics), the hypothesis of independent errors is invalid and the
covariance of the errors has to be taken into account (Section \ref{par:Data-covariance}).
\end{itemize}
Ignoring these limitations leads to unreliable parameters uncertainty. 

Based on $u_{i}$, and their possible dependence on the control variable,
a model of expected measurement uncertainty for new measurements has
to be designed for the estimation of prediction uncertainty by Eq.\,\ref{eq:pred}.

\subsubsection{Serial data\label{subsec:Serial-data}}

When considering series of calibration data, one has to assume that
they have been previously corrected for all known systematic errors
\cite{GUM}. In such conditions, one has no a priori information on
the value of possible residual systematic errors, $s$, and one makes
the hypothesis that they are null in mean, with an unknown dispersion
$\tau$:
\begin{equation}
s\sim\mathcal{N}(0,\tau^{2}).\label{eq:spdf}
\end{equation}
For predictions, information about the systematic errors should be
incorporated at the level of $p_{e}$ (Eq.\,\ref{eq:pred}).

In practice, systematic errors can be handled in several ways detailed
below and summarized in Table~\ref{tab:Methods-data}. The methods
of the type ``Correction'' are expected to improve the residuals
(lowering the $RMSD$), unlike those of the ``Statistical'' type,
improving only the Birge ratio $R_{B}$.

\begin{table}
\noindent \begin{centering}
\begin{tabular}{lllllll}
\hline 
Type &  & Action &  & Acronym &  & Section\tabularnewline
\cline{1-1} \cline{3-3} \cline{5-5} \cline{7-7} 
Correction &  & Optimize one shift parameter per series &  & Shift &  & \ref{par:Data-shifting}\tabularnewline
 &  & Optimize model parameters per series &  & Hier &  & \ref{par:Local-parameters}\tabularnewline
\noalign{\vskip\doublerulesep}
Statistical &  & Optimize data covariance matrix &  & Cov &  & \ref{par:Data-covariance}\tabularnewline
\noalign{\vskip\doublerulesep}
 &  & Reweight data &  & Wgt &  & \ref{par:Wgt}\tabularnewline
\hline 
\end{tabular}
\par\end{centering}
\caption{Methods\label{tab:Methods-data} to handle systematic errors in serial
data }
\end{table}

\paragraph{Shift: data shifting.\label{par:Data-shifting}}

In order to account for intra-series systematic errors, and assuming
that these are independent of $x,$ one can assign a shift parameter
to each series and modify the model and residuals accordingly:
\begin{eqnarray}
y_{i} & = & M(x_{i};\boldsymbol{\vartheta}_{M})+s^{(\kappa_{i})}+\epsilon_{i}\\
R_{i} & = & y_{i}-M(x_{i};\boldsymbol{\vartheta}_{M})-s^{(\kappa_{i})}.
\end{eqnarray}
The covariance matrix remains the same as above (Eq. \ref{eq:std-unc-var}). 

The shift parameters can be dealt with in two ways:
\begin{enumerate}
\item they can be directly optimized, adding therefore $N_{s}$ parameters
to the calibration set. In absence of any information on a global
shift, they are assigned independent normal prior pdfs $s^{(\kappa_{i})}\sim\mathcal{N}(0,\tau^{2})$.
The value of $\tau$ could be estimated a priori, from an analysis
of the residuals of the standard model. Estimation of the shift factors
requires that each data series contains a substantial number of points.
\item they can be treated as realizations of a random variable $s$ (Eq.~\ref{eq:spdf}),
and the hyperparameter $\tau$ is optimized in addition to the shifts.
This approach has several names in the literature: hierarchical model,
random effects model... \cite{McElreath2015} Estimation of $\tau$
requires a substantial number of data series. 
\end{enumerate}
When the shift parameters can be, even partly, compensated by the
physical model parameters $\boldsymbol{\vartheta}_{M}$, parameter
identification might benefit from additional constraints. For instance
a sum-to-zero constraint \cite{Toman2009}
\begin{equation}
\sum_{\kappa_{i}=1}^{N_{s}}s^{(\kappa_{i})}=0,\label{eq:sumToZero}
\end{equation}
or constraints on the range and sign of the shifts \cite{Cailliez2011}.

\paragraph{Hier: local parameters per series.\label{par:Local-parameters}}

Improved fits of inconsistent data series can be also achieved by
using individual values of the model parameters for each data series.
Using a hierarchical description of the parameters similar to the
one used for shifts, the residuals become
\begin{eqnarray}
R_{i} & = & y_{i}-M(x_{i};\boldsymbol{\vartheta}_{M}^{(\kappa_{i})}),
\end{eqnarray}
where, for each series, the parameters are drawn from a multivariate
distribution. Considering a normal distribution for the present study,
one has
\begin{equation}
\boldsymbol{\vartheta}_{M}^{(j)}\sim\mathcal{N}\left(\boldsymbol{\mu}_{M},\boldsymbol{V}_{\vartheta_{M}}\right),
\end{equation}
for which the mean values $\boldsymbol{\mu}_{M}$ and covariance matrix
$\boldsymbol{V}_{\vartheta_{M}}$ elements have to be estimated (hyperparameters). 

This model has been introduced as model $\mathcal{M}_{H1}$ in Wu
\emph{et al.}\cite{Wu2015}, where local sets of Lennard-Jones parameters
were estimated for each data series. Absorption of experimental biases
in physical parameters uncertainty somewhat conflicts with the clear
separation of error sources and has still to be validated. 

\paragraph{Cov: data covariance.\label{par:Data-covariance}}

Another approach to systematic errors is to design a covariance matrix
for the data. This approach is in principle similar to the data shifting
model with integration (marginalization) over the shift parameters
(see Appendix \ref{sec:Correlation-and-weighting}). Assuming intra-series
correlation and inter-series independence, one gets a $N\times N$
block-diagonal covariance matrix, with elements
\begin{equation}
V_{D,ij}=u_{i}^{2}\delta(i-j)+\tau^{2}\delta(\kappa_{i}-\kappa_{j}).\label{eq:like_var-cov}
\end{equation}
An advantage of this method is that it requires the identification
of a single additional parameter, $\tau$. A disadvantage is that
it does not provide improved residuals, which becomes a problem when
it is combined with the inference of model inadequacy.

\paragraph{Wgt: data reweighting.\label{par:Wgt}}

For the statistical treatment of multiple data series, a weighting
scheme has been proposed \cite{Turanyi2012} by introducing for each
point in a data series a weight equal to the square root of the series
cardinal $\sqrt{n^{(j)}}$. This approach simulates the effect on
parameters uncertainty of the data covariance matrix in the limit
of very strong intra-series correlation, \emph{i.e. $\tau\gg u_{i}$}
(see Appendix \ref{sec:Correlation-and-weighting}). 

This was designed to remove a bias towards data series with the highest
number of points \cite{Turanyi2012}. The covariance matrix in this
case is therefore diagonal, with elements
\begin{eqnarray}
V_{D,ij} & = & n^{(\kappa_{i})}(u_{i}^{2}+\tau^{2})\delta(i-j).
\end{eqnarray}
If $\tau$ is unknown, this weighting scheme leads to a severe underestimation
of its value through calibration, and to an enhancement of the relative
weight of random errors. The consequences are (i) large MPU, and (ii)
too small experiment prediction uncertainty. 

This weighting scheme is to be reserved for data with aggregated systematic
uncertainty into $u_{i}$, which do not require the estimation of
$\tau$. It is mentioned here for the sake of completeness, but will
not be considered further in this study.

\subsection{Model Errors }

Random errors of stochastic models can be taken into account in the
same way as for experimental uncertainty through a covariance matrix,
$\boldsymbol{V}_{M}$. In the following sections, we focus on deterministic
models, and, assuming that numerical errors are negligible, we consider
only errors due to model inadequacy.

\begin{figure}
\begin{centering}
\includegraphics[bb=0bp 0bp 1200bp 1200bp,clip,width=1\textwidth]{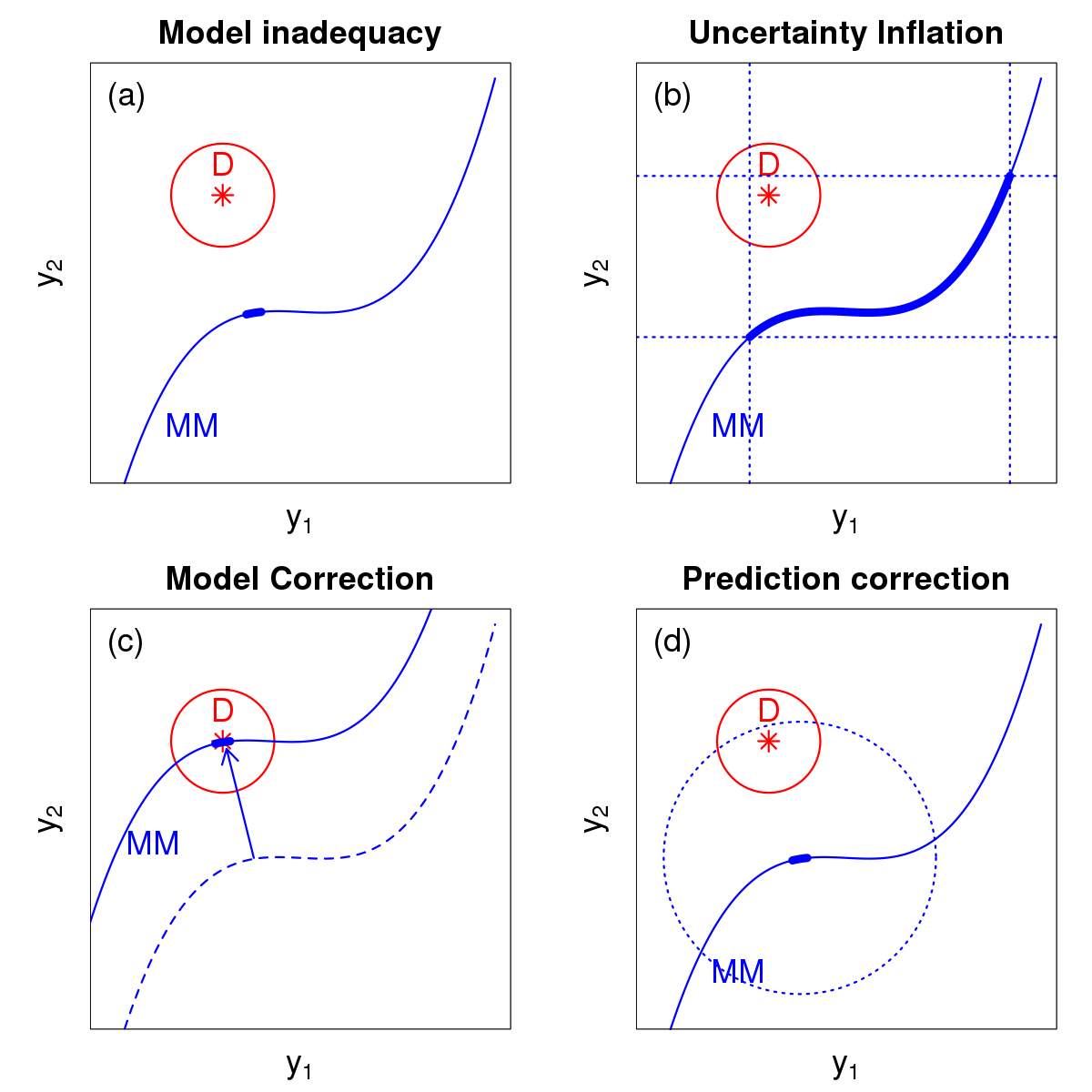}
\par\end{centering}
\caption{\label{fig:MM}Schematic representation of model inadequacy and correction
methods discussed in the text. A one-parameter model $M(x;\vartheta)$
is used to predict two values $y_{1}=M(x_{1};\vartheta)$ and $y_{2}=M(x_{2};\vartheta)$,
to be compared with reference data represented by a point in data
space, $D=\left\{ d_{1}\pm u_{1},d_{2}\pm u_{2}\right\} $ (red star).
\protect \\
(a) Model inadequacy occurs when the point in the model manifold (MM,
blue line) closest to $D$ (optimal parameter $\hat{\vartheta}\pm u_{\vartheta}$)
does not intersect with a high probability volume around $D$ (red
circle). \protect \\
(b) Without change in the model, one defines a space of parameters
values ($\boldsymbol{Q}=\left\{ \vartheta_{min}\le\vartheta\le\vartheta_{max}\right\} $)
which enables to reconcile model predictions with individual data
: $d_{1}\in M(x_{1};\boldsymbol{Q})$ and $d_{2}\in M(x_{2};\boldsymbol{Q})$
(PUI; Sect.~\ref{subsec:Model-variance-adaptation}). \protect \\
(c) The model can be corrected $M'=M+\delta M$ to remove inadequacy
(method GP; Sect.~\ref{par:Gaussian-Process}). \protect \\
(d) Model predictions can be statistically corrected to be compatible
with the data (method Disp; Sect.~\ref{par:Dispersion-parameter}). }
\end{figure}

In the $N$-dimensional \emph{data space,} the reference dataset is
represented by a point (Fig.~\ref{fig:MM}(a)). For a model depending
on $N_{\vartheta}$ parameters ($N_{\vartheta}<N$), the variation
of the parameters leads the model point in data space to explore a
``surface'' of dimension $N_{\vartheta}$, called the model manifold
(MM).\cite{Transtrum2011} The geometry of this surface depends on
the control points $x_{i}$. If one ignores prior information on the
parameters, the best fit is realized by the point on the model manifold
closest to the reference data point. In the presence of model inadequacy,
the distance between the data point and the best fit point is significantly
larger than data uncertainty, \emph{i.e.}, there is no intersection
between the model manifold and a high probability volume around the
data point (Fig.~\ref{fig:MM}(a)). 

Considering the structure of the problem, accounting for model errors
can be done along two directions, as depicted in Fig.~\ref{fig:MM}
and summarized in Table~\ref{tab:Methods-model}:
\begin{enumerate}
\item out of the model manifold, by adding a correction term, which might
either reduce the errors (GP; Fig.~\ref{fig:MM}(c)), or represent
them statistically (Disp; Fig.~\ref{fig:MM}(d)); and
\item within the model manifold, by increasing the MPU by adapting the covariance
matrix of its parameters (Fig.~\ref{fig:MM}(b)). In order to do
this, one has to replace the $N$-dimensional problem by a less-constrained
one, such as $N$ one-dimensional problems (Margin) or a statistics-matching
problem (ABC). One can also build a hierarchical model on data series
designed along the control space (HierC).
\end{enumerate}
\begin{table}
\noindent \begin{centering}
\begin{tabular}{lllllll}
\hline 
Type &  & Action &  & Acronym &  & Section\tabularnewline
\cline{1-1} \cline{3-3} \cline{5-5} \cline{7-7} 
Out of MM &  & Optimize dispersion parameter &  & Disp &  & \ref{par:Dispersion-parameter}\tabularnewline
 &  & Optimize discrepancy function &  & GP &  & \ref{par:Gaussian-Process}\tabularnewline
\noalign{\vskip\doublerulesep}
Within MM &  & Parameter uncertainty inflation &  &  &  & \tabularnewline
 &  & a/ Target prediction uncertainty &  &  &  & \tabularnewline
 &  & a.1/ Reweigh data (tweak $\boldsymbol{V}_{D}$) &  & VarInf &  & \ref{par:Variance-inflation}\tabularnewline
 &  & a.2/ Optimize $\boldsymbol{V}_{\vartheta_{M}}$ &  &  &  & \tabularnewline
 &  & a.2.i/ $N$ 1D problems &  & Margin &  & \ref{par:Margin}\tabularnewline
 &  & a.2.ii/ Statistics matching &  & ABC &  & \ref{subsec:ABC}\tabularnewline
 &  & b/ Optimize local parameters &  & HierC &  & \ref{subsec:Hier: local}\tabularnewline
\hline 
\end{tabular}
\par\end{centering}
\caption{\label{tab:Methods-model}Methods to handle model inadequacy. }
\end{table}

\subsubsection{Model correction out of the model manifold\label{subsec:Model-correction}}

\paragraph{GP: gaussian process.\label{par:Gaussian-Process}}

The model inadequacy can be treated by adding a correction term (the
so-called discrepancy function, $\delta M$) to the original model:
\begin{align}
y_{i} & =M(x_{i};\boldsymbol{\vartheta}_{M})+\delta M(x_{i};\boldsymbol{\vartheta}_{K})+\epsilon_{i}\\
R_{i} & =y_{i}-M(x_{i};\boldsymbol{\vartheta}_{M})-\delta M(x_{i};\boldsymbol{\vartheta}_{K})
\end{align}
This approach is subject to parameter identification problems if the
parameters of $M$ and $\delta M$ are optimized simultaneously without
strong prior information \cite{Arendt2012,OHagan2013,Brynjarsdottir2014}:
$\delta M$ can in principle correct any error due to misspecification
of $M$, which relaxes too strongly the constraints of the calibration
data on $\boldsymbol{\vartheta}_{M}$. 

The solution adopted in the present study is to constrain $\boldsymbol{\vartheta}_{M}$
with the posterior pdf resulting from an independent calibration of
$M$. In this case, $\delta M(.)$ is designed to fit the residuals
of $M(x;\boldsymbol{\vartheta}_{M})$ by a Gaussian Process (GP) \cite{Kennedy2001,Campbell2006}
of mean $0$ and covariance matrix
\begin{align}
V_{M,ij} & =\rho(x_{i},x_{j}),
\end{align}
based on a Gaussian kernel
\begin{equation}
\rho(u,v)=\alpha^{2}\exp(-\beta^{2}(u-v)^{2}).
\end{equation}
Both parameters $\alpha$ and $\beta$ have to be estimated in addition
to $\boldsymbol{\vartheta}_{M}$. 

The predictive posterior pdf can be derived in closed form expression
\begin{equation}
p_{M}(\tilde{y}|\tilde{x},\boldsymbol{D})=\mathcal{N}(\boldsymbol{\Omega}^{T}\boldsymbol{V}^{-1}\boldsymbol{y},\alpha^{2}-\boldsymbol{\Omega}^{T}\boldsymbol{V}^{-1}\boldsymbol{\Omega}),
\end{equation}
where $\boldsymbol{\Omega}_{i}=\rho(x_{i},\tilde{x})$ and $\boldsymbol{V}$
is defined by Eq.~\ref{eq:var}.

\paragraph{Disp: dispersion parameter.\label{par:Dispersion-parameter}}

One might also consider a simpler correction by taking 
\begin{eqnarray}
\delta M(x_{i};\sigma) & \sim & \mathcal{N}(0,\sigma^{2})\\
V_{M,ij} & = & \sigma^{2}\delta(i-j)\label{eq:stdUnc}
\end{eqnarray}
This version of model correction does not suffer from the parameters
identification problem of the GP approach. It is statistically justified
if there is no visible trend in the residuals (model errors randomly
distributed with respect to the control variable) \cite{Pernot2015},
but it can also be useful if one is interested in quantifying a representative
scale of model inadequacy without correcting it explicitly. 

For predictions, the dispersion parameter $\sigma$ has to be used
in Eq.\,\ref{eq:conf}, through the $f(\tilde{y}|\tilde{x},\boldsymbol{\vartheta}_{M})$
pdf. In this case, one has simply $f=\mathcal{N}(M(\tilde{x};\boldsymbol{\vartheta}_{M}),\sigma^{2})$.

\subsubsection{Parameter uncertainty inflation\label{subsec:Model-variance-adaptation}}

For deterministic models, the main source of prediction uncertainty
is parameter uncertainty. In the calibration process, parameter uncertainty
is a byproduct, but this can be transgressed by using it as a variable
to adjust MPU. In effect, parameter uncertainty can be tuned in order
to provide a prediction uncertainty $u_{M}$ which is large enough
to represent model inadequacy. We call this approach parameter uncertainty
inflation (PUI) \cite{Pernot2016}. As for the model correction approach,
this leads to a stochastic model, albeit with an internal/embedded
stochastic source \cite{Sargsyan2015}, opening the door to calibration
transferability. 

Several approaches have been proposed in the computational chemistry
literature: in one case, the covariance matrix $\boldsymbol{V}_{D}$
is scaled in order to produce suitable MPU (VarInf); in the other
cases, the covariance matrix of the parameters $\boldsymbol{V}_{\vartheta_{M}}$
is directly optimized, a more complex option with several variants
(Margin, ABC, HierC). Some of these PUI methods (VarInf, Margin \&
ABC) have been reviewed recently \cite{Pernot2016}, and are presented
here to ensure a self-contained framework. 

\paragraph{VarInf: variance inflation.\label{par:Variance-inflation}}

Scaling of the data covariance matrix is a simple way to increase
parameter uncertainty to a prescribed level. As discussed by Pernot
\cite{Pernot2016}, it has been used, for instance, in the calibration
of the mBEEF density functional,\cite{Brown2003,Frederiksen2004,Mortensen2005}
or, indirectly, in the estimation of the scaling factor uncertainty
for \emph{ab initio} properties statistical correction.\cite{Irikura2005,Irikura2009,Johnson2010} 

An \emph{empirical} likelihood is built by scaling the data covariance
matrix 
\begin{equation}
p(\boldsymbol{D}|\boldsymbol{\vartheta},T)\propto\left|T\boldsymbol{V}_{D}\right|^{-1/2}\exp\left(-\frac{1}{2T}\boldsymbol{R}^{T}\boldsymbol{V}_{D}^{-1}\boldsymbol{R}\right),\label{eq:likelihood-1}
\end{equation}
with a scaling factor $T$ to be determined.

Optimization of $T$ using this likelihood leads to the Birge ratio
procedure used in metrological inter-laboratory comparisons to reconcile
inconsistent data.\cite{Weise2000,Kacker2008,Bodnar2014} Assuming
an adequate model and misestimated data variances, this approach rescales
the data variances in order to get a valid statistical estimation
of the weighted data mean. In practice, $T$ is chosen so as to provide
a unit Birge ratio with Eq.~\ref{eq:likelihood-1}, \emph{i.e.},
$T=R_{B}^{0}$, where
\begin{equation}
R_{B}^{0}=\frac{1}{N-\nu}\boldsymbol{\hat{R}}^{T}\boldsymbol{V}_{D}^{-1}\boldsymbol{\hat{R}}.
\end{equation}

In the hypothesis of reliable data variances considered here, $T$
is chosen to compensate for model inadequacy and obtain valid prediction
statistics: 
\begin{itemize}
\item In the ensemble method,\cite{Brown2003,Frederiksen2004,Mortensen2005}
$T$ is chosen using a statistical mechanics analogy with a temperature,
leading to\cite{Pernot2016} 
\begin{align}
T & =\frac{N-\nu}{\nu}\,R_{B}^{0}.\label{eq:VarInf_Rb}
\end{align}
This approach is called VarInf\_Rb in the following.
\item Assuming a near-linear dependence of the model on its parameters in
their uncertainty range, an alternative estimation of $T$ can be
based on Eq.~\ref{eq:LUP-var}, in such a way that the mean variance
of model predictions reproduces the mean squared residuals \cite{Wellendorff2014}.
In the presence of non-negligible measurement uncertainty, one has
\begin{equation}
T\simeq\frac{MSR-\overline{u^{2}}}{MPV^{0}-\overline{u^{2}}},\label{eq:VarInf_MSE}
\end{equation}
using the mean prediction variance $MPV^{0}$ from a reference calibration
with $T=1$\emph{, }and $\overline{u^{2}}=\sum_{i}u_{i}^{2}/N$\emph{.
}This approach is called VarInf\_MSR in the following.
\end{itemize}
In all cases, prediction is done with Eq.~\emph{\ref{eq:conf}}.

\paragraph{Parameters covariance.}

In the direct approach, the model's parameters are considered as random
variables, with a pdf conditioned by a set of hyperparameters, typically
their mean values $\boldsymbol{\mu}_{\vartheta}$ and a covariance
matrix $\boldsymbol{V}_{\vartheta}$, defining a multivariate normal
distribution $p(\boldsymbol{\vartheta}_{M}|\boldsymbol{\mu}_{\vartheta_{M}},\boldsymbol{V}_{\vartheta_{M}})$. 

Such stochastic parameters can be handled in the bayesian inference
problem, either at the model level, leading to a stochastic model
within the standard likelihood framework (Eq.~\ref{eq:likelihood}),
or at the likelihood level. An indirect way to estimate $p(\boldsymbol{\vartheta}_{M}|\boldsymbol{\mu}_{\vartheta_{M}},\boldsymbol{V}_{\vartheta_{M}})$
is to use a hierarchical model with local sets of parameters $\boldsymbol{\vartheta}_{M}^{(\kappa_{i})}$
to represent complementary sections of the control parameter space
(HierC) \cite{Wu2015}. 

\subparagraph{Margin: Model level.\label{par:Margin}}

At the model level, one estimates the impact of stochastic parameters
on model predictions by uncertainty propagation\cite{GUMSupp1}
\begin{equation}
f_{M}(\boldsymbol{\xi};\boldsymbol{x},\boldsymbol{\mu}_{\vartheta_{M}},\boldsymbol{V}_{\vartheta_{M}})=\int\thinspace\prod_{i=1}^{N}\delta\left(\xi_{i}-M(x_{i};\boldsymbol{\vartheta}_{M})\right)\,p(\boldsymbol{\vartheta}_{M}|\boldsymbol{\mu}_{\vartheta_{M}},\boldsymbol{V}_{\vartheta_{M}})\,d\boldsymbol{\vartheta}_{M},
\end{equation}
where $f_{M}(.;\boldsymbol{x},\boldsymbol{\mu}_{\vartheta_{M}},\boldsymbol{V}_{\vartheta_{M}})$
is the multivariate pdf of the model's predictions at the vector of
control points $\boldsymbol{x}$. 

Inserting the resulting stochastic model in Eq.~\ref{eq:likelihood}
can be done by replacing $M(x_{i};\boldsymbol{\vartheta}_{M})$ by
the mean predictions (Eq.~\ref{eq:LUP-mean}) and their covariance
matrix $\boldsymbol{V}_{M}$ 
\begin{equation}
p(\boldsymbol{D}|\boldsymbol{\mu}_{\vartheta_{M}},\boldsymbol{V}_{\vartheta_{M}},\boldsymbol{\vartheta}_{V})\propto\left|\boldsymbol{V}_{D}+\boldsymbol{V}_{M}\right|^{-1/2}\exp\left(-\frac{1}{2}\boldsymbol{R}^{T}\left(\boldsymbol{V}_{D}+\boldsymbol{V}_{M}\right)^{-1}\boldsymbol{R}\right),\label{eq:likelihood-2}
\end{equation}
where
\begin{align}
\boldsymbol{V}_{M,ij} & \equiv u_{M}^{2}(x_{i},x_{j})=\boldsymbol{J}^{T}(x_{i};\boldsymbol{\mu}_{\vartheta_{M}})\boldsymbol{V}_{\vartheta_{M}}\boldsymbol{J}(x_{j};\boldsymbol{\mu}_{\vartheta_{M}}),\\
R_{i} & =y_{i}-\mu_{M|D}(x_{i}).
\end{align}
Note that using the full variance matrix of Eq.~\ref{eq:likelihood-2}
in the calculation of the Birge ratio (Eq.~\ref{eq:birge}) should
enable to validate the model with $R_{B}\simeq1$ by increasing the
variance without affecting the residuals.

For a deterministic model $M$, when the number of parameters is smaller
than the number of data points, $\boldsymbol{V}_{M}$ is singular
(non positive-definite), causing the likelihood to be degenerate,
and the calibration to be intractable \cite{Sargsyan2015}. By definition,
for inadequate models, the data covariance matrix is numerically too
small to alleviate the degeneracy problem. As all data points cannot
be reproduced \emph{simultaneously} by the model (Fig.~\ref{fig:MM}(b)),
one has to replace the multivariate problem by a set of univariate
problems (marginal likelihoods \cite{Sargsyan2015,Cheung2016}), \emph{i.e.},
one ignores the covariance structure of model predictions by taking
\begin{equation}
\boldsymbol{V}_{M,ij}=u_{M}^{2}(x_{i},x_{j})\delta(i-j).
\end{equation}

\subparagraph{ABC: Likelihood level.\label{subsec:ABC}}

A new likelihood, conditioned on the hyperparameters to be inferred,\cite{Oliver2015,Sargsyan2015}
is obtained by integration of the standard likelihood (Eq.~\ref{eq:likelihood})
over the possible values of the parameters (marginalization) 
\begin{equation}
p(\boldsymbol{D}|\boldsymbol{\mu}_{\vartheta_{M}},\boldsymbol{V}_{\vartheta_{M}},\boldsymbol{\vartheta}_{V})=\int p(\boldsymbol{D}|\boldsymbol{\vartheta}_{M},\boldsymbol{\vartheta}_{V})p(\boldsymbol{\vartheta}_{M}|\boldsymbol{\mu}_{\vartheta_{M}},\boldsymbol{V}_{\vartheta_{M}})\,d\boldsymbol{\vartheta}_{M}.\label{eq:integLik}
\end{equation}

As in the previous case, it is pointed out by Sargsyan \emph{et al.}\cite{Sargsyan2015}
that this likelihood is in general degenerate, so that the inference
problem has to be solved by alternative methods, such as Approximate
Bayesian Computation (ABC).\cite{Csillery2010,Sunnaker2013} In this
case, the full likelihood (Eq.~\ref{eq:integLik}) is replaced by
a tractable expression, based on summary statistics of the model predictions,
to be compared with similar statistics of the data. An example is
provided in Sargsyan \emph{et al.,}\cite{Sargsyan2015} where the
mean value of the model and its prediction uncertainty are used. A
version adapted to the present problem, with an explicit treatment
of experimental uncertainty is used here:
\begin{equation}
p_{ABC}(\boldsymbol{D}|\boldsymbol{\mu}_{\vartheta_{M}},\boldsymbol{V}_{\vartheta_{M}},\boldsymbol{\vartheta}_{V})\propto\exp\left(-\frac{1}{2}\boldsymbol{R}^{T}\boldsymbol{V}_{D}^{-1}\boldsymbol{R}\right)\times p_{reg}(\boldsymbol{D}|\boldsymbol{\mu}_{\vartheta_{M}},\boldsymbol{V}_{\vartheta_{M}})
\end{equation}
where the first term has the same expression as the standard likelihood
(Eq.~\ref{eq:likelihood}) using residuals evaluated at the mean
of the model prediction, and the second term ensures that the predicted
model uncertainty $u_{M}(x_{i})$, combined with experimental uncertainty
$u_{i}$, is of a magnitude compatible with the residuals
\begin{equation}
p_{reg}(\boldsymbol{D}|\boldsymbol{\mu}_{\vartheta_{M}},\boldsymbol{V}_{\vartheta_{M}})=\exp\left(-\sum_{i=1}^{N}\frac{\left(\sqrt{u_{M}^{2}(x_{i})+u_{i}^{2}}-|R_{i}|\right){}^{2}}{2u_{i}^{2}}\right).\label{eq:ABC}
\end{equation}
As evidenced in our notation, this term can also be seen as a regularization
function, necessary to constrain the parameters covariance matrix
$\boldsymbol{V}_{\vartheta_{M}}$ in the inference process. The constraint
imposed here is a statistical variant of Eq.~\ref{eq:valid}.

\subparagraph{HierC: local parameters in control space\label{subsec:Hier: local}}

Model inadequacy can sometimes be seen as a consequence of the use
of a unique parameter set in different experimental conditions: a
``better'' model can indeed be obtained by using different values
of the parameters along the control space. This can be achieved by
modeling the dependence of the physical parameters on the control
variable \cite{Zarkova1996,Zarkova2006,Zarkova2009}, or by splitting
the data in series along the control space and using a hierarchical
model identical to the one in Section~\ref{par:Local-parameters}
for inference of the hyperparameters describing the model's parameters
distribution (model $\mathcal{M}_{H2}$ in Wu \emph{et al.} \cite{Wu2015}).
To differentiate both hierarchical schemes, the present one is named
HierC.

\subsection{Combined methods}

Dealing with systematic measurement errors and model inadequacy errors
requires to combine the methods exposed above. Combined methods are
noted by joining the acronyms of the components, \emph{e.g.} GP-Shift,
Hier-Cov...

Not all combinations are favorable: for methods like GP and ABC, which
rely on a good estimation of model inadequacy residuals, the explicit
correction of systematic measurement errors by the Shift method is
the most coherent option. 

The Hier method can accommodate the Shift and Cov approaches, but
care has to be taken with identification issues between shift and
model parameters. This will be illustrated in Section~\ref{subsec:SD-3}.

\section{Applications\label{sec:Applications}}

Before applying the statistical calibration/prediction models exposed
above to reference data, we adopt a step-wise approach on synthetic
data to illustrate their advantages and drawbacks. The synthetic data
are designed to mimic the main features of a dataset of temperature-dependent
measurements of Krypton viscosity. 

\subsection{Viscosity model\label{sec:Viscosity-model}}

The following Chapman-Enskog model provides the viscosity $\eta$
in $\mu$Pa.s \cite{Cailliez2011}, when the inputs are in K ($T$
and $\epsilon_{LJ}$) and $\textrm{\AA}$ ($\sigma_{LJ}$): 
\begin{eqnarray}
\eta & = & M(T;\epsilon_{LJ},\sigma_{LJ})=2.6693\frac{\sqrt{mT}}{\sigma_{LJ}^{2}\Omega}\\
\Omega & = & \frac{A}{(T^{*})^{B}}+\frac{C}{\exp(DT^{*})}+\frac{E}{\exp(FT^{*})}
\end{eqnarray}
with $T^{*}=T/\epsilon_{LJ}$, $m=83.978$~g/mol, the molar mass
of Kr, and dimensionless coefficients $A=1.16145$, $B=0.14874$,
$C=0.52487$, $D=0.77320$, $E=2.16178$, and $F=2.43787$.

\subsection{Synthetic data}

Synthetic data are generated by random sampling of systematic errors,
$s$, and random errors, $\epsilon$, in the generative model
\begin{equation}
y_{i}=M(x_{i};\boldsymbol{\vartheta}_{0})+s^{(\kappa_{i})}+\epsilon_{i}\label{eq:Generative-SD}
\end{equation}
with reference LJ parameters $\boldsymbol{\vartheta}_{0}=\left\{ \epsilon_{LJ}:\thinspace195\,\textrm{K},\thinspace\sigma_{LJ}:\thinspace3.6\,\lyxmathsym{\AA}\right\} $. 

Random errors are modeled as an additive heteroscedastic noise 
\begin{equation}
\epsilon_{i}\sim\mathcal{N}(0,f_{i}\times y_{i}),
\end{equation}
with uncertainty factor
\begin{equation}
f_{i}=f_{0}*\exp\left(g\left|\frac{1}{x_{i}}-\frac{1}{300}\right|\right)
\end{equation}
where $f_{0}=0.001$ and $g=50$\,K. This functional shape enforces
that the relative measurement uncertainties are smaller at room temperature.

Systematic errors are generated from a normal distribution of mean
$0$ and standard deviation to be specified.

As the generative model for experimental uncertainty is unknown in
the analysis of experimental data, one has to build an approximation
for prediction of experimental data. In the following, one uses the
mean relative uncertainty calculated on the calibration dataset $\overline{u}_{r}=1/N\sum u_{i}/y_{i}$,
multiply it by the model value at the prediction point and ensures
that the result does not get below the minimal experimental uncertainty
in the dataset, $u_{min}$:
\begin{equation}
p(\tilde{y_{e}}|\tilde{y})=\mathcal{N}(\tilde{y},\mathrm{max}(u_{min},\,\tilde{y}\times\overline{u}_{r})).
\end{equation}

\subsubsection{SD-1: model inadequacy\label{subsec:SD-1:-model-inadequacy}}

A first dataset of size $N=100$ is generated by simulating 10 series
of data with cardinals between 5 and 15 and with overlapping temperature
ranges\emph{. }The control points $x_{i}$ are located between $T=120$
and $2000$~K.\emph{ }The series are consistent\emph{, i.e.} $s^{(\kappa_{i})}=0$
in the generative model, Eq. \ref{eq:Generative-SD}. The standard
deviation of the random noise in this set is about $0.07$\,$\mu$Pa.s.

To simulate model inadequacy, one uses for data analysis a modified
version of the Chapman-Enskog model (Section \ref{sec:Viscosity-model}),
in which the constant $C$ was replaced by $C'=C*0.5$. The standard
deviation from the true model due to this perturbation before model
calibration is $0.55$\,$\mu$Pa.s.

\begin{figure}
\begin{centering}
\begin{tabular}{ccc}
\includegraphics[bb=0bp 0bp 1200bp 1200bp,clip,height=5cm]{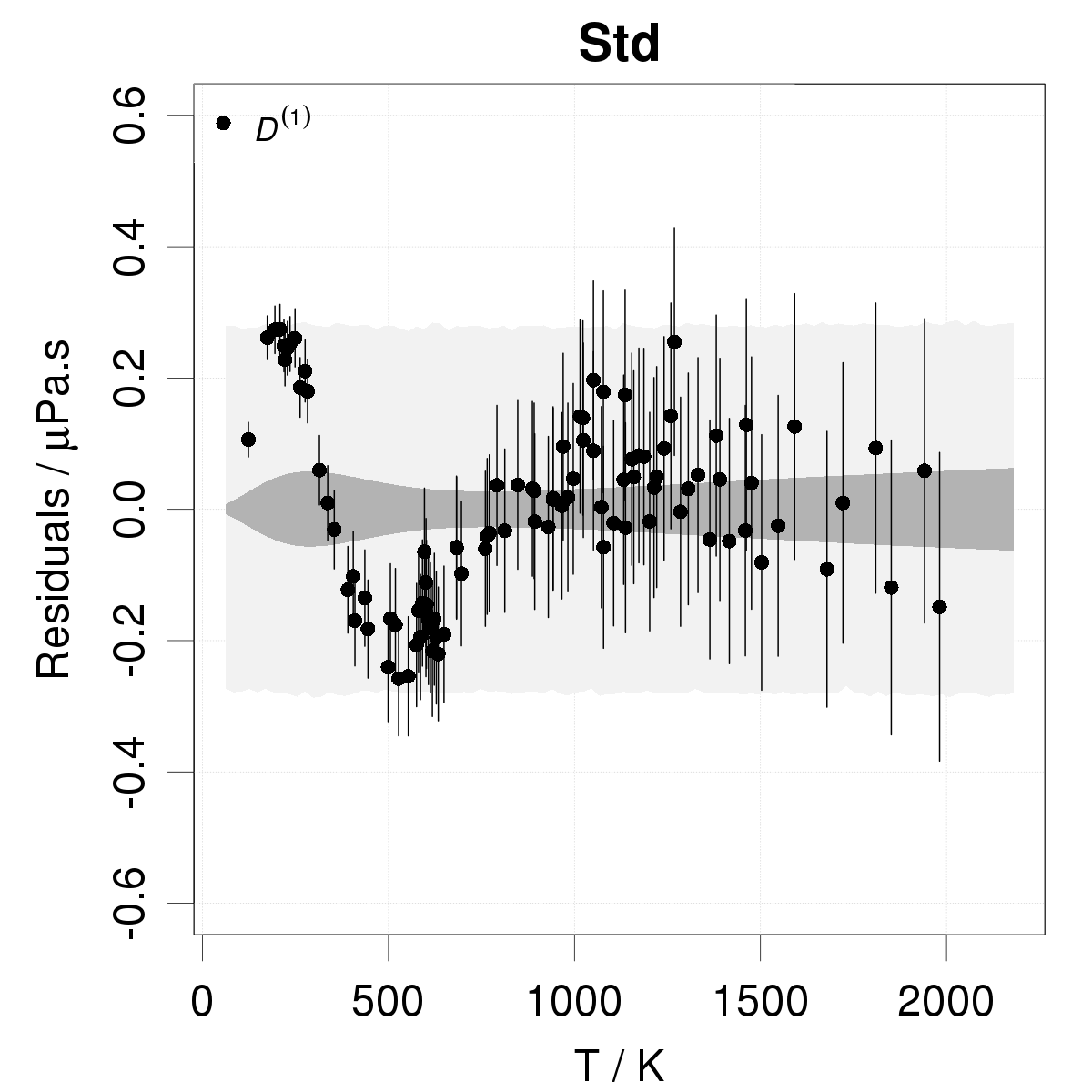} & \includegraphics[bb=0bp 0bp 1200bp 1200bp,clip,height=5cm]{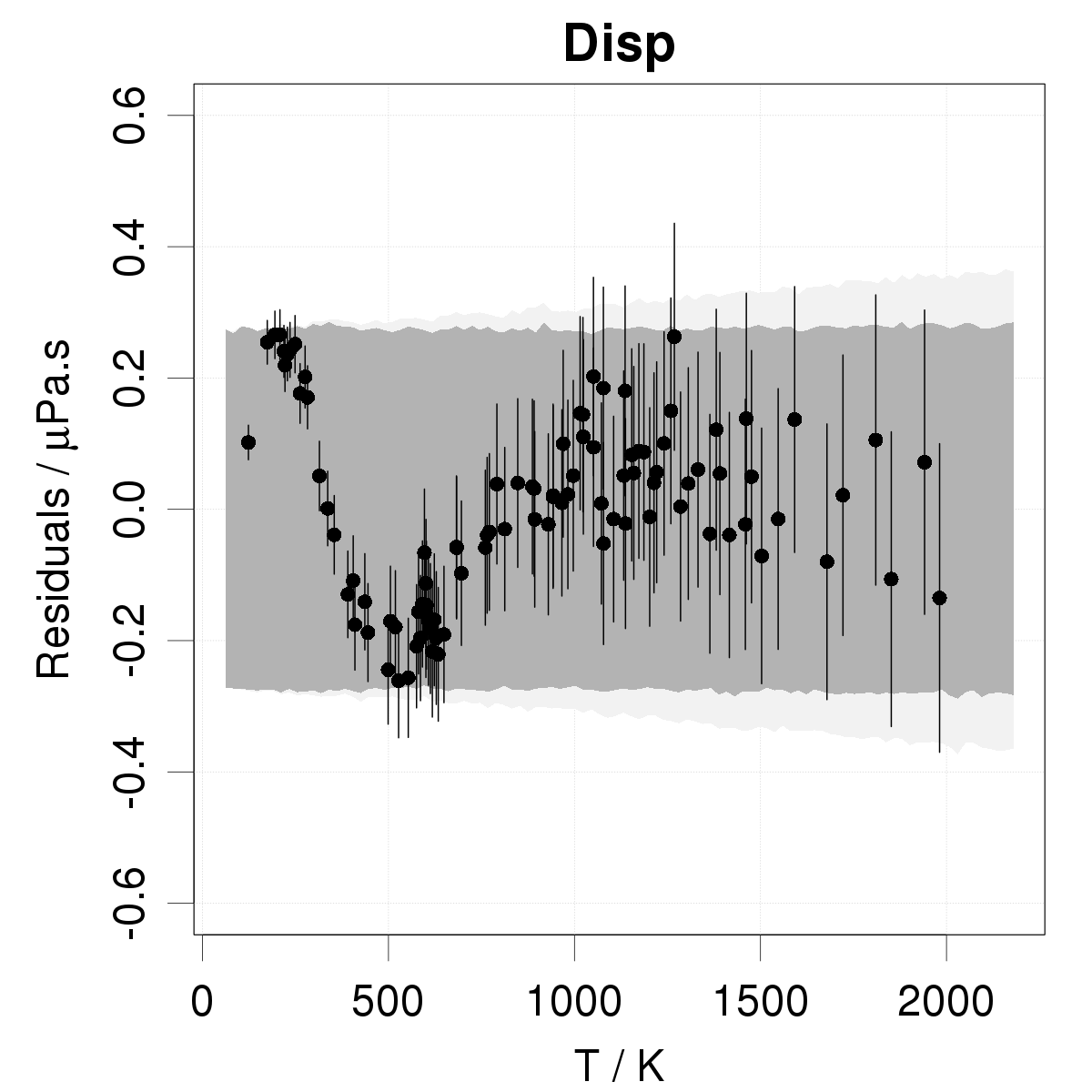} & \includegraphics[bb=0bp 0bp 1200bp 1200bp,clip,height=5cm]{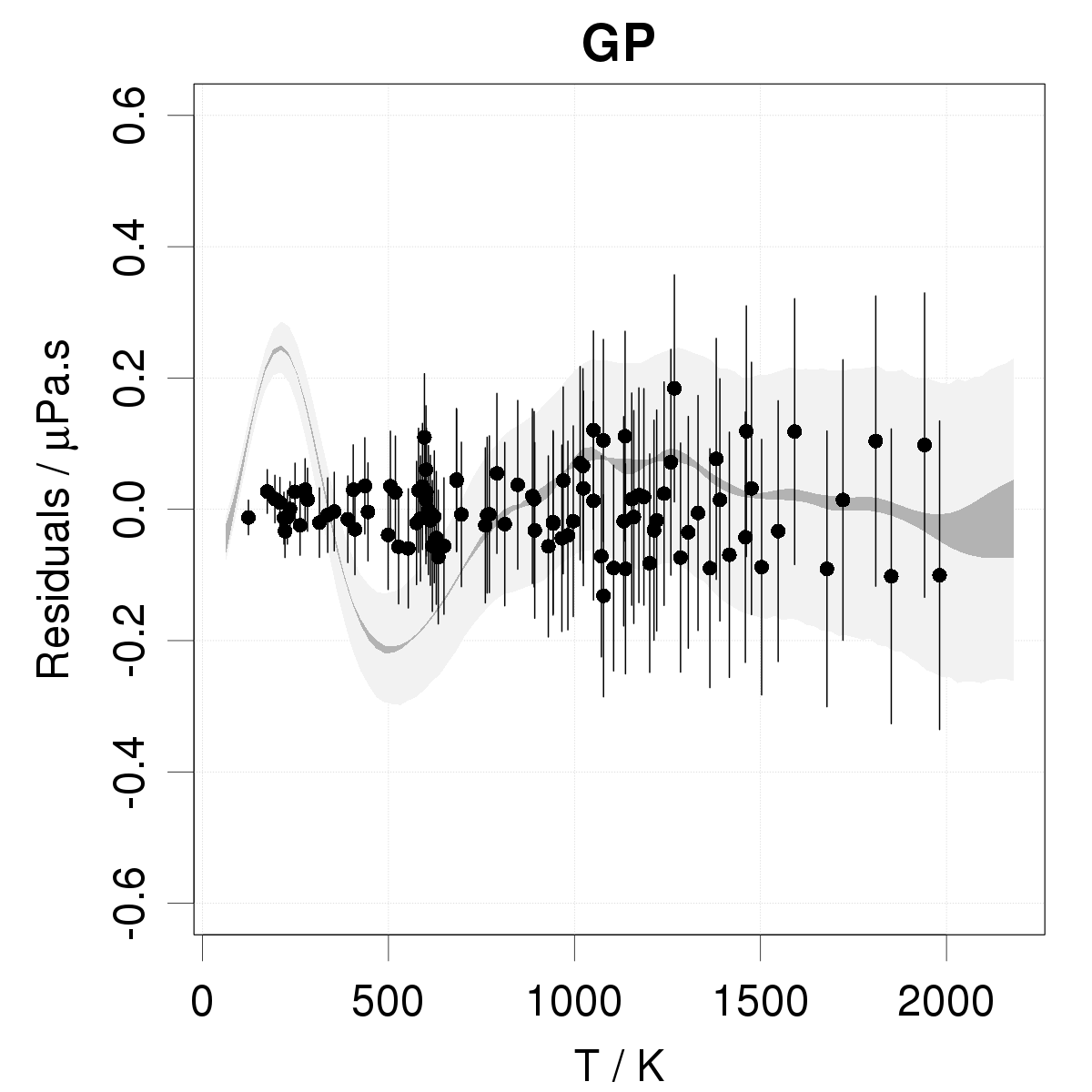}\tabularnewline
\includegraphics[bb=0bp 0bp 1200bp 1200bp,clip,height=5cm]{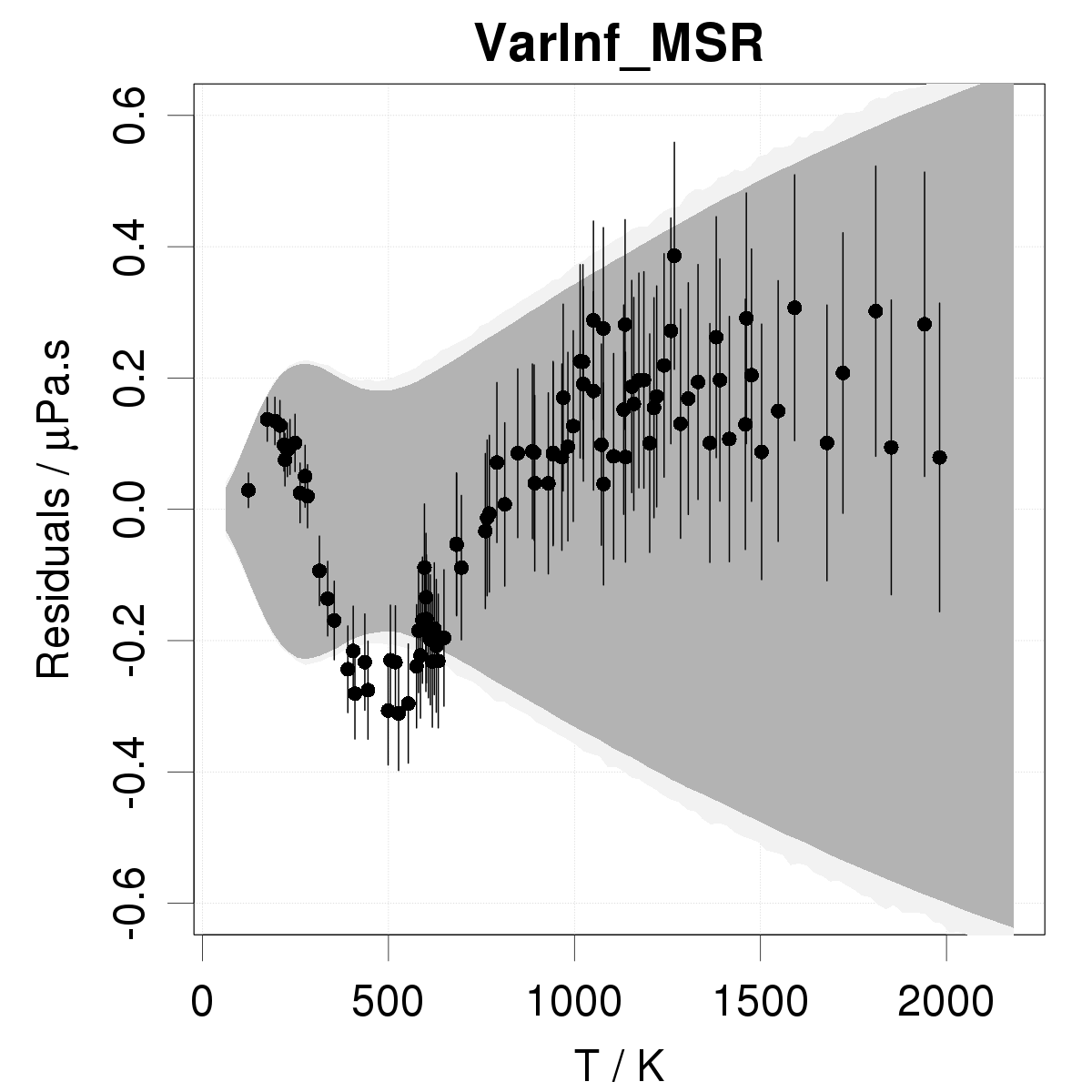} & \includegraphics[bb=0bp 0bp 1200bp 1200bp,clip,height=5cm]{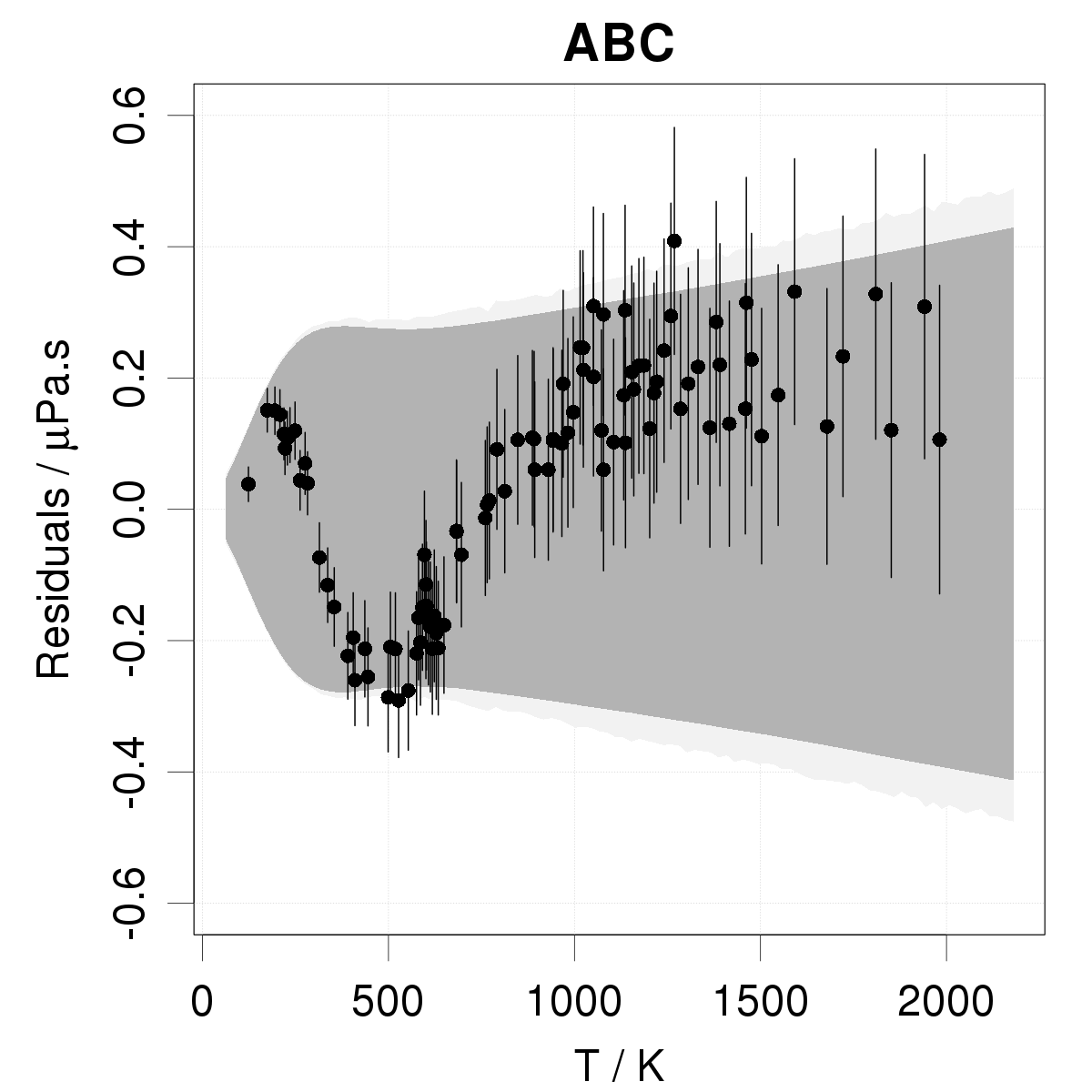} & \includegraphics[bb=0bp 0bp 1200bp 1200bp,clip,height=5cm]{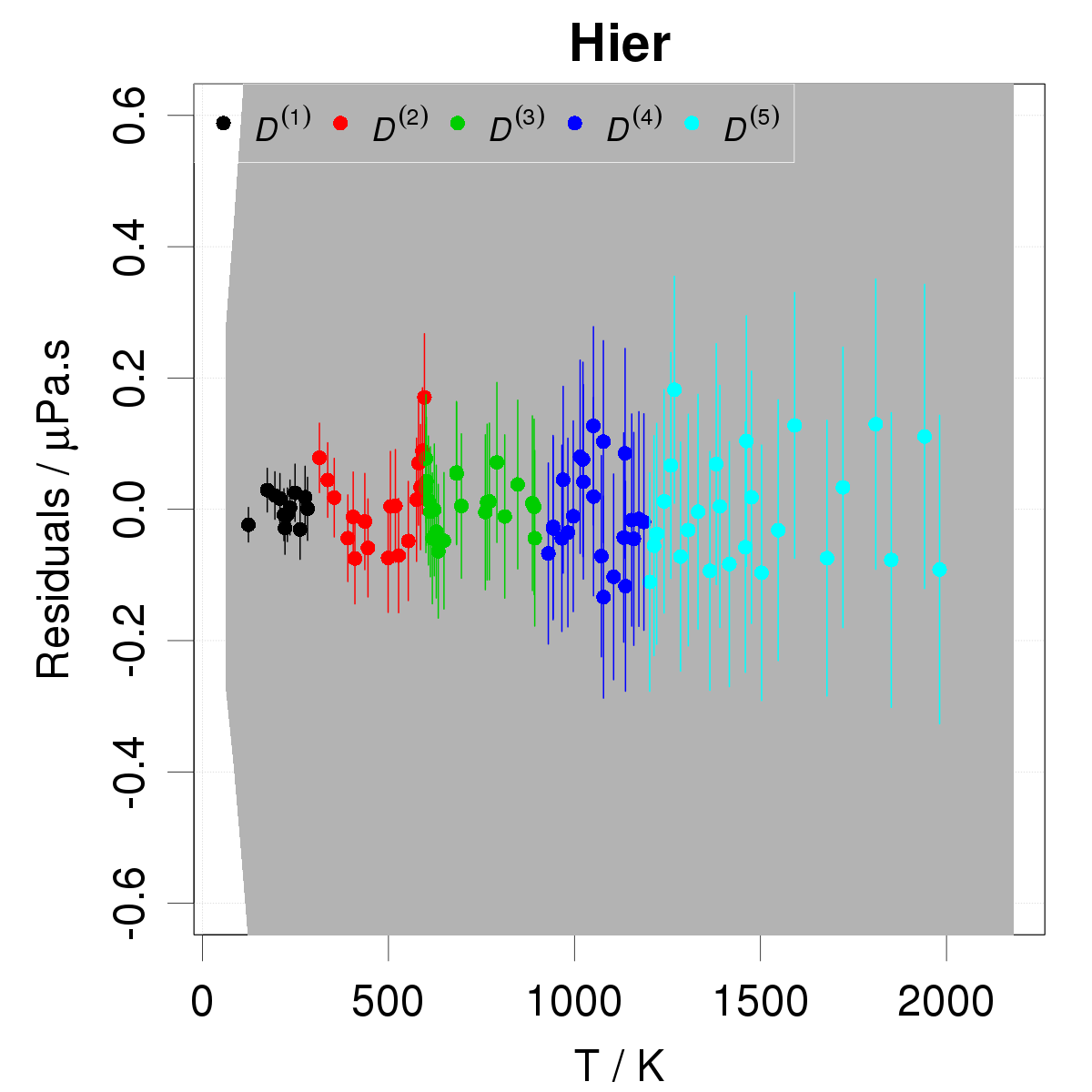}\tabularnewline
\includegraphics[bb=0bp 0bp 1200bp 1200bp,clip,height=5cm]{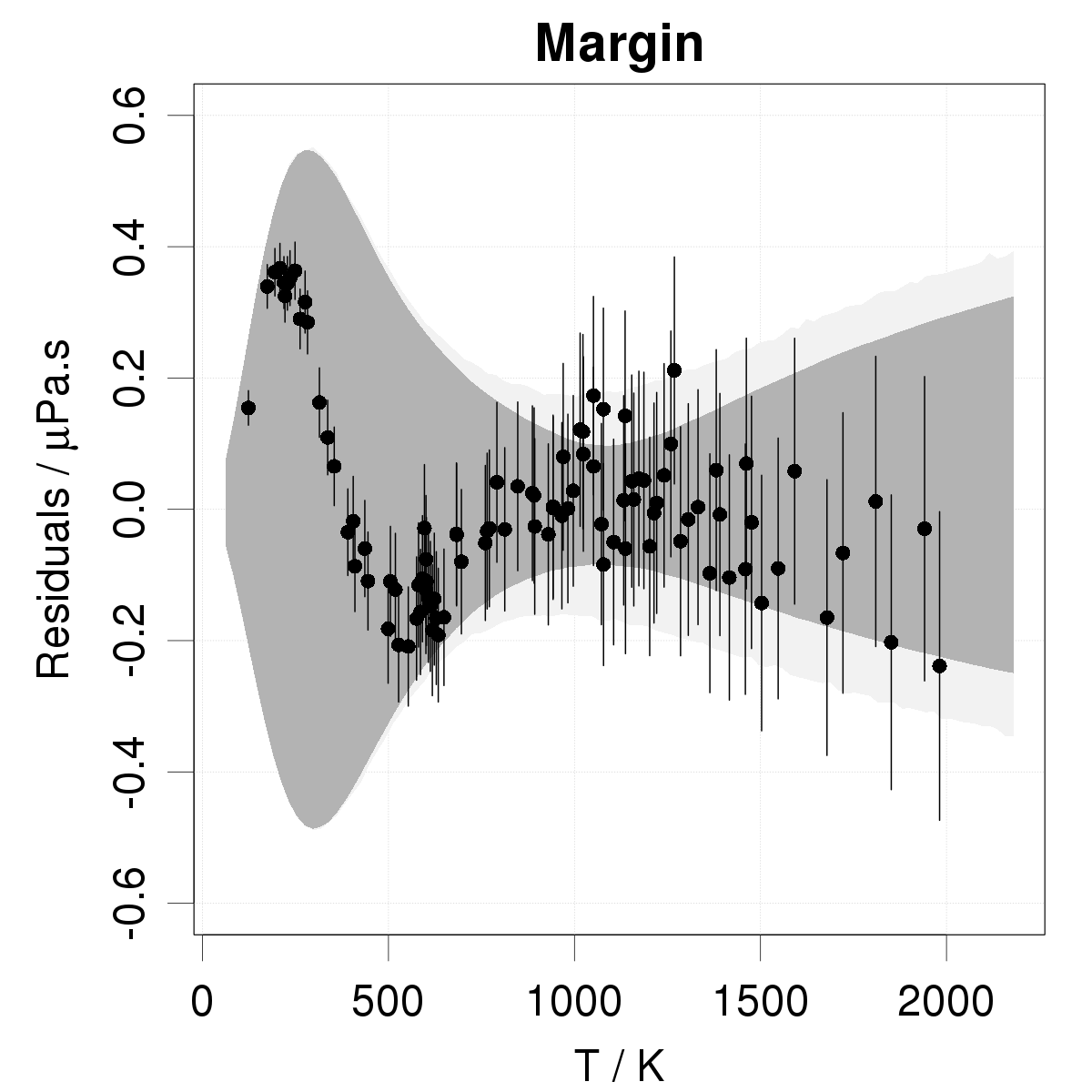} &  & \tabularnewline
\end{tabular}
\par\end{centering}
\caption{\label{fig:Predict-Synth1}Calibration residuals and posterior prediction
intervals for data SD-1. For each model, the results are centered
on the maximum a posteriori results (MAP). The dark gray band represents
the model prediction 95\,\% probability interval ($p_{M}$) and the
light gray area the experiment prediction 95\,\% probability interval
($p_{e}$). }
\end{figure}

\begin{figure}
\begin{centering}
\begin{tabular}{ccl}
\includegraphics[bb=0bp 0bp 1200bp 1200bp,clip,height=5cm]{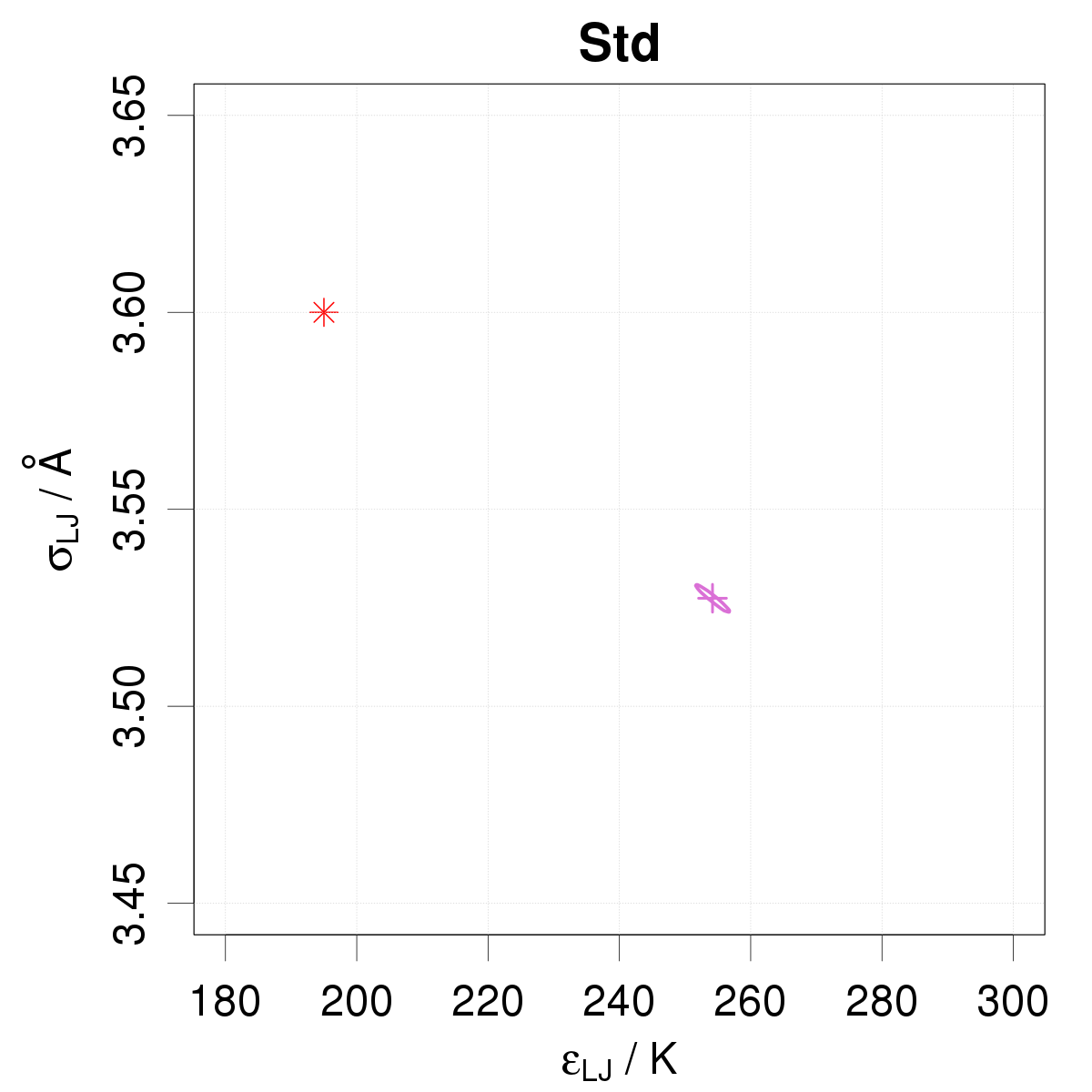} & \includegraphics[bb=0bp 0bp 1200bp 1200bp,clip,height=5cm]{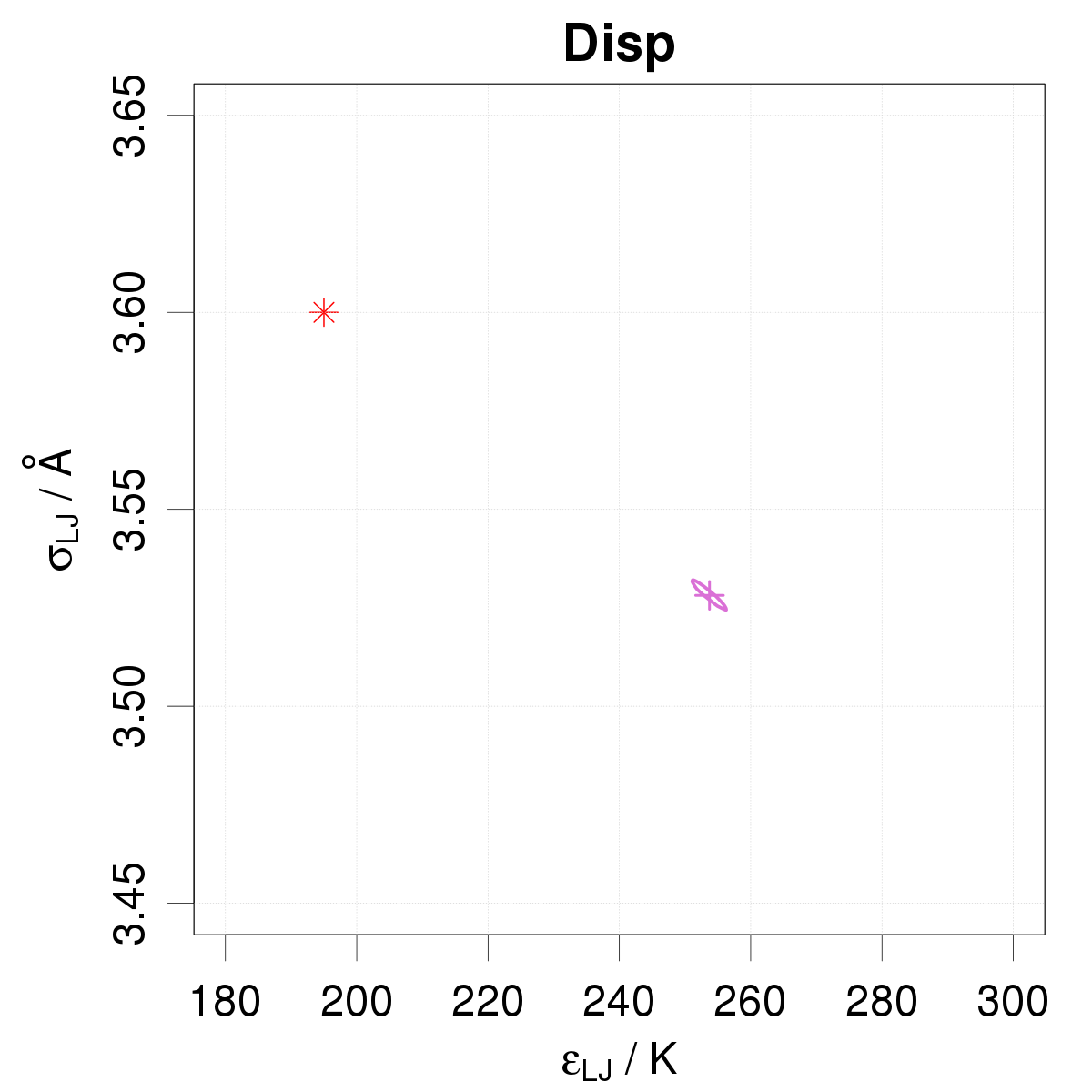} & \includegraphics[bb=0bp 0bp 1200bp 1200bp,clip,height=5cm]{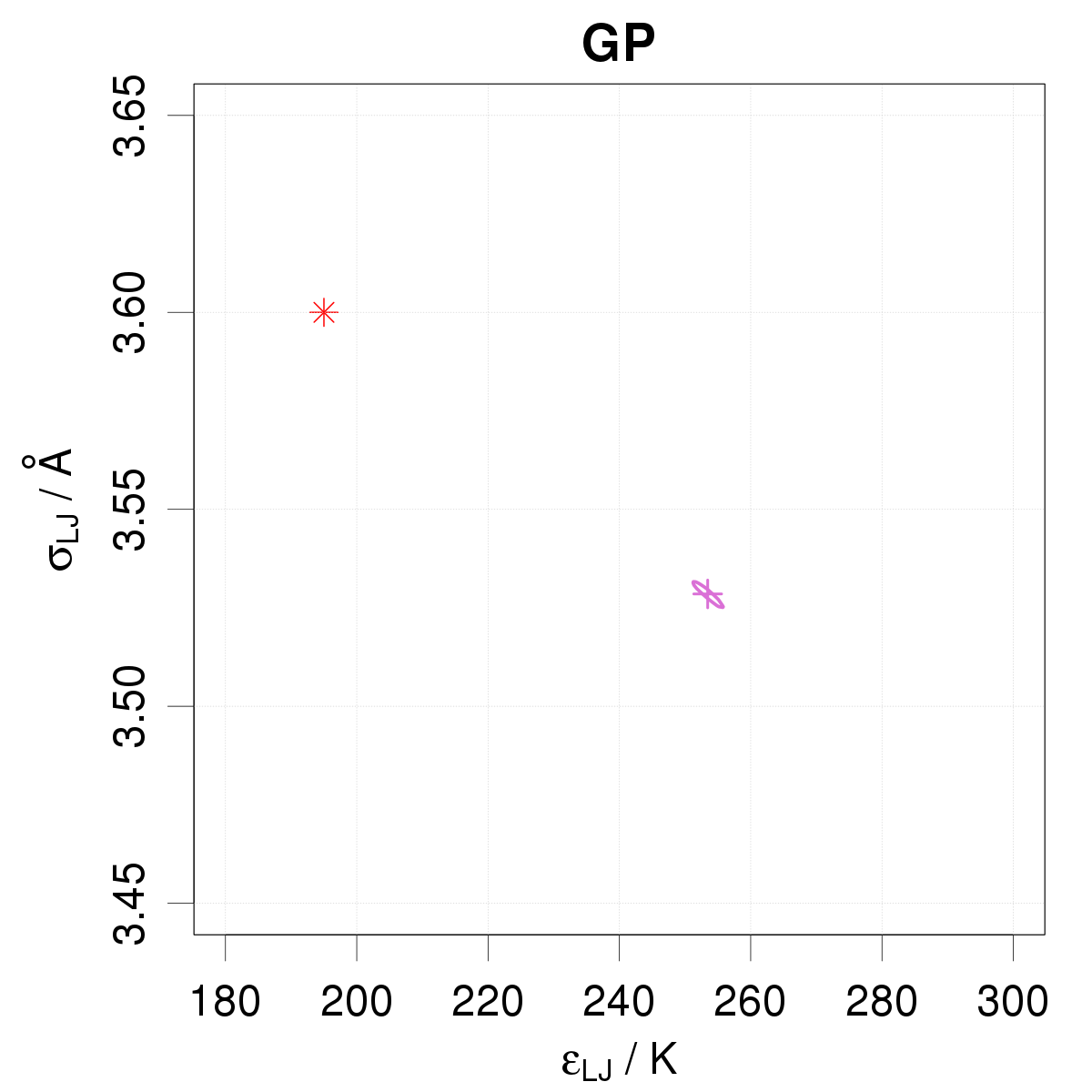}\tabularnewline
\includegraphics[bb=0bp 0bp 1200bp 1200bp,clip,height=5cm]{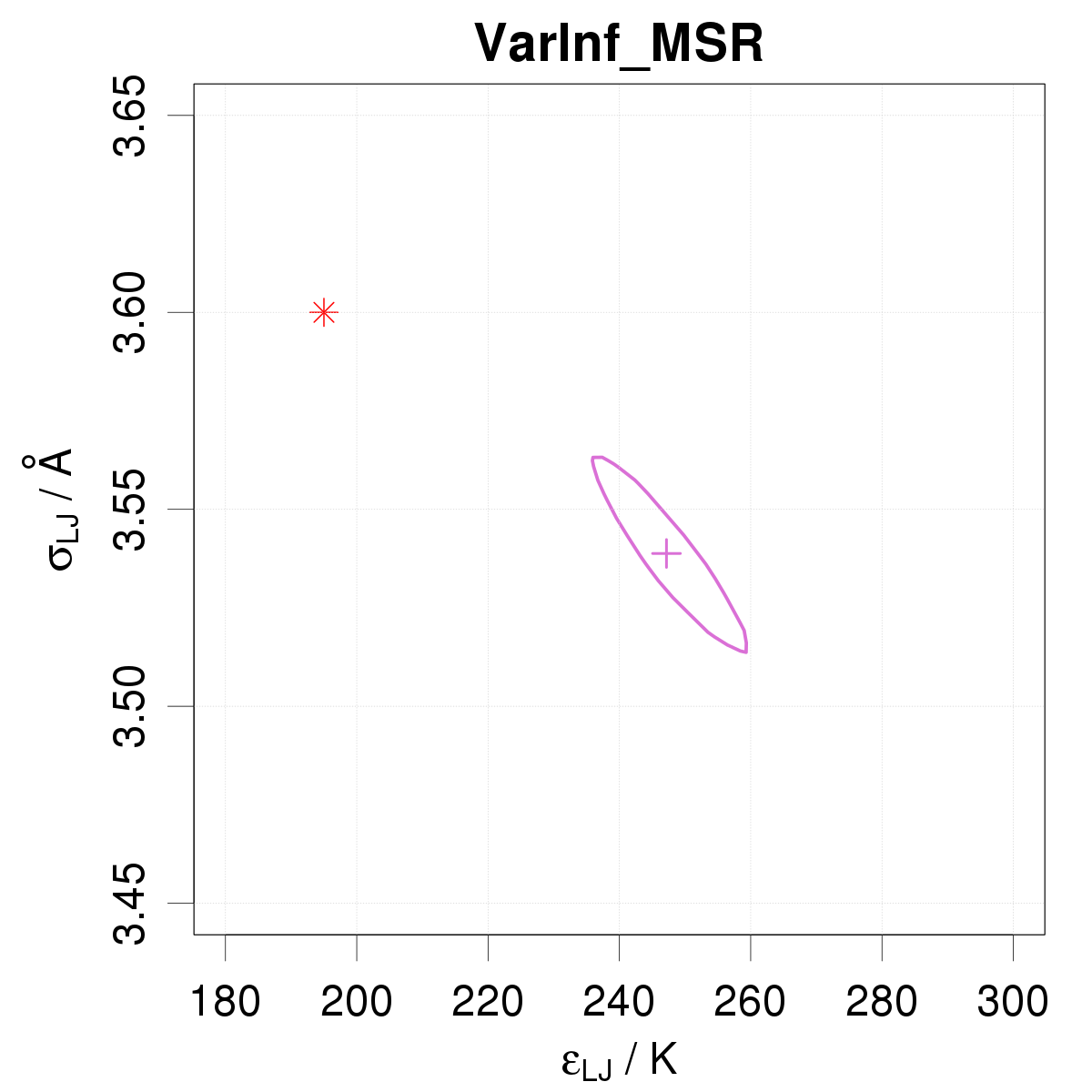} & \includegraphics[bb=0bp 0bp 1200bp 1200bp,clip,height=5cm]{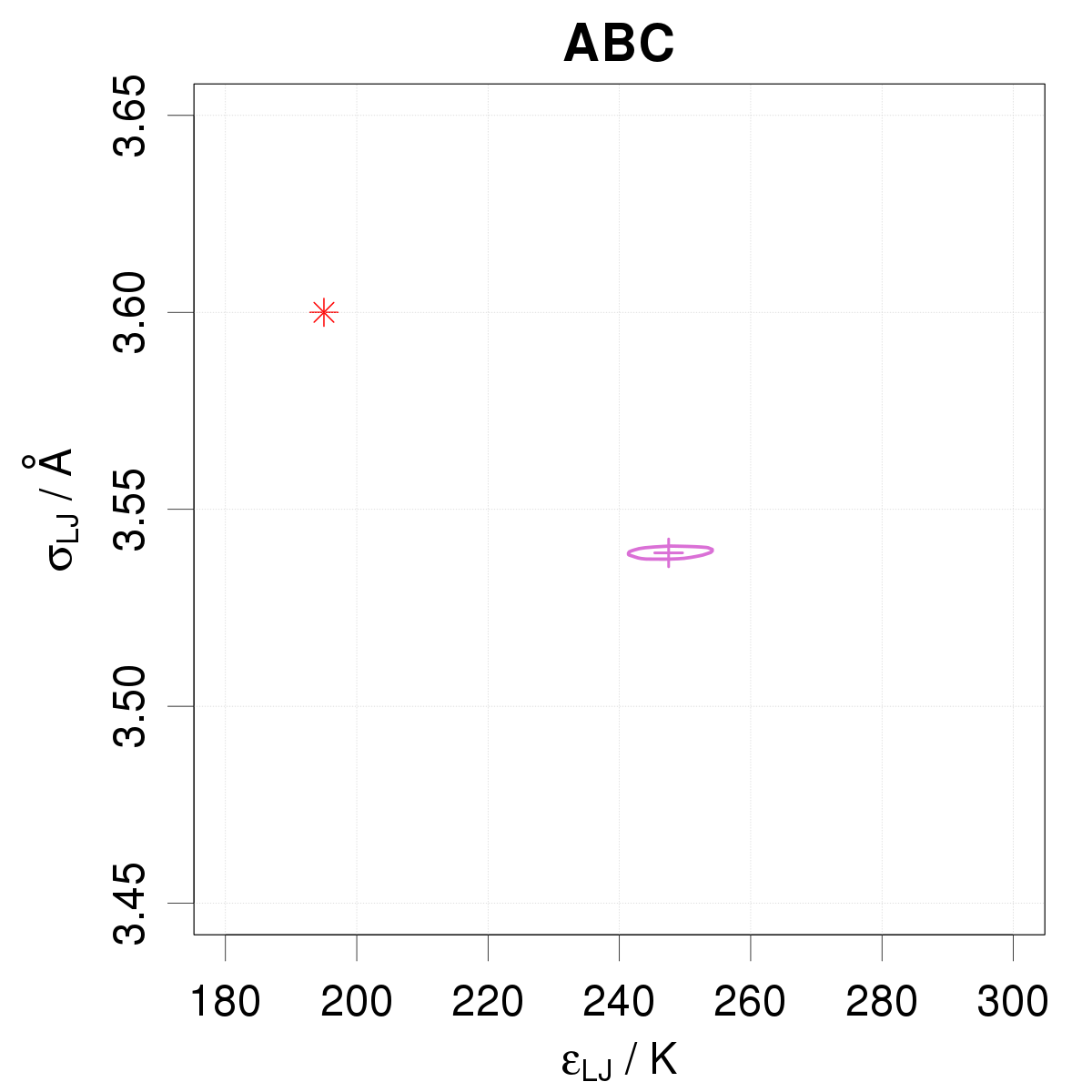} & \includegraphics[bb=0bp 0bp 1200bp 1200bp,clip,height=5cm]{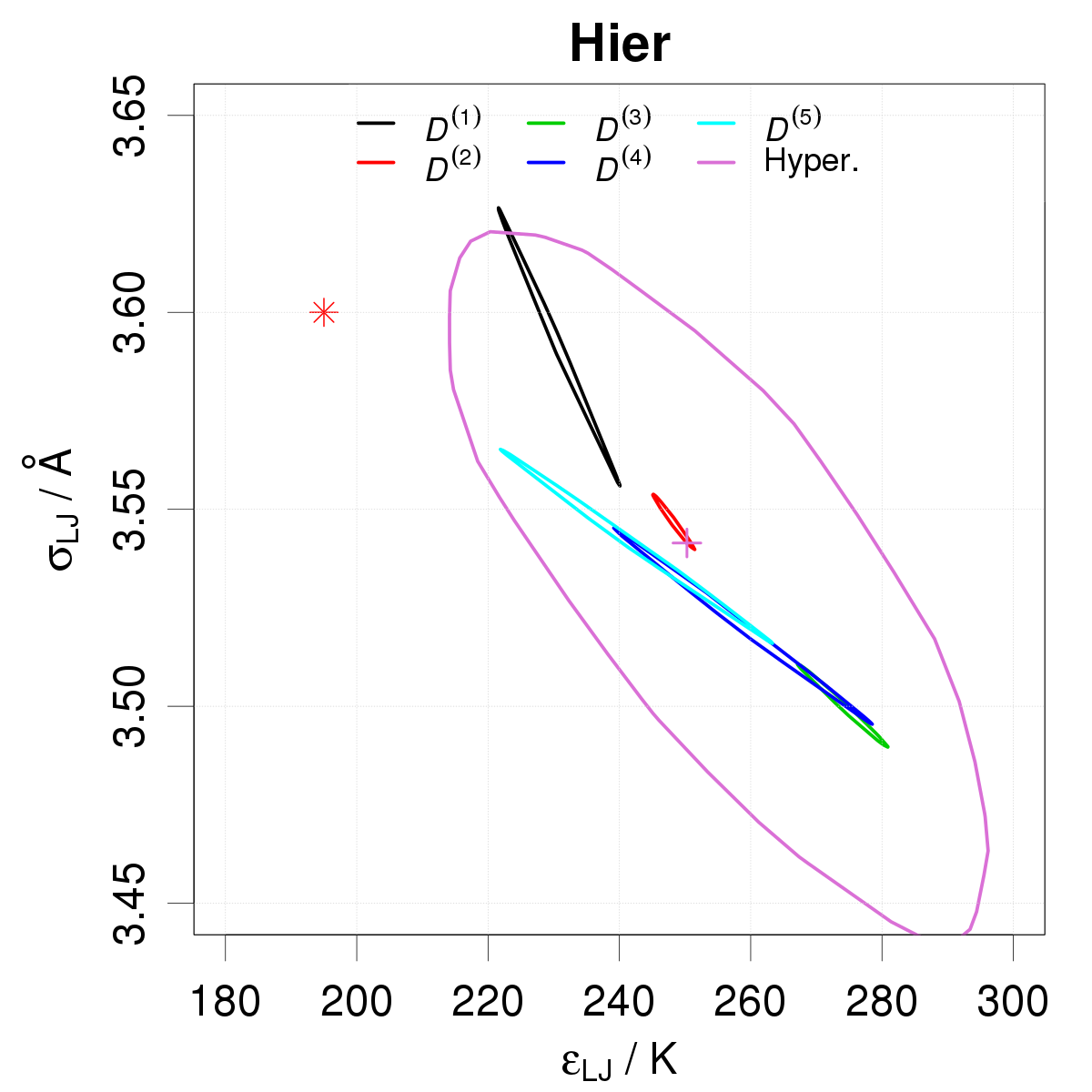}\tabularnewline
\includegraphics[bb=0bp 0bp 1200bp 1200bp,clip,height=5cm]{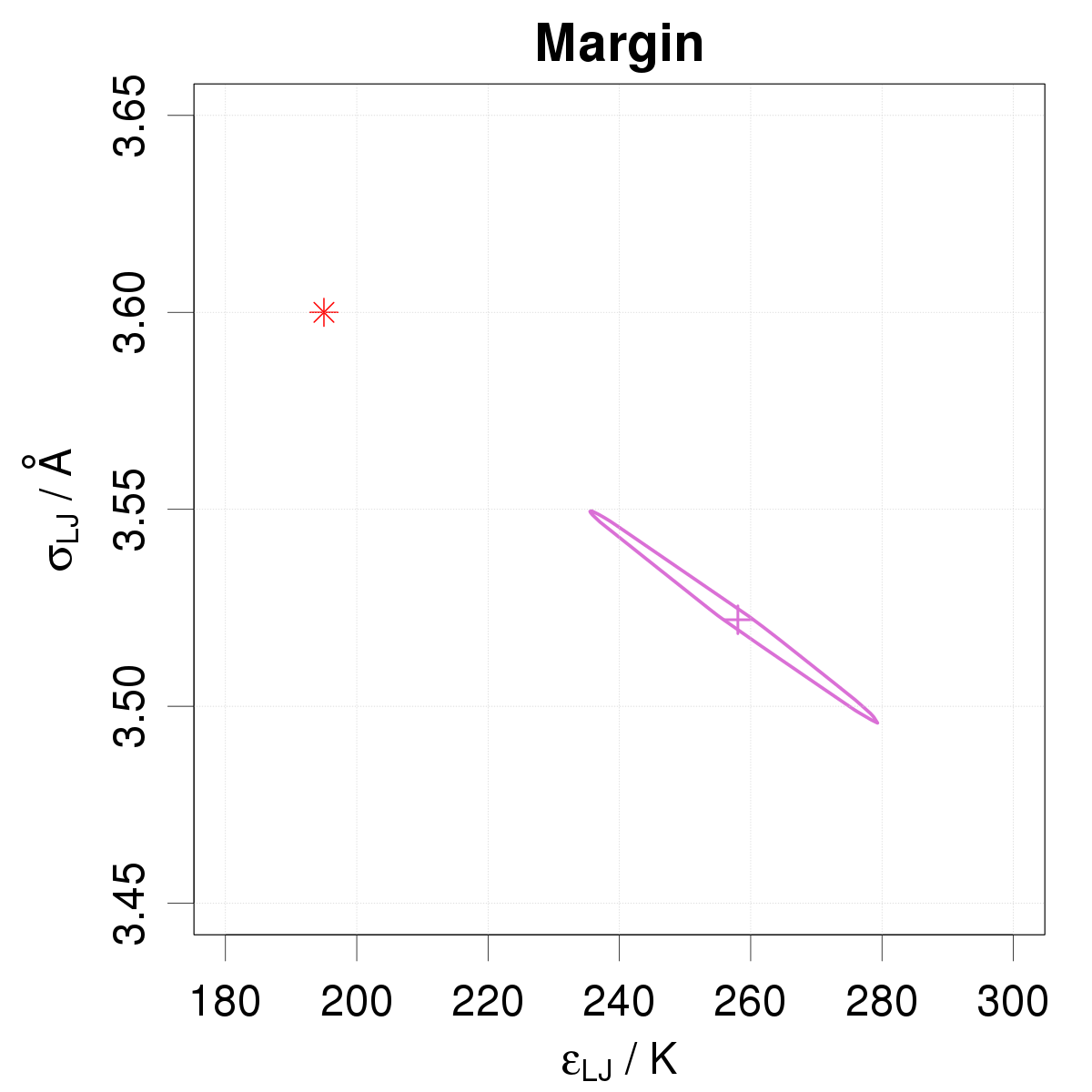} &  & \tabularnewline
\end{tabular}
\par\end{centering}
\caption{\label{fig:Sample-Synth1}Posterior samples of LJ parameters calibrated
on data SD-1: the contours represent 95~\% probability intervals;
the red cross depicts the reference value of the parameters. }
\end{figure}

The following statistical models were run on the SD-1 dataset:
\begin{itemize}
\item Std, with parameters $\epsilon_{LJ}$, $\sigma_{LJ}$ and $\sigma$;
\item Disp, with the same parameters (different interpretation for $\sigma$);
\item GP, with parameters $\epsilon_{LJ}$, $\sigma_{LJ}$, $\alpha$ and
$\beta$ \textendash{} in order to prevent identification problems,
the posterior pdf of the ``Disp'' model is used as prior for $\epsilon_{LJ}$
and $\sigma_{LJ}$;
\item WLS, with parameters $\epsilon_{LJ}$, $\sigma_{LJ}$; 
\item VarInf\_Rb and VarInf\_MSR, with prescribed value of $T$ (Eq.~\ref{eq:VarInf_Rb}
and Eq.~\ref{eq:VarInf_MSE}, respectively) and parameters $\epsilon_{LJ}$,
$\sigma_{LJ}$;
\item Margin and ABC, with parameters $\epsilon_{LJ}$, $\sigma_{LJ}$,
$u_{\epsilon}$, $u_{\sigma}$ and $\rho$, such as
\begin{equation}
\boldsymbol{V}_{\vartheta_{M}}=\left(\begin{array}{cc}
u_{\epsilon}^{2} & u_{\epsilon}u_{\sigma}\rho\\
u_{\epsilon}u_{\sigma}\rho & u_{\sigma}^{2}
\end{array}\right).
\end{equation}
The validity of the linear approximation for the estimation of model
uncertainty in the Margin and ABC methods has been checked by calculating
the relative error of a linear approximation of the model over the
whole range of temperature and a sample of LJ parameters drawn from
the posterior pdf (1000 points). The maximal relative error is about
$10^{-4}$, validating the linear approximation. 
\item HierC, based on a serialization of the dataset along the control variable,
with parameters $\epsilon_{LJ}$, $\sigma_{LJ}$, $u_{\epsilon}$,
$u_{\sigma}$, $\rho$ and $\sigma$. The data were arbitrarily split
into $5$ series at $T=300$, $600$, $900$ and $1200$\,K. 
\end{itemize}
A summary of the parameters of all models is reported in Table~(\ref{tab:Fit-results-SD-1}).
The residuals at the MAP for all methods and the prediction bands
are plotted in Fig.~\ref{fig:Predict-Synth1}, and the summaries
of the posterior samples of the LJ parameters are shown in Fig.~\ref{fig:Sample-Synth1}.

The structure and serial correlation of the residuals of the Std method
reveal the amplitude and shape of model inadequacy after calibration
(Fig.~\ref{fig:Predict-Synth1}). The $RMSD$ of $0.14$\,$\mu$Pa.s
shows that parameter optimization is able to correct a large fraction
of the original inadequacy ($0.55$\,$\mu$Pa.s). This is done however
at the expense of a large bias in parameters recovery, as can be seen
in Table~\ref{tab:Fit-results-SD-1} and Fig.~\ref{fig:Sample-Synth1}. 

Only two models are able to improve over the $RMSD$ of the Std method
and recover the level of random experimental noise ($0.07$\,$\mu$Pa.s):
the gaussian process GP, which completes the model with an adapted
correction term, and the hierarchical model HierC, which uses \emph{local}
LJ parameters adapted for different temperature zones. The WLS, VarInf-type
and ABC methods achieve slightly higher $RMSD$ ($0.18$\,$\mu$Pa.s).
The first ones have one less parameters than the Std method. In the
case of the ABC method, this is due to the additional constraint in
the likelihood. 

Several methods have Birge ratio differing notably from the unit value:
WLS ($R_{B}=13.0$) has no provision to compensate for model inadequacy;
ABC ($1.3$) for the same reason as above; GP ($0.87$) might present
some level of overfitting; and the VarInf-type methods ($0.02-0.03$)
are based on exalted data variance. It is not possible to validate
the latter methods by such statistics at the calibration level. 

\begin{table}
\noindent \begin{centering}
\begin{tabular}{rlllr@{\extracolsep{0pt}.}llr@{\extracolsep{0pt}.}llr@{\extracolsep{0pt}.}llllr@{\extracolsep{0pt}.}llr@{\extracolsep{0pt}.}l}
\hline 
Model &  & $\epsilon_{LJ}$  &  & \multicolumn{2}{c}{$\sigma_{LJ}$} &  & \multicolumn{2}{c}{$\sigma$ } &  & \multicolumn{2}{c}{$MR$} &  & $RMSD$ &  & \multicolumn{2}{c}{$R_{B}$} &  & \multicolumn{2}{c}{$u_{e|D}$}\tabularnewline
 &  & (K) &  & \multicolumn{2}{c}{($\text{\AA}$)} &  & \multicolumn{2}{c}{($\mu$Pa$.$s)} &  & \multicolumn{2}{c}{($\mu$Pa$.$s)} &  & ($\mu$Pa$.$s) &  & \multicolumn{2}{c}{} &  & \multicolumn{2}{c}{($\mu$Pa$.$s)}\tabularnewline
\cline{1-1} \cline{3-3} \cline{5-6} \cline{8-9} \cline{11-12} \cline{14-14} \cline{16-17} \cline{19-20} 
Ref &  & 195 &  & 3&6 &  & \multicolumn{2}{c}{} &  & \multicolumn{2}{c}{} &  &  &  & 1&0 &  & \multicolumn{2}{c}{}\tabularnewline
\noalign{\vskip\doublerulesep}
Std  &  & 254(1)  &  & 3&527(2)  &  & 0&14(1)  &  & 0&00  &  & 0.14  &  & 1&0  &  & 0&14 \tabularnewline
\noalign{\vskip\doublerulesep}
Disp  &  & 254(1)  &  & 3&528(2)  &  & 0&14(1)  &  & 0&00  &  & 0.14  &  & 0&94  &  & 0&16 \tabularnewline
\noalign{\vskip\doublerulesep}
GP  &  & 253(1)  &  & 3&529(2)  &  & \multicolumn{2}{c}{- } &  & 0&00  &  & 0.06  &  & 0&87  &  & 0&07 \tabularnewline
\noalign{\vskip\doublerulesep}
WLS  &  & 247.0(2)  &  & 3&5391(5)  &  & \multicolumn{2}{c}{- } &  & 0&03  &  & 0.18  &  & 13&0  &  & 0&07 \tabularnewline
\noalign{\vskip\doublerulesep}
VarInf\_Rb$^{*}$  &  & 247(6)  &  & 3&54(1)  &  & \multicolumn{2}{c}{- } &  & 0&03  &  & 0.18  &  & 0&02  &  & 0&23 \tabularnewline
\noalign{\vskip\doublerulesep}
VarInf\_MSR$^{*}$  &  & 247(5)  &  & 3&54(1)  &  & \multicolumn{2}{c}{- } &  & 0&03  &  & 0.17  &  & 0&02  &  & 0&18 \tabularnewline
\noalign{\vskip\doublerulesep}
Margin  &  & 258(2)  &  & 3&522(2)  &  & \multicolumn{2}{c}{- } &  & 0&01  &  & 0.15  &  & 0&91  &  & 0&15 \tabularnewline
$u_{\epsilon},\,u_{\sigma}$ &  & 10(1)  &  & 0&012(2)  &  & \multicolumn{2}{c}{} &  & \multicolumn{2}{c}{} &  &  &  & \multicolumn{2}{c}{} &  & \multicolumn{2}{c}{}\tabularnewline
$\rho$ &  & \multicolumn{4}{c}{-0.997(3)} &  & \multicolumn{2}{c}{} &  & \multicolumn{2}{c}{} &  &  &  & \multicolumn{2}{c}{} &  & \multicolumn{2}{c}{}\tabularnewline
\noalign{\vskip\doublerulesep}
ABC  &  & 247.5(2)  &  & 3&5389(4)  &  & \multicolumn{2}{c}{- } &  & 0&06  &  & 0.18  &  & 1&3 &  & 0&17 \tabularnewline
$u_{\epsilon},\,u_{\sigma}$ &  & 3.0(2)  &  & 0&0007(6)  &  & \multicolumn{2}{c}{} &  & \multicolumn{2}{c}{} &  &  &  & \multicolumn{2}{c}{} &  & \multicolumn{2}{c}{}\tabularnewline
$\rho$ &  & \multicolumn{4}{c}{0.2(5)} &  & \multicolumn{2}{c}{} &  & \multicolumn{2}{c}{} &  &  &  & \multicolumn{2}{c}{} &  & \multicolumn{2}{c}{}\tabularnewline
\noalign{\vskip\doublerulesep}
Hier  &  & 251(10)  &  & 3&54(2)  &  & 0&02(1)  &  & 0&00  &  & 0.06  &  & 1&1  &  & 1&20 \tabularnewline
$u_{\epsilon},\,u_{\sigma}$ &  & 19(9)  &  & 0&04(2)  &  & \multicolumn{2}{c}{} &  & \multicolumn{2}{c}{} &  &  &  & \multicolumn{2}{c}{} &  & \multicolumn{2}{c}{}\tabularnewline
$\rho$ &  & \multicolumn{4}{c}{-0.7(3)} &  & \multicolumn{2}{c}{} &  & \multicolumn{2}{c}{} &  &  &  & \multicolumn{2}{c}{} &  & \multicolumn{2}{c}{}\tabularnewline
\hline 
\end{tabular}
\par\end{centering}
\noindent \begin{centering}
\par\end{centering}
\caption{\label{tab:Fit-results-SD-1}Summary of the main parameters and statistics
for the methods tested on the SD-1 set. The first line (labeled Ref)
provides reference value of the parameters and statistics. \protect \\
$^{*}$ The scale factors $T$ for VarInf\_MSR and VarInf\_Rb are
respectively 435 and 643. }
\end{table}

The posterior samples show large differences (Table~\ref{tab:Fit-results-SD-1}
and Fig.~\ref{fig:Sample-Synth1}). The Std and Disp models have
similar outputs, the main difference lying in the signification attributed
to the residual dispersion parameter $\sigma$. In the Disp method,
$\sigma$ is part of the model, and therefore contributes to MPU,
which ensures that the model prediction band covers correctly the
part of the residuals due to model systematic errors (Fig.~\ref{fig:Predict-Synth1}). 

The posterior pdf of the GP method for the LJ parameters is identical
to its prior pdf which was taken from the Disp posterior pdf. This
constraint is intended to solve the parameters identifiability problem
of this approach. 

Methods based on the PUI strategy (VarInf-type, Margin, ABC) display
larger dispersion of the LJ parameters, while the Hier method produces
the posterior sample with the largest dispersion. 

Considering the prediction statistics $u_{e|D}$, it is close to the
$RMSD$ for all methods, except: WLS, for the same reasons as above;
VarInf\_Rb, which has overestimated parameter uncertainty; and Hier,
which has a strongly over-estimated prediction uncertainty, due to
the large dispersion of the LJ parameters necessary to their adaptation
to the full temperature range.

The impact of the posterior samples on the predictions is observable
in Fig.~\ref{fig:Predict-Synth1}. In the present scenario, one expects
that the model prediction confidence band (from distribution $p_{M}$,
Eq.~\ref{eq:conf}) accounts for model inadequacy. The Std model
cannot be successful because of the ambiguity on the meaning of parameter
$\sigma$. This ambiguity is solved in Disp, as this parameters enters
the stochastic part of the model describing model errors. In this
case, the prediction confidence band has a constant width over the
control parameter range and covers correctly model inadequacy. 

By construction, the GP model has prediction bands which follow perfectly
the structure of model inadequacy. In contrast, the Hier model, which
achieved also very good residuals, has strongly over-estimated prediction
bands.

Among the remaining PUI methods, VarInf\_MSR and Margin have prediction
bands which cover well the residuals, but with a structure which does
not correlate well with model inadequacy. The ABC method presents
more regular prediction bands. 

Note that Pernot\cite{Pernot2016} observed a multimodality in the
posterior pdf of the hyperparameters for the Margin and ABC distribution
on a similar problem. Each mode corresponded to a minimal or near-minimal
value for the $u_{\epsilon}$, $u_{\sigma}$ and $\rho$ hyperparameters.
In the present case, no multimodality is observed, but the posterior
pdf for Margin is localized at a VarInf-like solution corresponding
to $\rho\simeq-1$ (Table~\ref{tab:Fit-results-SD-1} and Fig.~\ref{fig:Sample-Synth1}),
whereas the ABC method adopts a mode where $u_{\sigma}$ is very small.

Considering the results of this test, the methods best adapted to
quantify model inadequacy errors would be Disp, GP and ABC. 

\subsubsection{SD-2: data inconsistency\label{subsec:SD-2:-data-inconsistency}}

The SD-2 dataset of size $N=100$ is generated by simulating 10 series
of data with cardinals between 5 and 15, shifted by a random factor
$s^{(\kappa_{i})}\sim\mathcal{N}(0,0.5)$, and with overlapping temperature
ranges. The standard deviation of the random noise in this set is
about $0.06$\,$\mu$Pa.s, and the systematic errors contribute for
$0.42$\,$\mu$Pa.s.

There is no model inadequacy in this scenario, so the relevant models
are:
\begin{itemize}
\item a variant of Disp, with parameters $\epsilon_{LJ}$, $\sigma_{LJ}$
and $\tau$, the latter describing inter-series dispersion instead
of model inadequacy; 
\item Shift, where the serial shifts are treated explicitly as realizations
of a random process with dispersion $\tau$ (hierarchical model); 
\item Cov, with parameters $\epsilon_{LJ}$, $\sigma_{LJ}$ and $\tau$;
\item Hier, a hierarchical model with parameters $\epsilon_{LJ}$, $\sigma_{LJ}$,
$u_{\epsilon}$, $u_{\sigma}$, and $\rho$, as proposed by Wu \emph{et
al.} \cite{Wu2015} ($\mathcal{M}_{H1}$ model), and in which the
LJ parameters are expected to compensate for systematic measurement
errors; and
\item Hier-Shift and Hier-Cov, combinations of the Hier model and the Shift
and Cov descriptions of systematic data errors. These combined model
are used here to show the interplay of model parameters and experimental
systematic errors. 
\end{itemize}
All methods recover correctly the reference LJ parameters (Table~\ref{tab:Fit-results-SD-2},
Figs~\ref{fig:Predict-Synth2} \& \ref{fig:Sample-Synth2}). The
Shift, Hier and Hier-Shift models are able to efficiently correct
the systematic errors and produce very small residuals, comparable
with the random errors ($RMSD=0.06$\,$\mu$Pa.s). The other methods
reach a level compatible with the systematic errors ($RMSD\sim0.4$\,$\mu$Pa.s).
Similarly, all methods reach adequate values of $R_{B}$.

When comparing the posterior pdfs of the parameters (Fig.~\ref{fig:Sample-Synth2}),
it is striking that models accounting explicitly for data inconsistency
achieve smaller parameter uncertainty (\emph{e.g.}, Shift \emph{vs.}
Disp or Hier-Shift \emph{vs.} Hier). In the Disp case, the hypothesis
of independent data unduly increases parameter uncertainty. When the
covariance structure of the data is taken into account (model Hier-Cov),
the hierarchical model does not attempt to adapt the LJ parameters
to compensate for experimental systematic errors and provides a much
more concentrated posterior sample (Fig.~\ref{fig:Sample-Synth2}).

For the Hier method, the mean prediction error $u_{e|D}$ ($0.5$\,$\mu$Pa.s)
is considerably larger than the $RMSD$ ($0.06$\,$\mu$Pa.s), which
is due to the choice of absorbing systematic experimental errors into
the parameters. It is also the method with the largest prediction
bands. The Hier-Shift method has also an excess of prediction uncertainty,
although at a lesser level.

Fig.~\ref{fig:Shifts-Synth2} shows that the Shift and Hier-Shift
methods are able to recover correctly the shifts used to generate
the data. However, due to the interactions with the local LJ-parameters,
the uncertainties on the recovered shifts are much larger in the Hier-Shift
method. The slight bias observed in the case of Hier-Shift is due
to the sum-to-zero constraint (Eq.~\ref{eq:sumToZero}), which is
not necessary for the Shift method.

The main conclusion of this experiment is that taking the statistical
structure of the dataset into account, notably intra-series correlation,
is an important step to free the LJ parameters from exogenous constraints.
In fact, all methods estimate correctly the LJ parameters with less
uncertainty than the Disp and Hier methods. The latter, trying to
explain experimental inconsistency with the LJ parameters is particularly
detrimental to their estimation, and produces exaggerated prediction
uncertainties. 
\noindent \begin{flushleft}
\begin{figure}
\begin{centering}
\begin{tabular}{ccc}
\includegraphics[bb=0bp 0bp 1200bp 1200bp,clip,height=5cm]{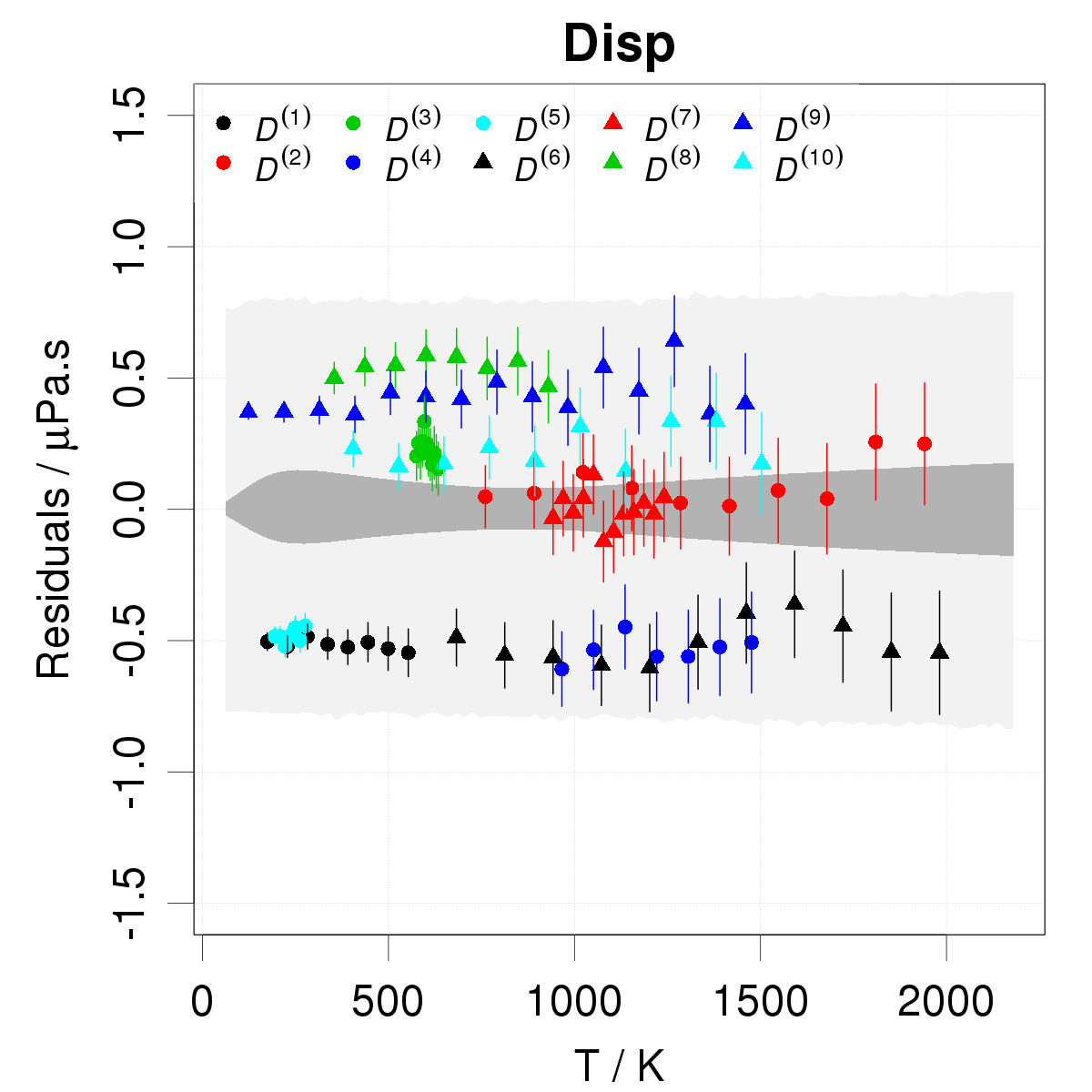} & \includegraphics[bb=0bp 0bp 1200bp 1200bp,clip,height=5cm]{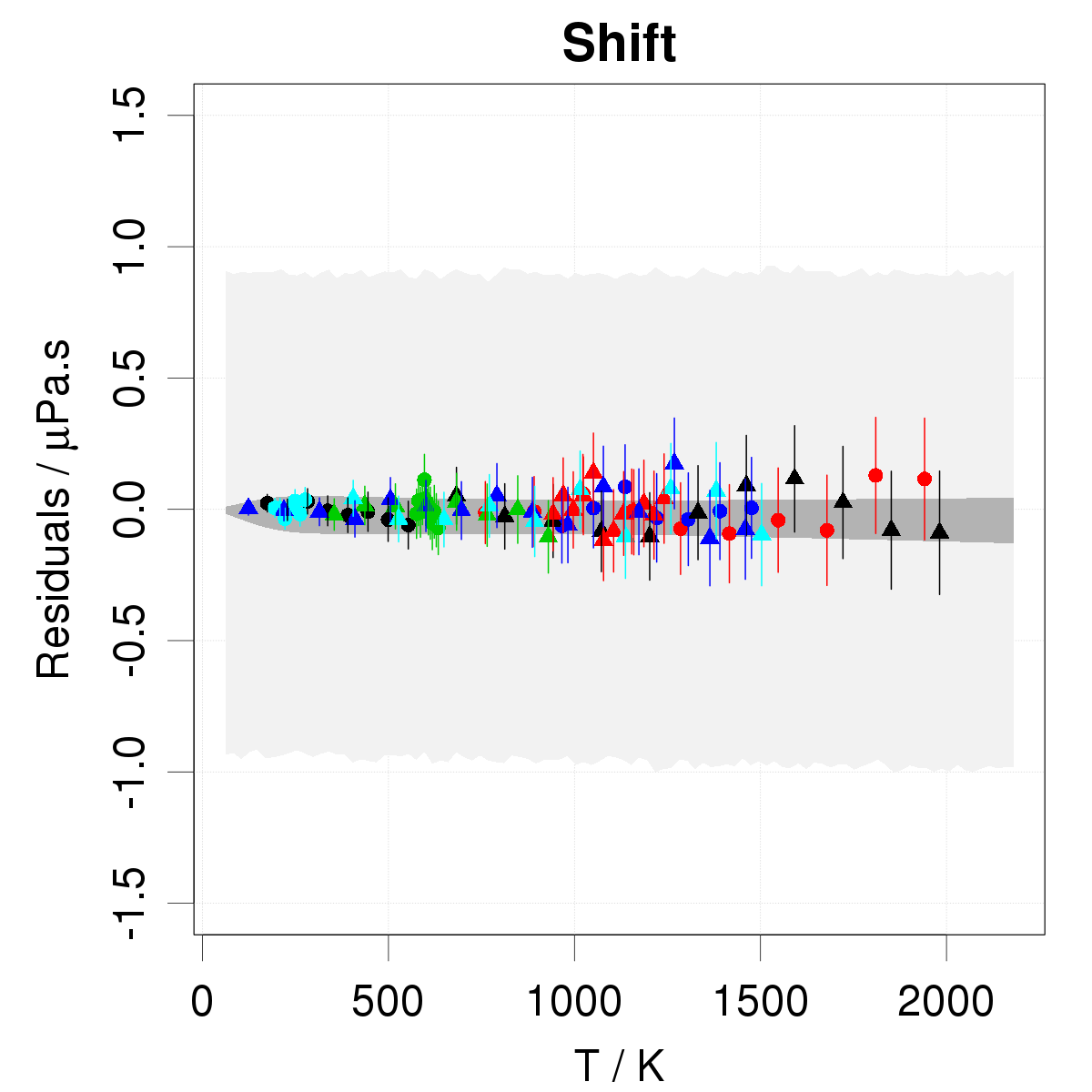} & \includegraphics[bb=0bp 0bp 1200bp 1200bp,clip,height=5cm]{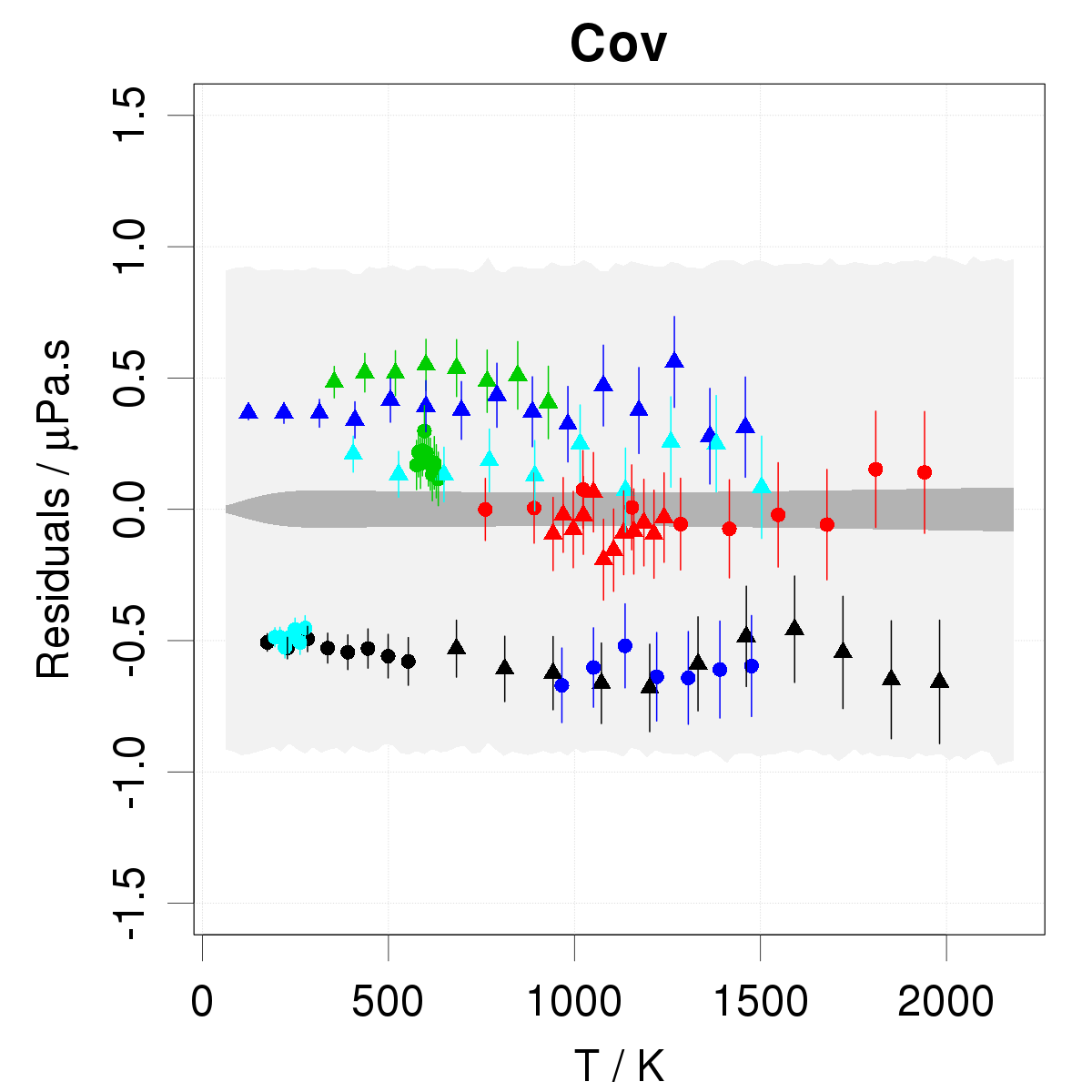}\tabularnewline
\includegraphics[bb=0bp 0bp 1200bp 1200bp,clip,height=5cm]{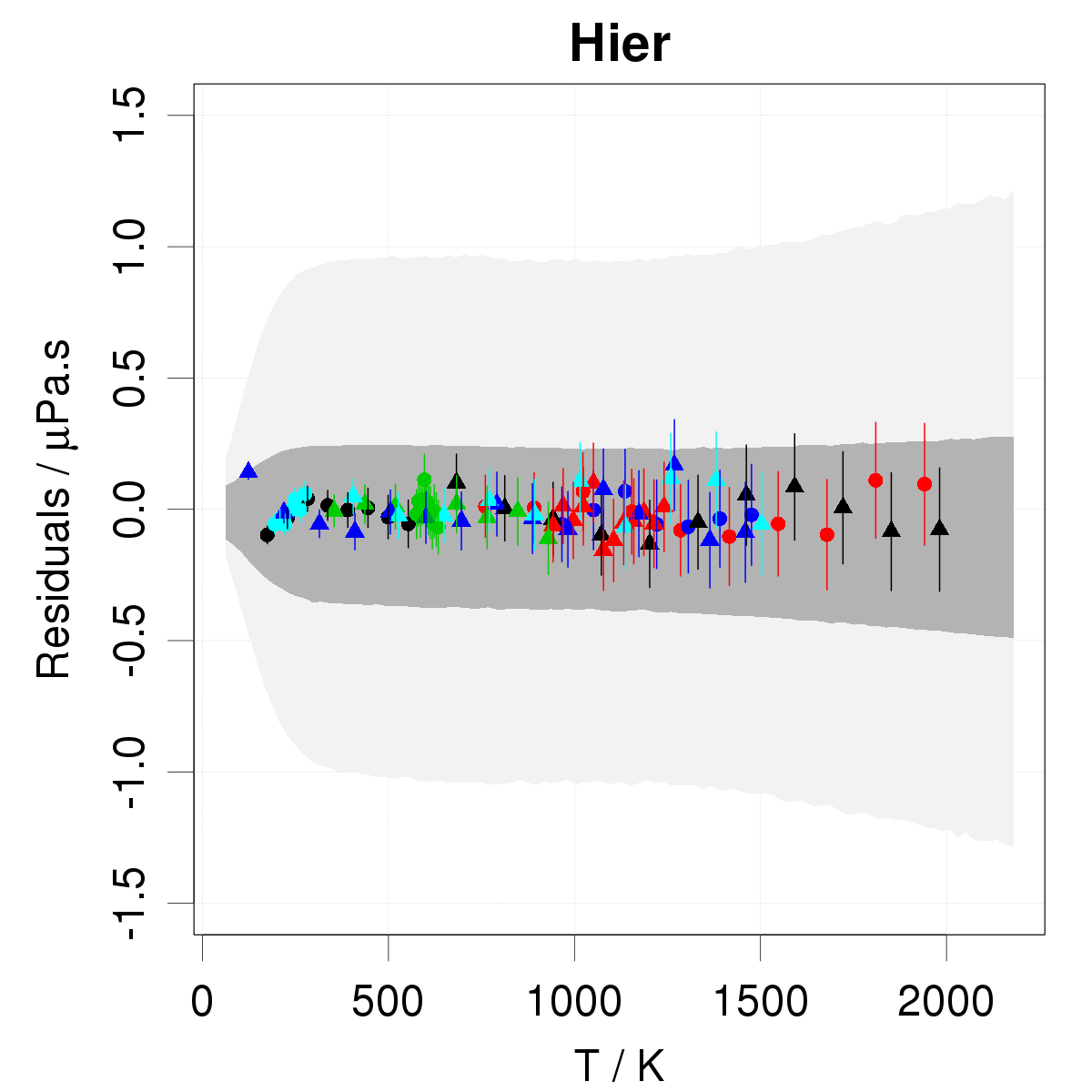} & \includegraphics[bb=0bp 0bp 1200bp 1200bp,clip,height=5cm]{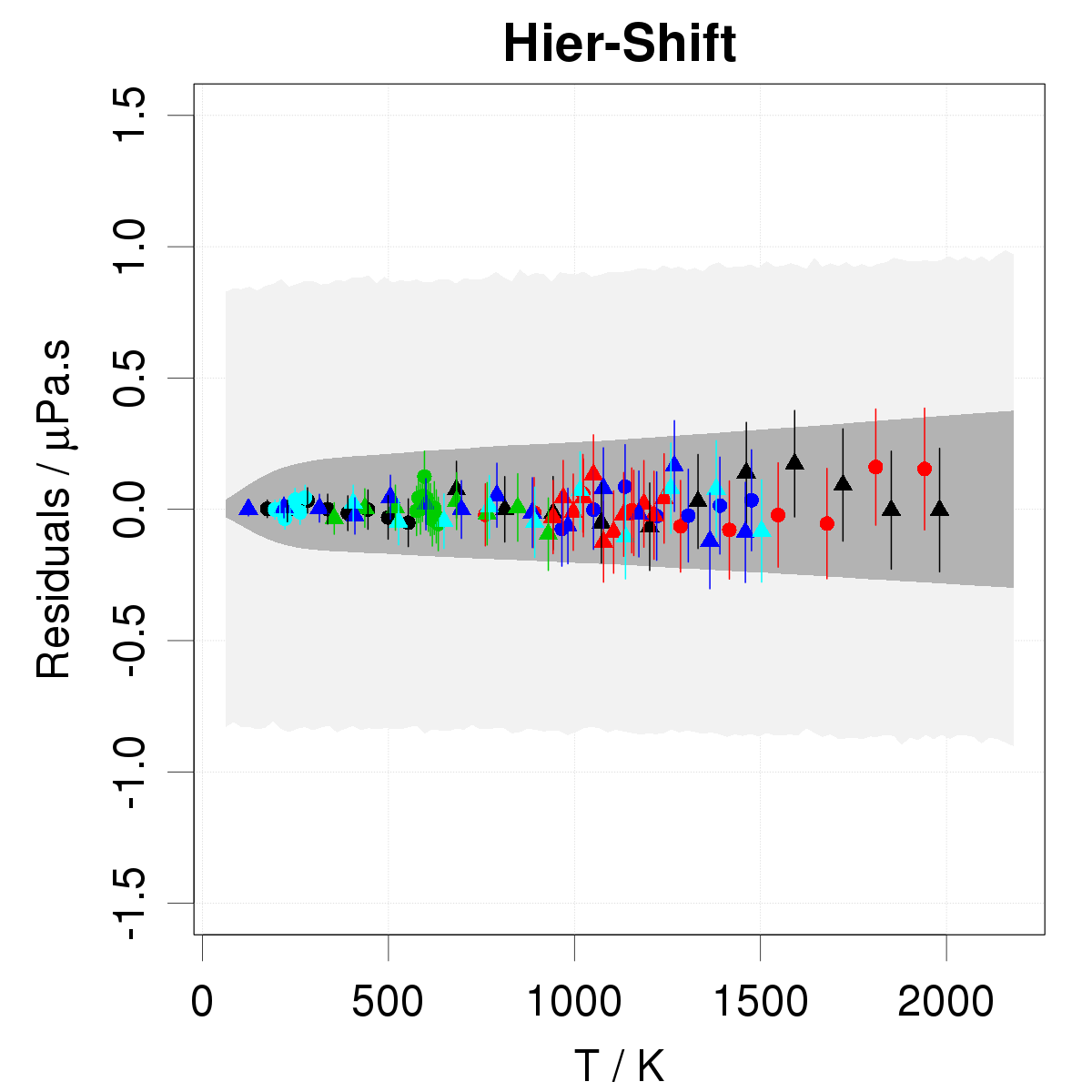} & \includegraphics[bb=0bp 0bp 1200bp 1200bp,clip,height=5cm]{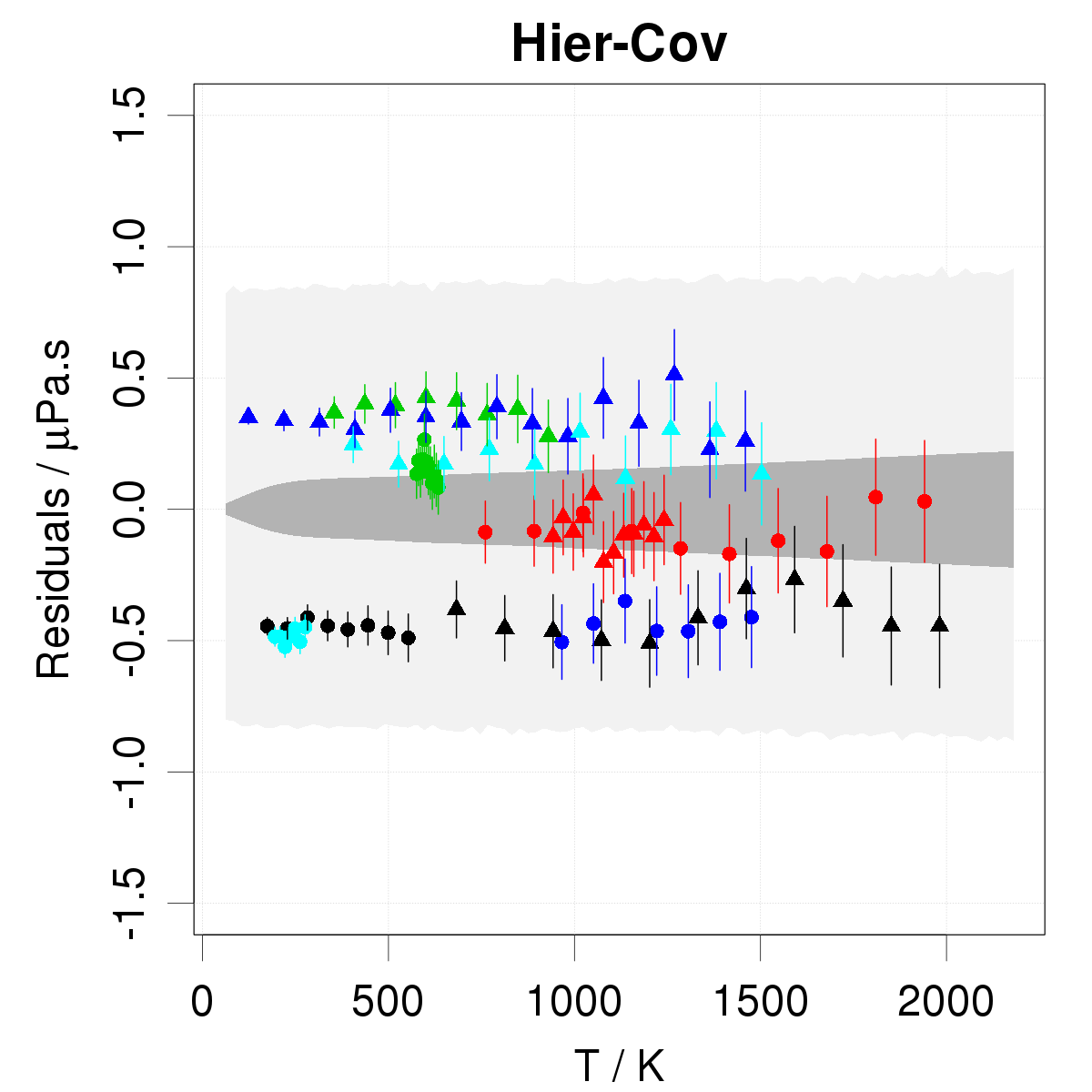}\tabularnewline
\end{tabular}
\par\end{centering}
\caption{\label{fig:Predict-Synth2}Calibration residuals and posterior prediction
intervals for data SD-2. For each model, the results are centered
on the maximum a posteriori results (MAP). The dark gray band represents
the model prediction 95\,\% probability interval ($p_{M}$) and the
light gray area the experiment prediction 95\,\% probability interval
($p_{e}$). }
\end{figure}
\par\end{flushleft}

\noindent \begin{flushleft}
\begin{table}
\noindent \begin{centering}
\begin{tabular}{rrr@{\extracolsep{0pt}.}lcr@{\extracolsep{0pt}.}lcccccccccc}
\hline 
Model &  & \multicolumn{2}{c}{$\epsilon_{LJ}$} &  & \multicolumn{2}{c}{$\sigma_{LJ}$} &  & $\tau$ &  & $MR$ &  & $RMSD$ &  & $R_{B}$ &  & $u_{e|D}$\tabularnewline
 &  & \multicolumn{2}{c}{(K)} &  & \multicolumn{2}{c}{($\text{\AA}$)} &  & ($\mu$Pa.s) &  & ($\mu$Pa.s) &  & ($\mu$Pa.s) &  &  &  & ($\mu$Pa.s)\tabularnewline
\cline{1-1} \cline{3-4} \cline{6-7} \cline{9-9} \cline{11-11} \cline{13-13} \cline{15-15} \cline{17-17} 
\noalign{\vskip\doublerulesep}
Ref &  & \multicolumn{2}{c}{195} &  & 3&6 &  & 0.42 &  &  &  &  &  & 1.00 &  & \tabularnewline
\noalign{\vskip\doublerulesep}
Disp &  & \multicolumn{2}{c}{195(2) } &  & 3&602(4) &  & 0.39(3) &  & 0.00  &  & 0.39  &  & 1.00  &  & 0.40\tabularnewline
\noalign{\vskip\doublerulesep}
Shift &  & 195&5(9)  &  & 3&599(1)  &  & 0.4(1)  &  & 0.00  &  & 0.06  &  & 0.85  &  & 0.07 \tabularnewline
\noalign{\vskip\doublerulesep}
Cov &  & 195&5(9)  &  & 3&599(1)  &  & 0.4(1)  &  & -0.05  &  & 0.40  &  & 0.85  &  & 0.46 \tabularnewline
\noalign{\vskip\doublerulesep}
Hier &  & \multicolumn{2}{c}{198(3) } &  & 3&599(2)  &  & - &  & -0.01  &  & 0.06  &  & 0.99  &  & 0.50 \tabularnewline
$u_{\epsilon},\,u_{\sigma}$ &  & \multicolumn{2}{c}{9(2) } &  & 0&004(3) &  &  &  &  &  &  &  &  &  & \tabularnewline
$\rho$ &  & \multicolumn{5}{c}{-0.7(3)} &  &  &  &  &  &  &  &  &  & \tabularnewline
\noalign{\vskip\doublerulesep}
Hier-Shift &  & 197&3(8)  &  & 3&599(1)  &  & 0.4(1)  &  & 0.01  &  & 0.06  &  & 1.10  &  & 0.13\tabularnewline
$u_{\epsilon},\,u_{\sigma}$ &  & \multicolumn{2}{c}{1(1)} &  & 0&001(1) &  &  &  &  &  &  &  &  &  & \tabularnewline
$\rho$ &  & \multicolumn{5}{c}{0.0(6)} &  &  &  &  &  &  &  &  &  & \tabularnewline
\noalign{\vskip\doublerulesep}
Hier-Cov &  & \multicolumn{2}{c}{196(1) } &  & 3&599(1) &  & 0.4(1) &  & -0.04  &  & 0.39  &  & 0.98  &  & 0.43 \tabularnewline
$u_{\epsilon},\,u_{\sigma}$ &  & 0&3(7) &  & 0&0008(8) &  &  &  &  &  &  &  &  &  & \tabularnewline
$\rho$ &  & \multicolumn{5}{c}{0.0(6)} &  &  &  &  &  &  &  &  &  & \tabularnewline
\hline 
\end{tabular}
\par\end{centering}
\caption{\label{tab:Fit-results-SD-2}Summary of the main parameters and statistics
for the methods tested on the SD-2 set. The first line (labeled Ref)
provides reference value of the parameters and statistics.}
\end{table}
\begin{figure}
\begin{centering}
\begin{tabular}{ccc}
\includegraphics[bb=0bp 0bp 1200bp 1200bp,clip,height=5cm]{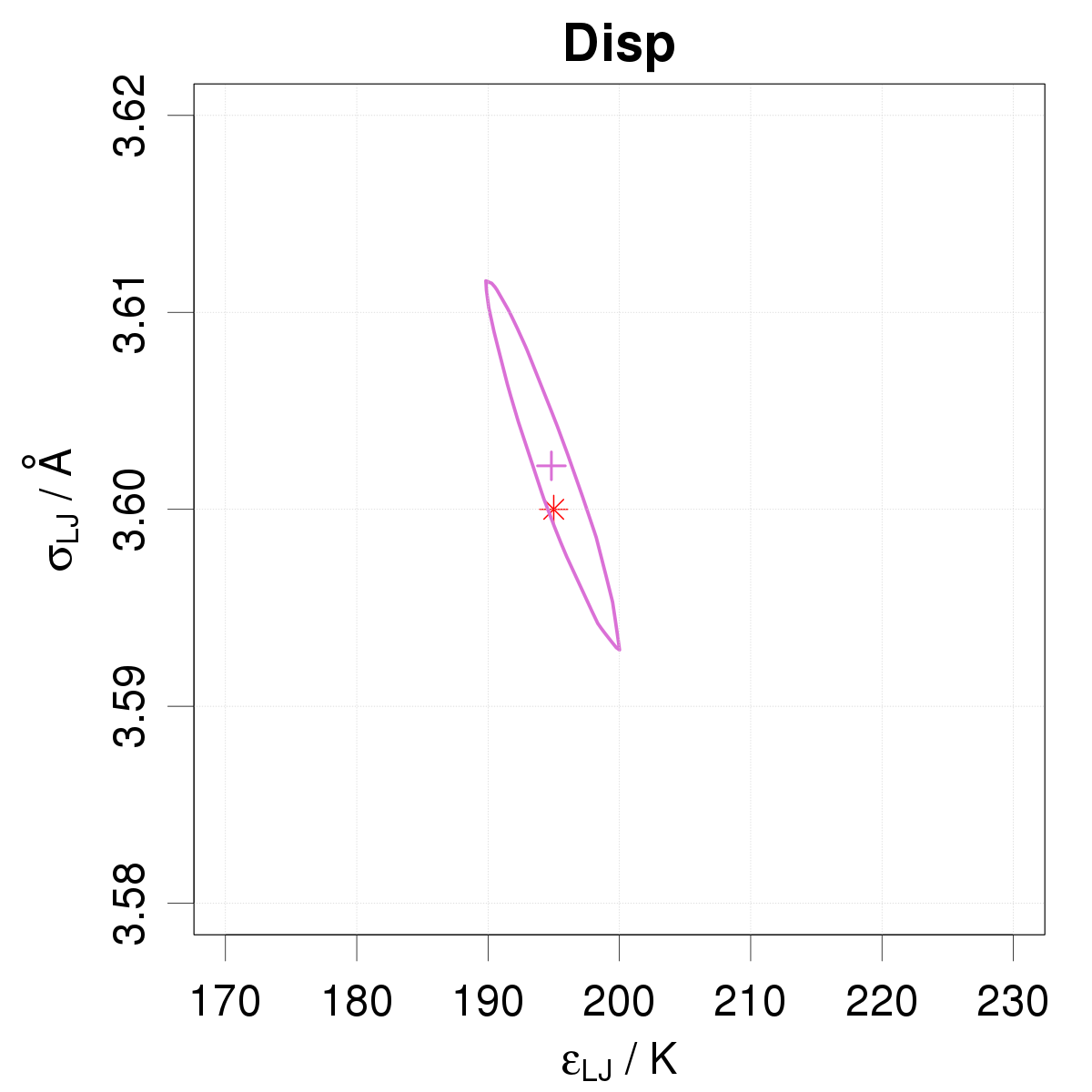} & \includegraphics[bb=0bp 0bp 1200bp 1200bp,clip,height=5cm]{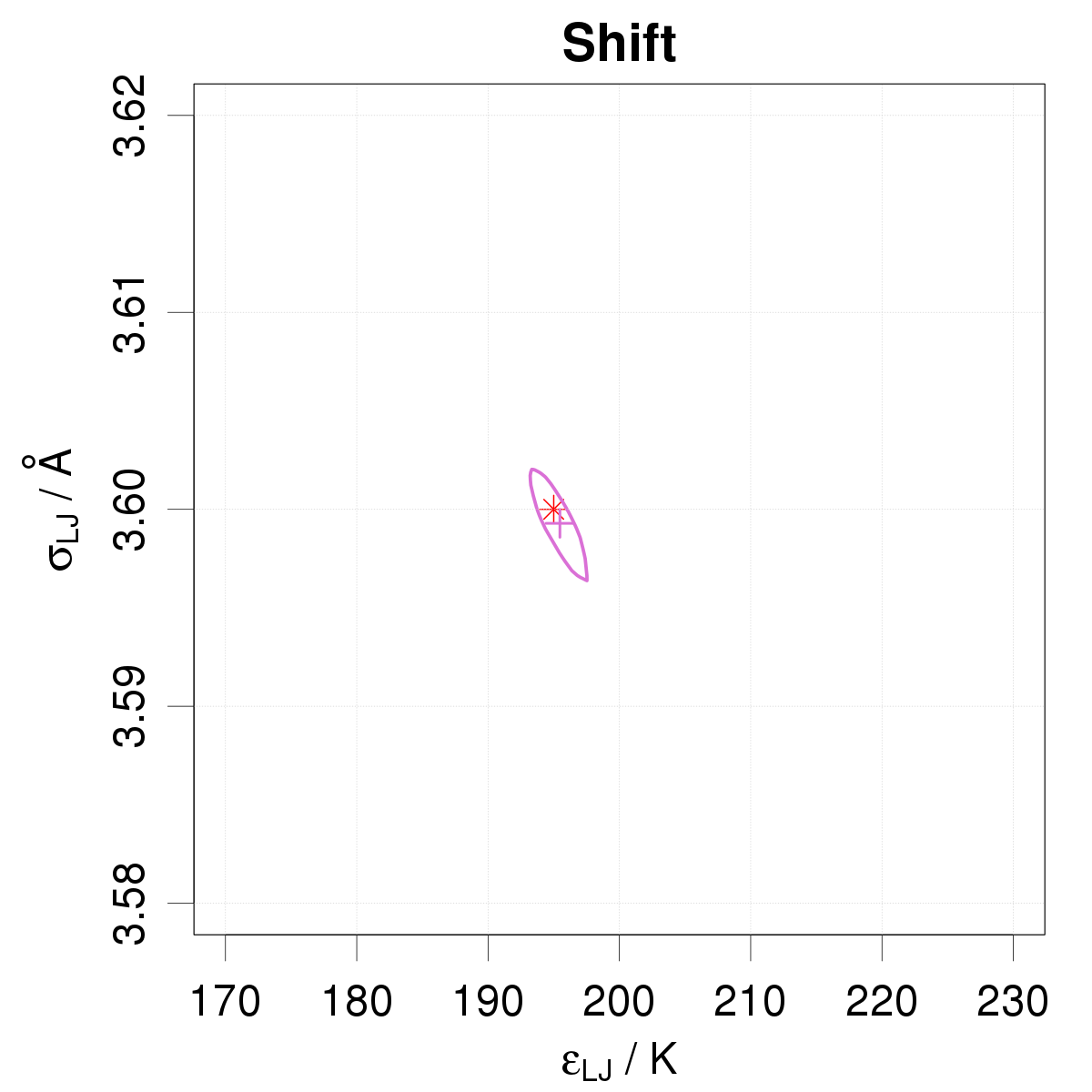} & \includegraphics[bb=0bp 0bp 1200bp 1200bp,clip,height=5cm]{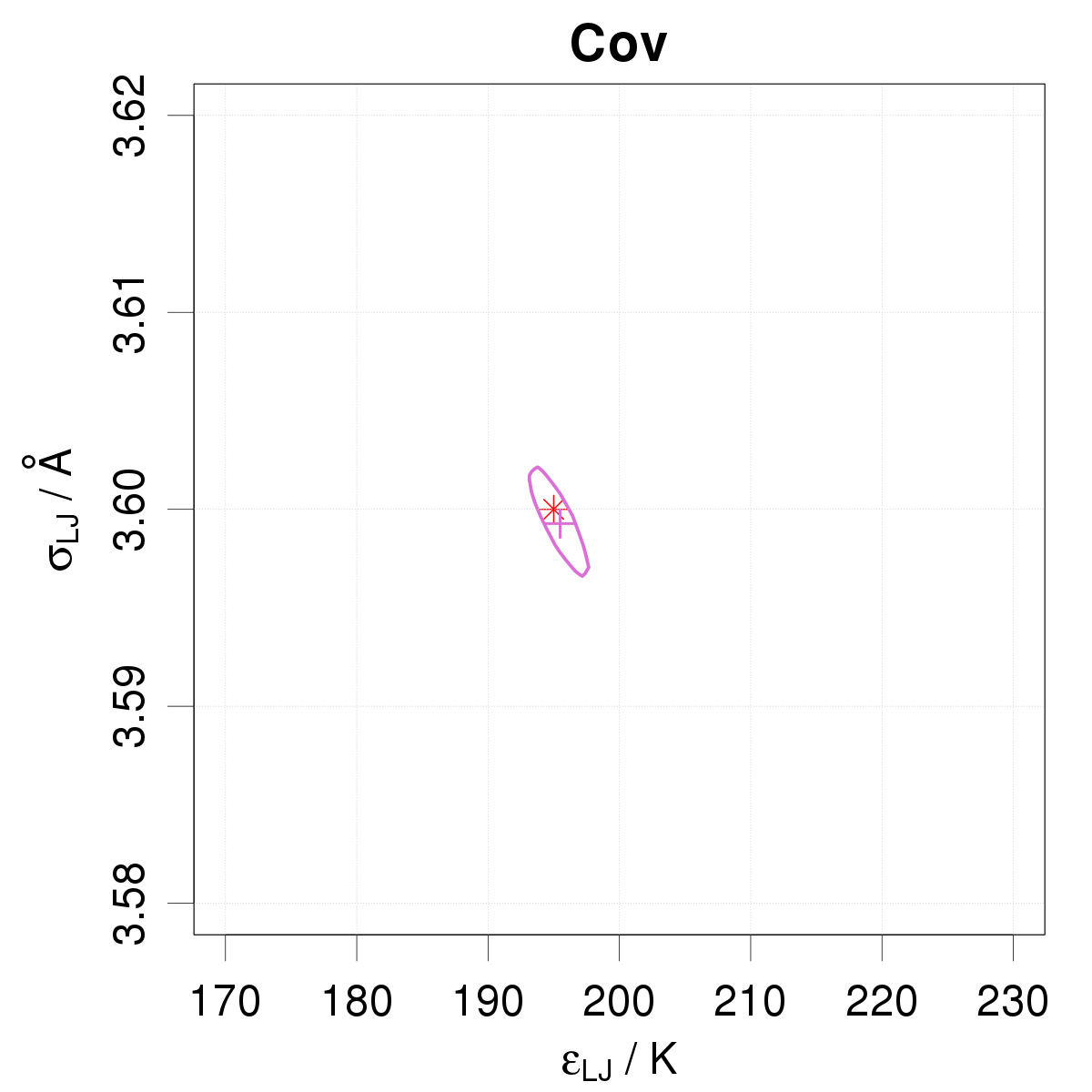}\tabularnewline
\includegraphics[bb=0bp 0bp 1200bp 1200bp,clip,height=5cm]{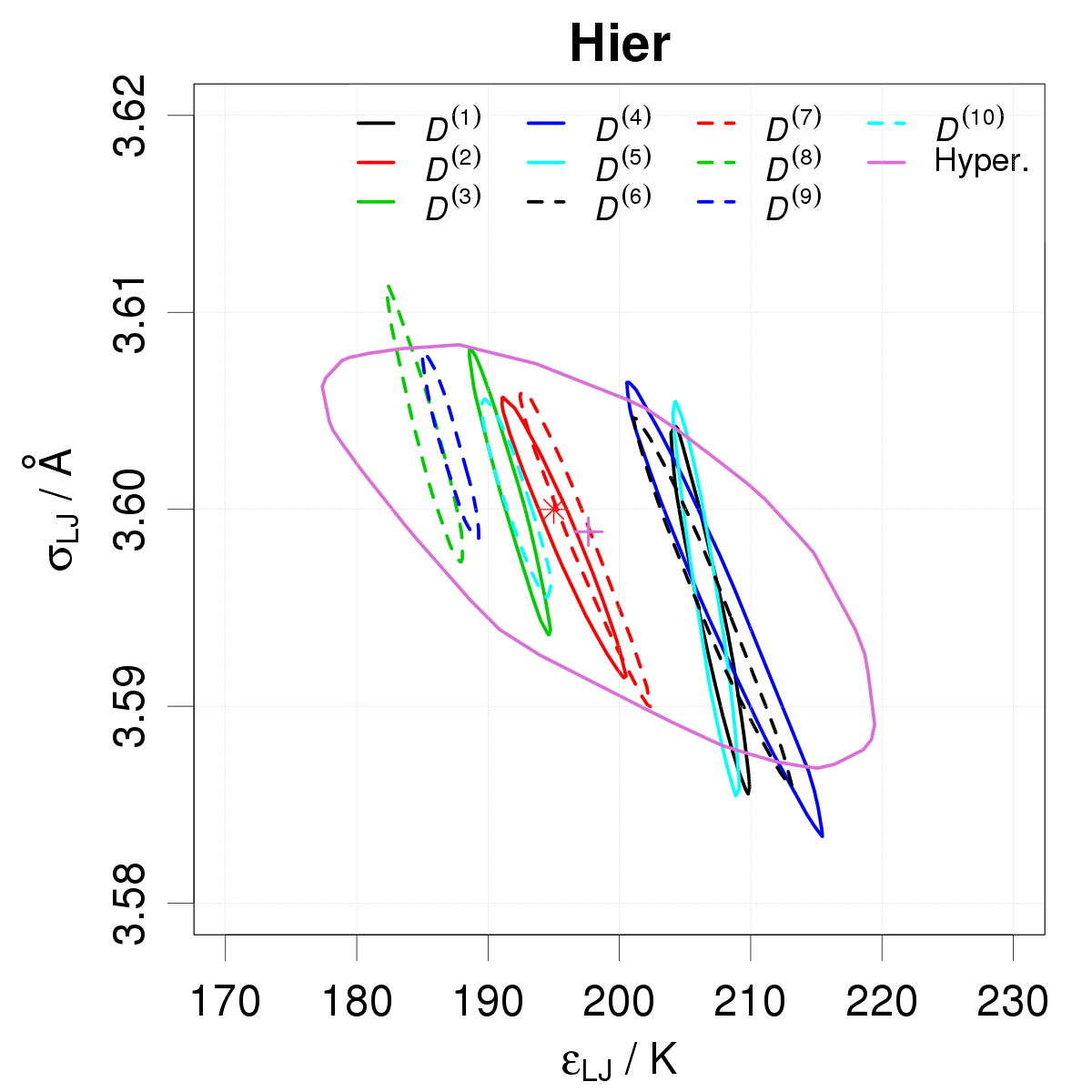} & \includegraphics[bb=0bp 0bp 1200bp 1200bp,clip,height=5cm]{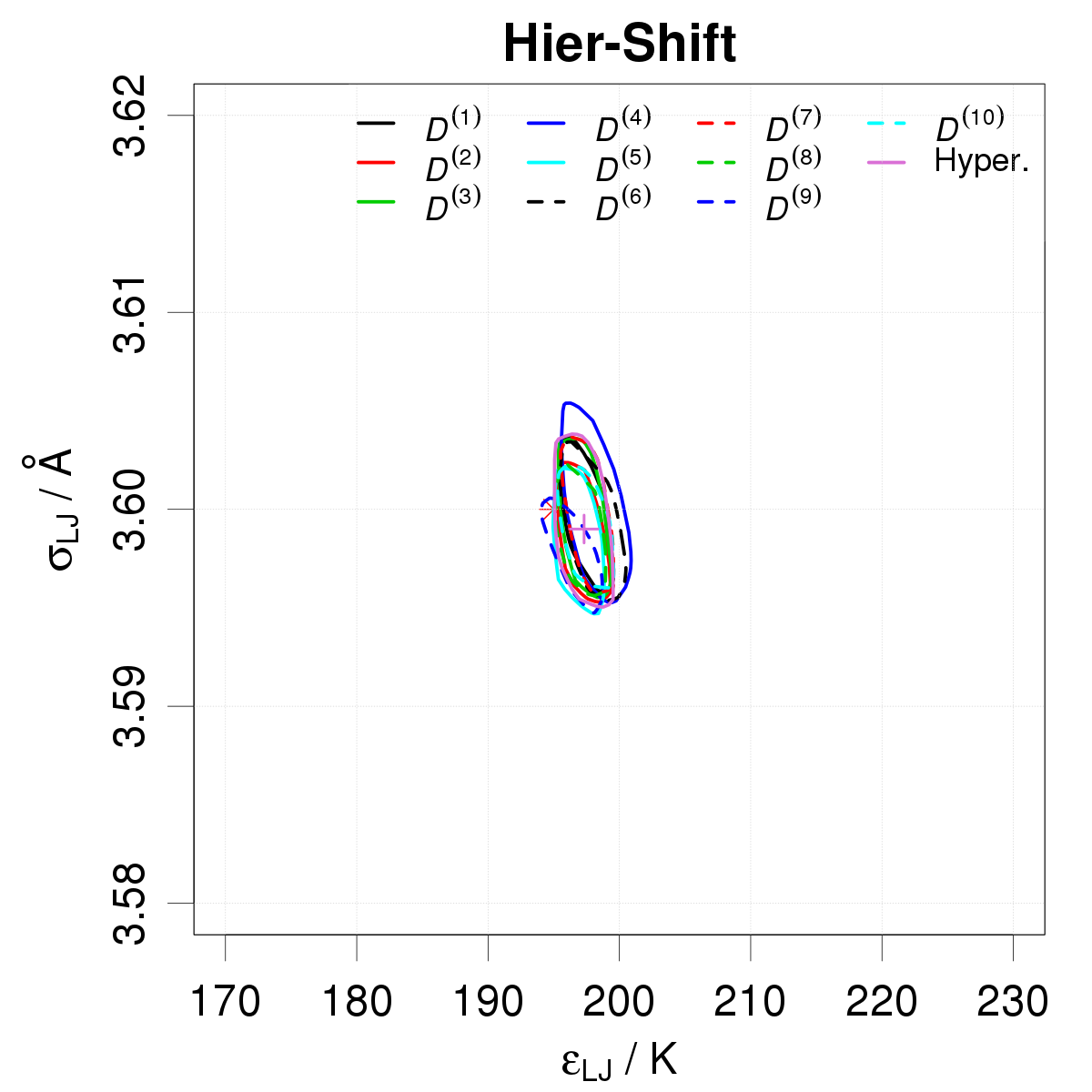} & \includegraphics[bb=0bp 0bp 1200bp 1200bp,clip,height=5cm]{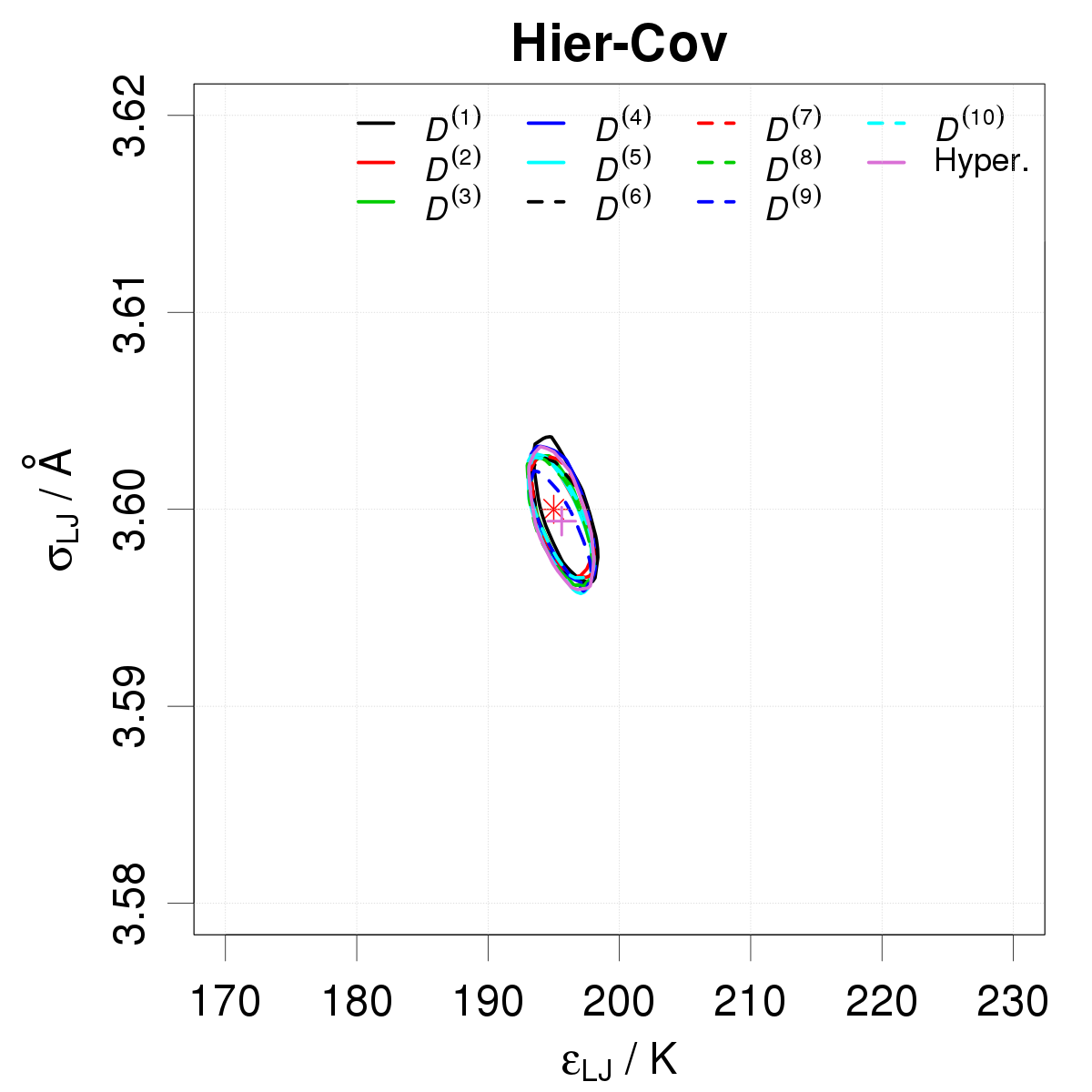}\tabularnewline
\end{tabular}
\par\end{centering}
\caption{\label{fig:Sample-Synth2}Posterior samples of LJ parameters calibrated
on data SD-2: the elliptic contours represent 95~\% probability intervals;
the red cross depicts the reference value of the parameters. }
\end{figure}
\par\end{flushleft}

\noindent \begin{flushleft}
\begin{figure}
\begin{centering}
\includegraphics[bb=0bp 0bp 1800bp 1200bp,clip,width=1\textwidth]{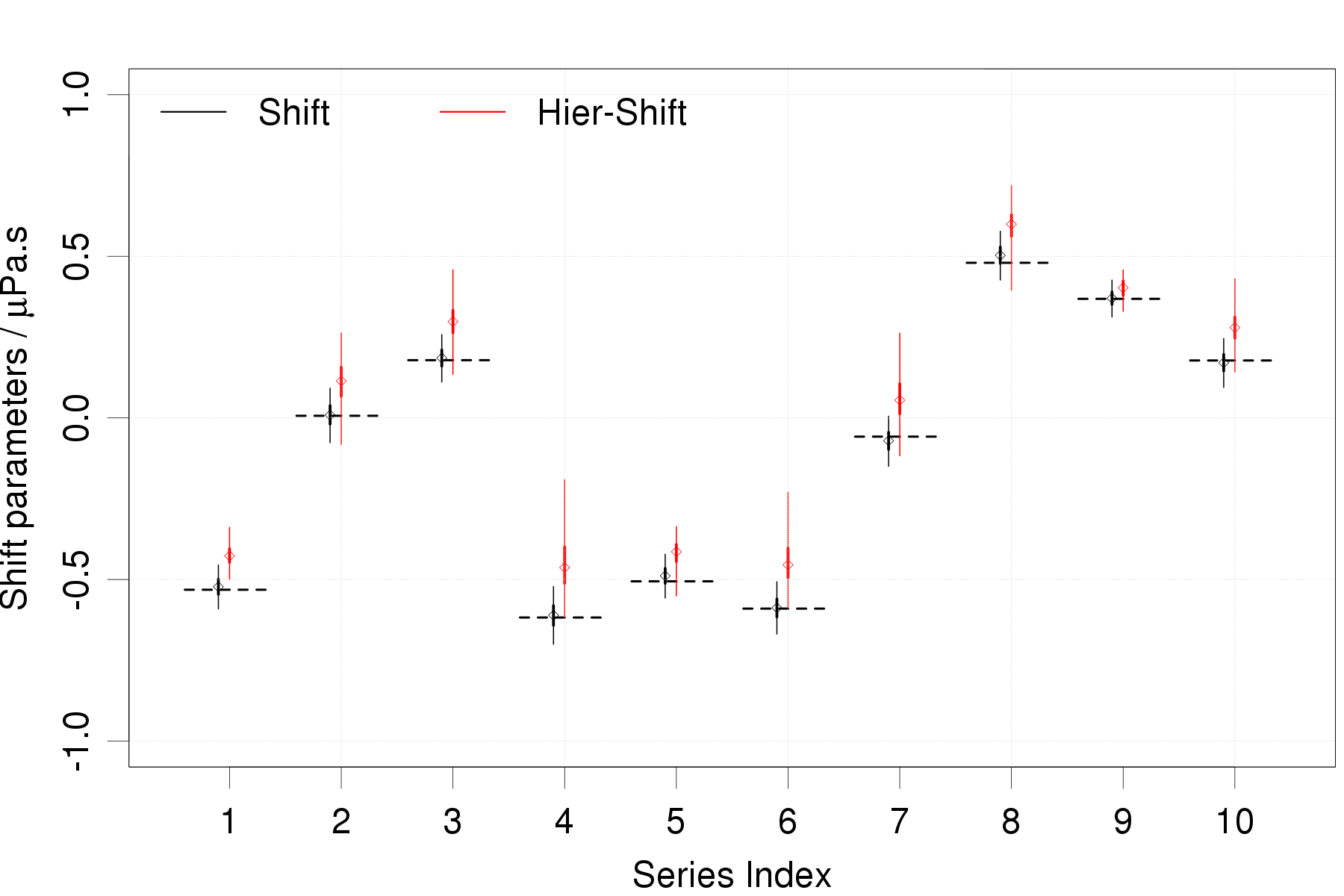}
\par\end{centering}
\caption{\label{fig:Shifts-Synth2}95\% confidence intervals for the shift
parameters recovered by two methods from the SD-2 dataset. The horizontal
bars represent the true values. }
\end{figure}
\par\end{flushleft}

\subsubsection{SD-3: model discrepancy \& data inconsistency\label{subsec:SD-3}}

We now use a dataset combining the two error sources of sets SD-1
and SD-2. This demands for a combination of the corresponding statistical
models. 

Simple models, such Disp are a priori unable to discriminate between
the experimental and model systematic error sources. More promising
methods can be obtained by combining the GP, ABC and Hier approaches
of model discrepancy with Shift and Cov treatments of systematic experimental
errors. The GP , Margin and ABC methods require a good estimation
of model errors to be efficient. We have seen above (Section~\ref{subsec:SD-2:-data-inconsistency})
that only the Shift method was able to provide residuals corrected
for the systematic experimental errors. It is therefore the choice
complement for the GP and ABC approaches. On the other hand, as shown
in case SD-2, the Hier method can benefit from the Cov model, taking
advantage of its smaller number of parameters. We consider therefore
the following models: Disp for reference; Disp-Shift; GP-Shift; Margin-Shift;
ABC-Shift; Hier-Shift; and Hier-Cov.

The models involving the determination of experimental shifts might
present a parameter identification problem. If one does not constrain
the shifts, the correction term $\delta M$ of the GP model attempts
to compensate for them, and the model prediction is contaminated by
the variance for systematic measurement errors. The problem occurs
also for the ABC and Hier approaches, more strongly for the latter,
as it uses individual LJ parameters sets for each data series.\textcolor{purple}{{}
}In order to avoid this problem, the sum-to zero constraint (Eq.~\ref{eq:sumToZero})
has been imposed in all shift-based methods. 

The residuals of the Disp-Shift method (Table~\ref{tab:Fit-results-SD-3}),
with a $RMSD$ of $0.10$\,$\mu$Pa.s, provide a scale for model
inadequacy. Despite the additional constraint on the shift parameters,
the GP-Shift method achieves a $RMSD$ value close to the random noise
limit ($0.06$\,$\mu$Pa.s). Comparing the residuals of the Disp
and Hier-Cov methods, one sees how the Hier method attempts to compensate
for the systematic experimental errors by adjusting the local LJ parameters.

All methods have adequate mean prediction uncertainty, except Hier-Shift
and Hier-Cov for which it exceeds strongly the $RMSD$, as observed
for the Hier method in case SD-1. This is visible in the large dispersion
of the LJ parameters for these methods, notably $\sigma_{LJ}$ (Fig.~(\ref{fig:Sample-Synth3})).

In terms of prediction bands, the Disp-Shift method provides adequate
uncertainty bands for the model and experiment predictions (Fig.~\ref{fig:Predict-Synth3}).
The GP-Shift and Hier-Shift methods both correct well for model inadequacy,
at least in large part, but provide very different MPUs. In fact,
both the Hier-Shift and Hier-Cov methods produce unacceptably large
prediction bands. In contrast, the model prediction band for ABC-Shift
seems too narrow, notably at low temperature. The Margin-Shift method
provides a structured prediction band with adequate covering of the
residuals.

Distributions of shift parameters recovered by the five pertinent
methods are shown in Fig.~\ref{fig:Shifts-Synth3}, and compared
with the exact values. The sum-to-zero constraint introduces a slight
bias in the recovered values, but they are well identified. As seen
for case SD-2, only the Hier-Shift method provides large uncertainties
on these shifts, due to their strong interaction with the local LJ-parameters
of the Hier approach.

From this dataset with mixed experimental and model systematic errors,
only the Disp-Shift, GP-Shift, Margin-Shift and ABC-Shift methods
provide adequate prediction intervals. The Hier-based methods are
strongly overestimating parameter uncertainty and prediction uncertainty.
\noindent \begin{flushleft}
\begin{figure}
\begin{centering}
\begin{tabular}{ccc}
\includegraphics[bb=0bp 0bp 1200bp 1200bp,clip,height=5cm]{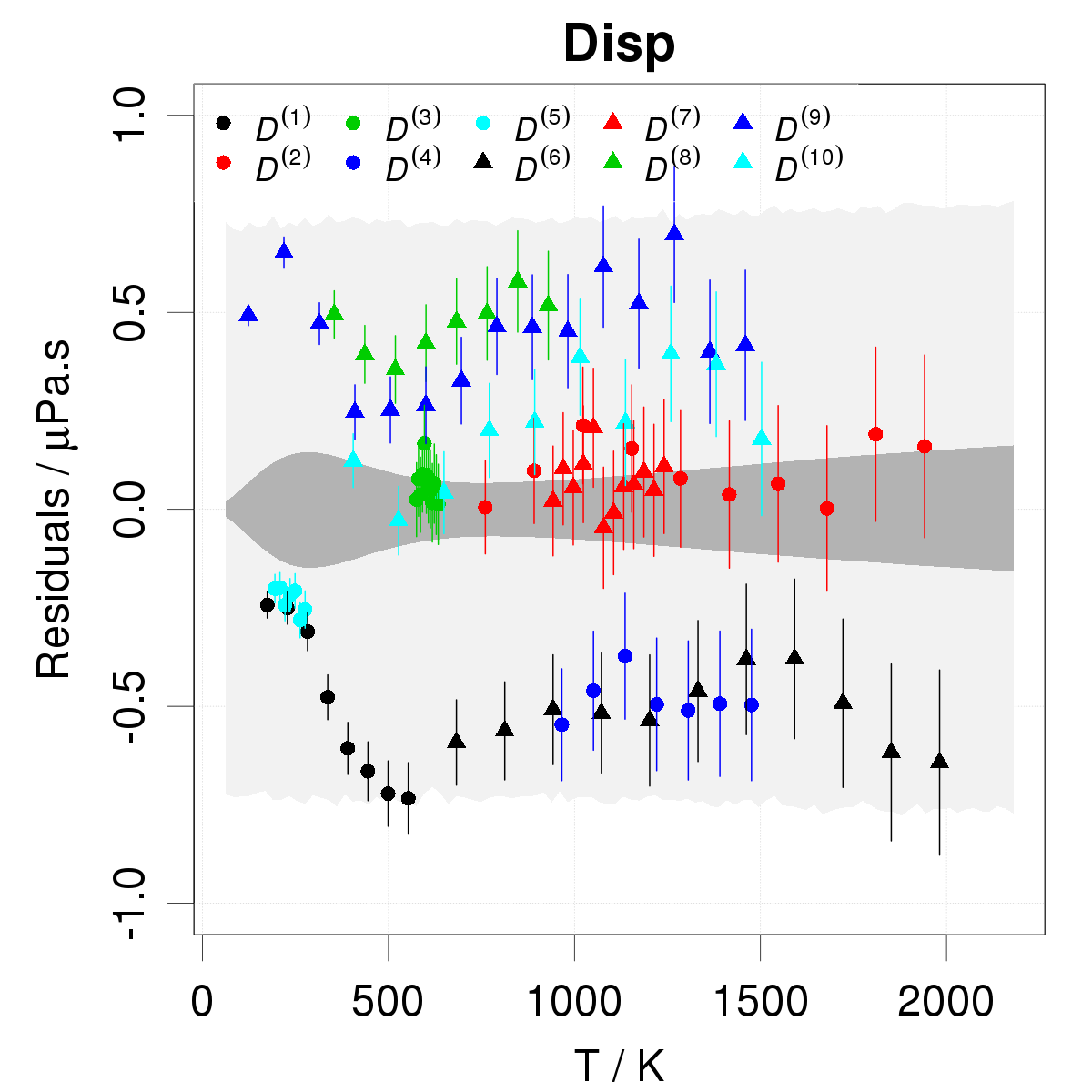} & \includegraphics[bb=0bp 0bp 1200bp 1200bp,clip,height=5cm]{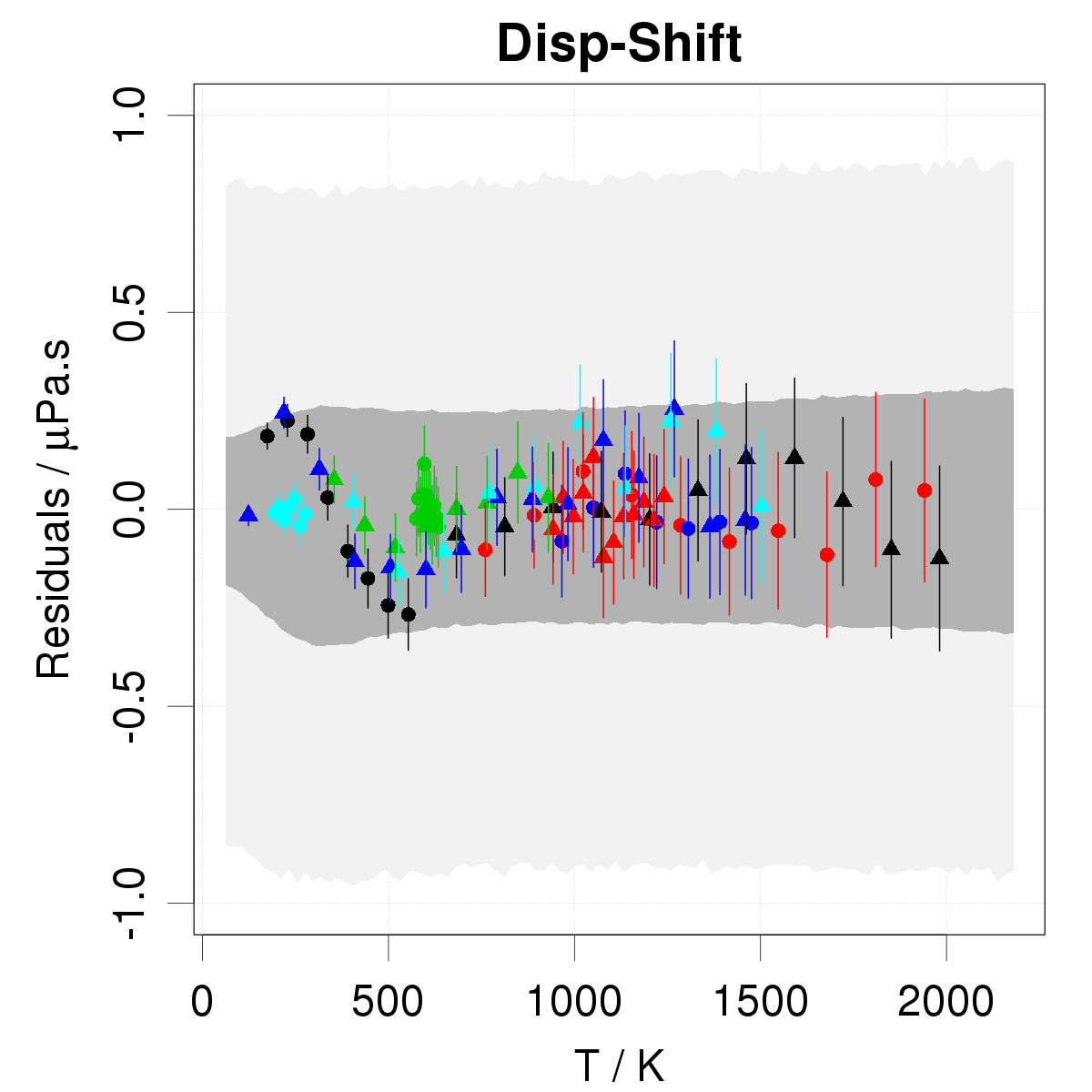} & \includegraphics[bb=0bp 0bp 1200bp 1200bp,clip,height=5cm]{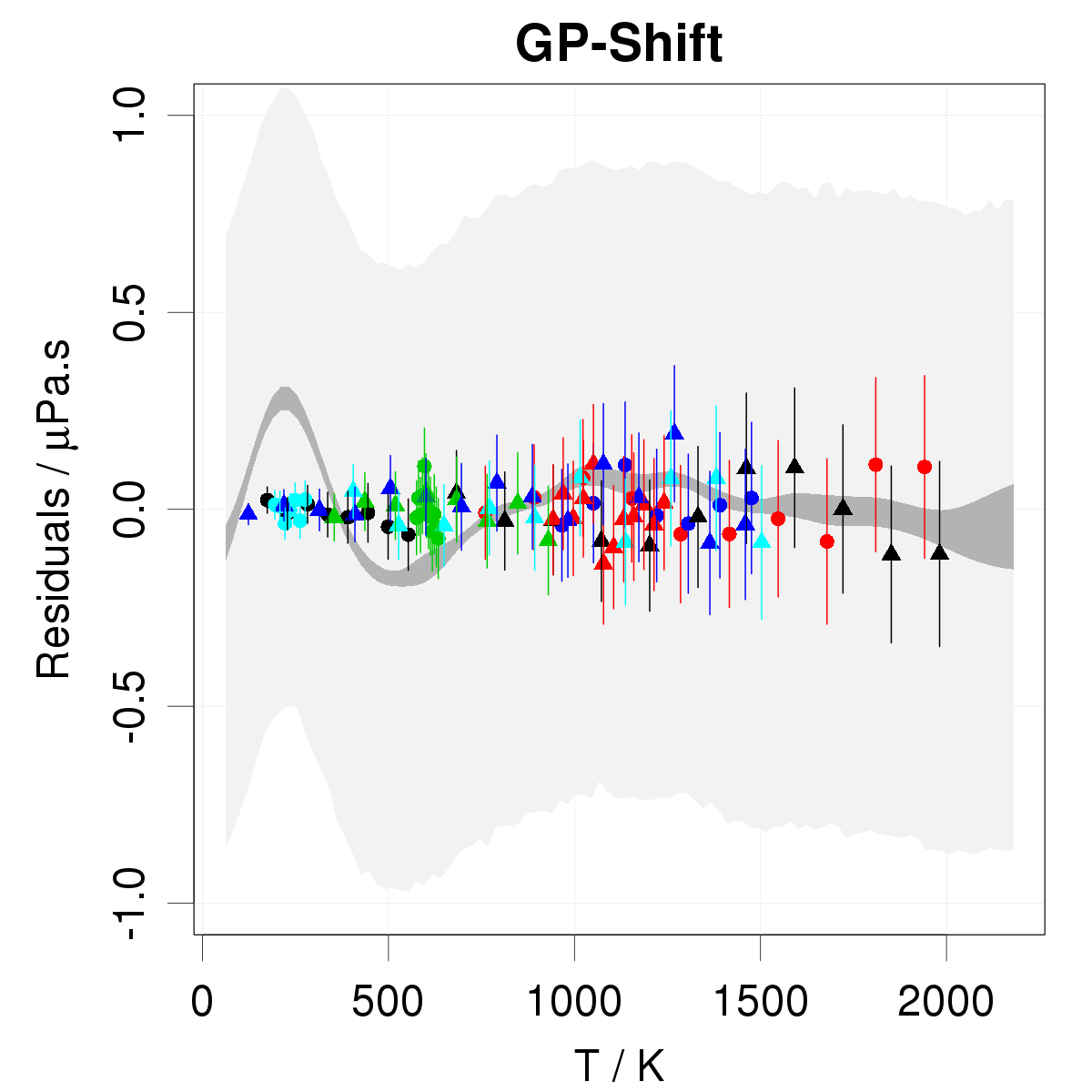}\tabularnewline
\includegraphics[bb=0bp 0bp 1200bp 1200bp,clip,height=5cm]{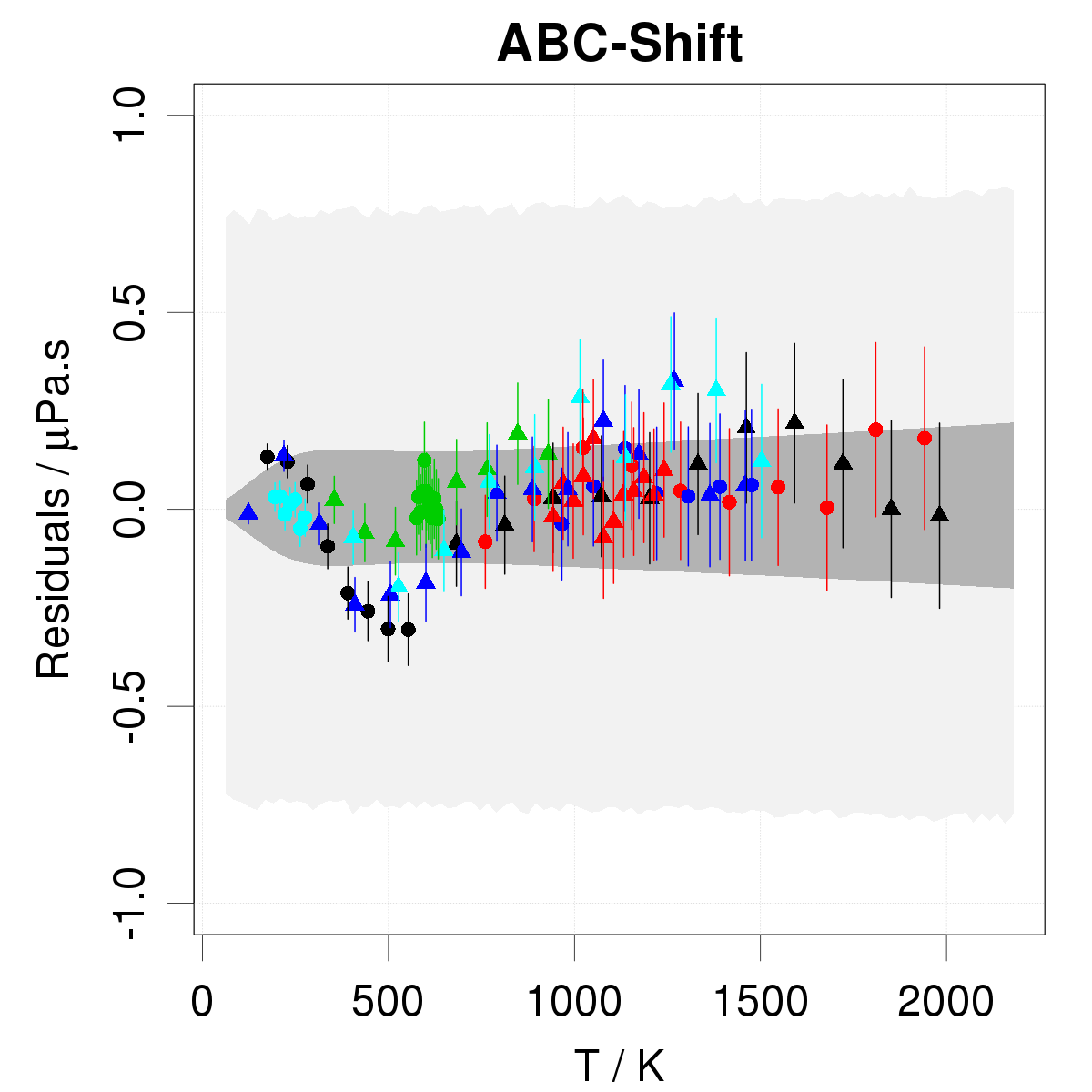} & \includegraphics[bb=0bp 0bp 1200bp 1200bp,clip,height=5cm]{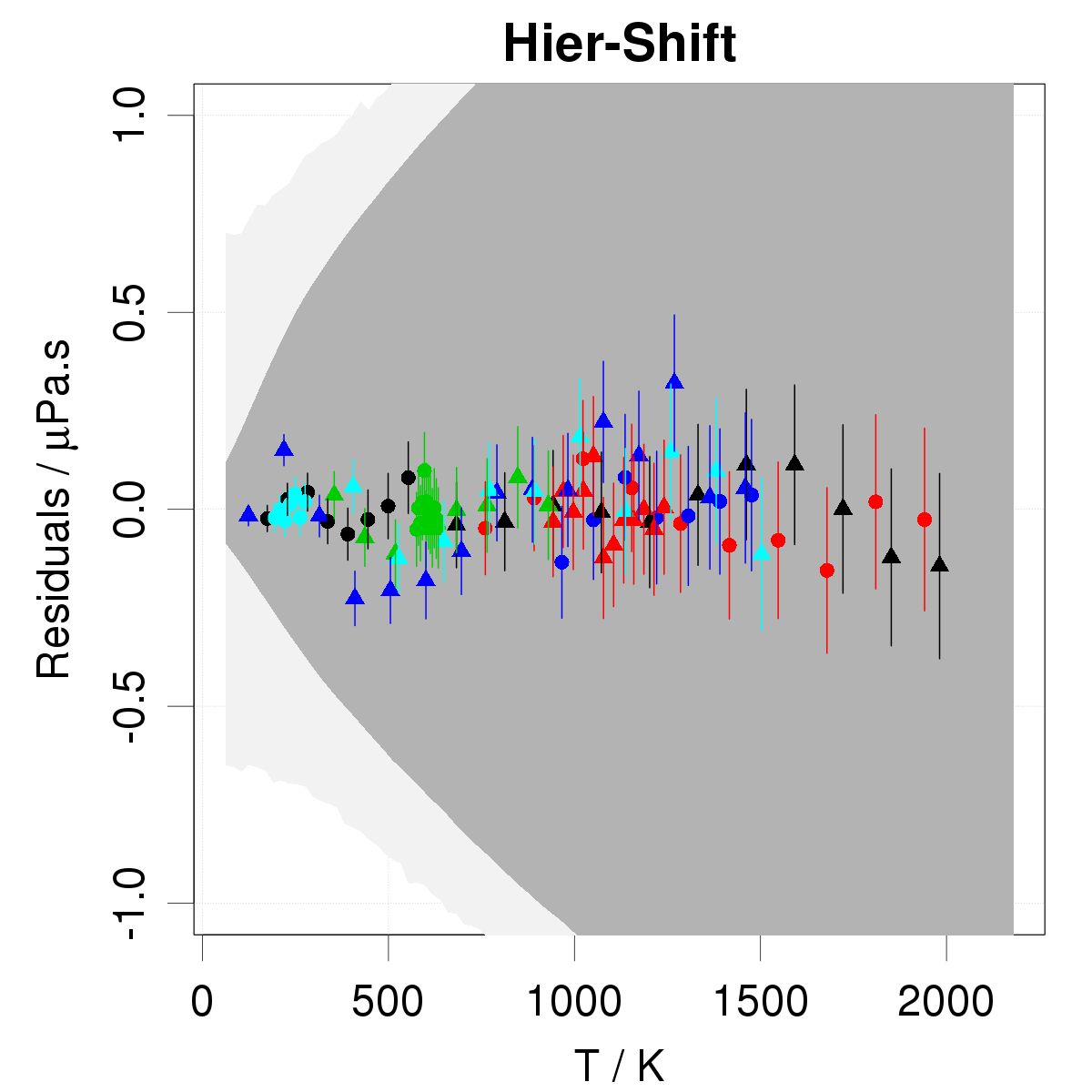} & \includegraphics[bb=0bp 0bp 1200bp 1200bp,clip,height=5cm]{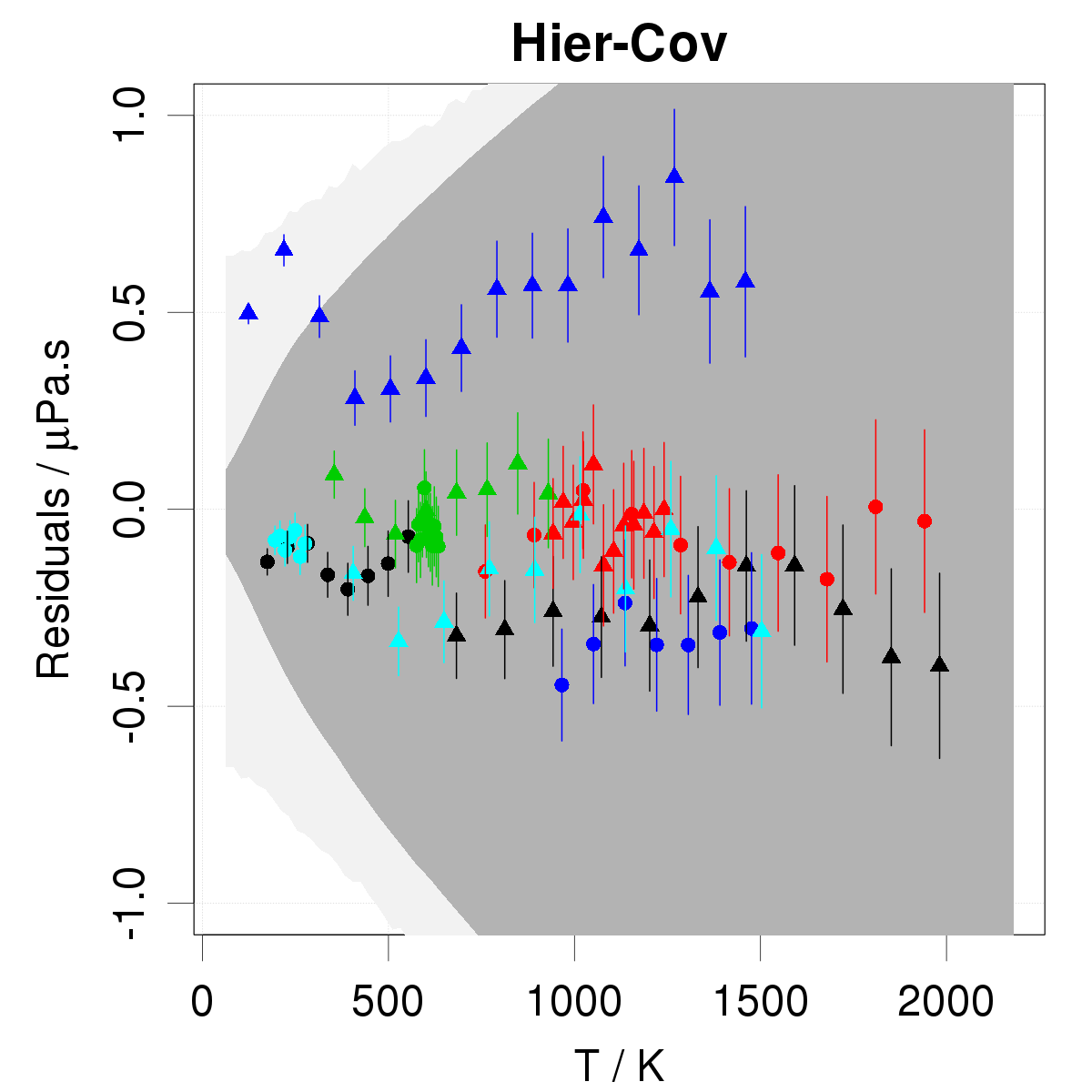}\tabularnewline
\includegraphics[bb=0bp 0bp 1200bp 1200bp,clip,height=5cm]{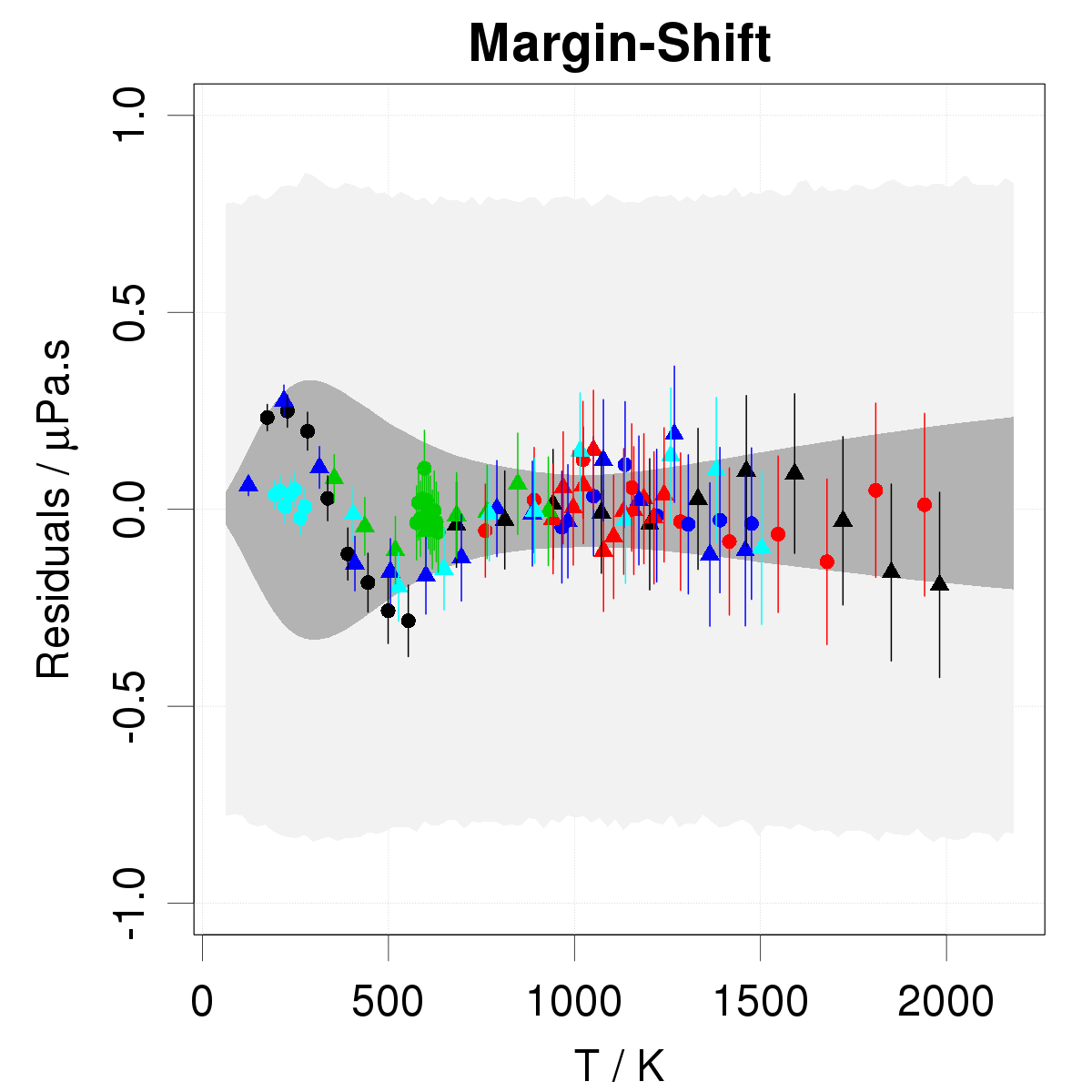} &  & \tabularnewline
\end{tabular}
\par\end{centering}
\caption{\label{fig:Predict-Synth3}Calibration residuals and posterior prediction
intervals for data SD-3. For each model, the results are centered
on the maximum a posteriori results (MAP). The dark gray band represents
the model prediction 95\,\% probability interval ($p_{M}$) and the
light gray area the experiment prediction 95\,\% probability interval
($p_{e}$). }
\end{figure}
\par\end{flushleft}

\noindent \begin{flushleft}
\begin{table}
\noindent \begin{centering}
\begin{tabular}{rrccccccccccccccc}
\hline 
Model &  & $\epsilon_{LJ}$ &  & $\sigma_{LJ}$ &  & $\sigma$ &  & $\tau$  &  & $MR$ &  & $RMSD$ &  & $R_{B}$ &  & $u_{e|D}$\tabularnewline
 &  & (K) &  & ($\text{\AA}$) &  & ($\mu$Pa.s) &  & ($\mu$Pa.s) &  & ($\mu$Pa.s) &  & ($\mu$Pa.s) &  &  &  & ($\mu$Pa.s)\tabularnewline
\cline{1-1} \cline{3-3} \cline{5-5} \cline{7-7} \cline{9-9} \cline{11-11} \cline{13-13} \cline{15-15} \cline{17-17} 
\noalign{\vskip\doublerulesep}
Ref &  & 195 &  & 3.6 &  &  &  & 0.42 &  &  &  &  &  & 1.00 &  & \tabularnewline
\noalign{\vskip\doublerulesep}
\noalign{\vskip\doublerulesep}
Disp &  & 255(3)  &  & 3.528(4)  &  & 0.37(3)  &  & -  &  & 0.00  &  & 0.37  &  & 1.00  &  & 0.38 \tabularnewline
\noalign{\vskip\doublerulesep}
\noalign{\vskip\doublerulesep}
Disp-Shift &  & 263(3)  &  & 3.523(2)  &  & 0.094(9)  &  & 0.4(1)  &  & 0.00  &  & 0.10  &  & 1.00  &  & 0.12 \tabularnewline
\noalign{\vskip\doublerulesep}
GP-Shift &  & 260(3)  &  & 3.524(2)  &  & -  &  & 0.40(6)  &  & 0.00  &  & 0.06  &  & 0.81  &  & 0.07 \tabularnewline
\noalign{\vskip\doublerulesep}
Margin-Shift &  & 260(2)  &  & 3.522(2)  &  & -  &  & 0.38(9)  &  & -0.01  &  & 0.10  &  & 1.00  &  & 0.12 \tabularnewline
$u_{\epsilon},\,u_{\sigma}$ &  & 6(1)  &  & 0.007(2) &  &  &  &  &  &  &  &  &  &  &  & \tabularnewline
$\rho$ &  & \multicolumn{3}{c}{-0.99(1)} &  &  &  &  &  &  &  &  &  &  &  & \tabularnewline
\noalign{\vskip\doublerulesep}
ABC-Shift &  & 252.7(5) &  & 3.5335(8)  &  & -  &  & 0.36(9)  &  & 0.03  &  & 0.12  &  & 2.00  &  & 0.11 \tabularnewline
$u_{\epsilon},\,u_{\sigma}$ &  & 1.7(2) &  & 0.0005(4) &  &  &  &  &  &  &  &  &  &  &  & \tabularnewline
$\rho$ &  & \multicolumn{3}{c}{-0.2(6)} &  &  &  &  &  &  &  &  &  &  &  & \tabularnewline
\noalign{\vskip\doublerulesep}
Hier-Shift &  & 253(2)  &  & 3.535(6)  &  & -  &  & 0.32(9) &  & 0.01  &  & 0.08  &  & 3.90  &  & 0.57 \tabularnewline
$u_{\epsilon},\,u_{\sigma}$ &  & 2(3)  &  & 0.016(5) &  &  &  &  &  &  &  &  &  &  &  & \tabularnewline
$\rho$ &  & \multicolumn{3}{c}{-0.3(5) } &  &  &  &  &  &  &  &  &  &  &  & \tabularnewline
\noalign{\vskip\doublerulesep}
Hier-Cov &  & 253(2) &  & 3.535(6) &  & -  &  & 0.31(8)  &  & 0.07  &  & 0.31  &  & 3.50  &  & 0.66 \tabularnewline
$u_{\epsilon},\,u_{\sigma}$ &  & 2(3)  &  & 0.016(5) &  &  &  &  &  &  &  &  &  &  &  & \tabularnewline
$\rho$ &  & \multicolumn{3}{c}{-0.3(5) } &  &  &  &  &  &  &  &  &  &  &  & \tabularnewline
\hline 
\end{tabular}
\par\end{centering}
\caption{\label{tab:Fit-results-SD-3}Summary of the main parameters and statistics
for the methods tested on the SD-3 set. The first line (labeled Ref)
provides reference value of the parameters and statistics.}
\end{table}
\begin{figure}
\begin{centering}
\begin{tabular}{ccc}
\includegraphics[bb=0bp 0bp 1200bp 1200bp,clip,height=5cm]{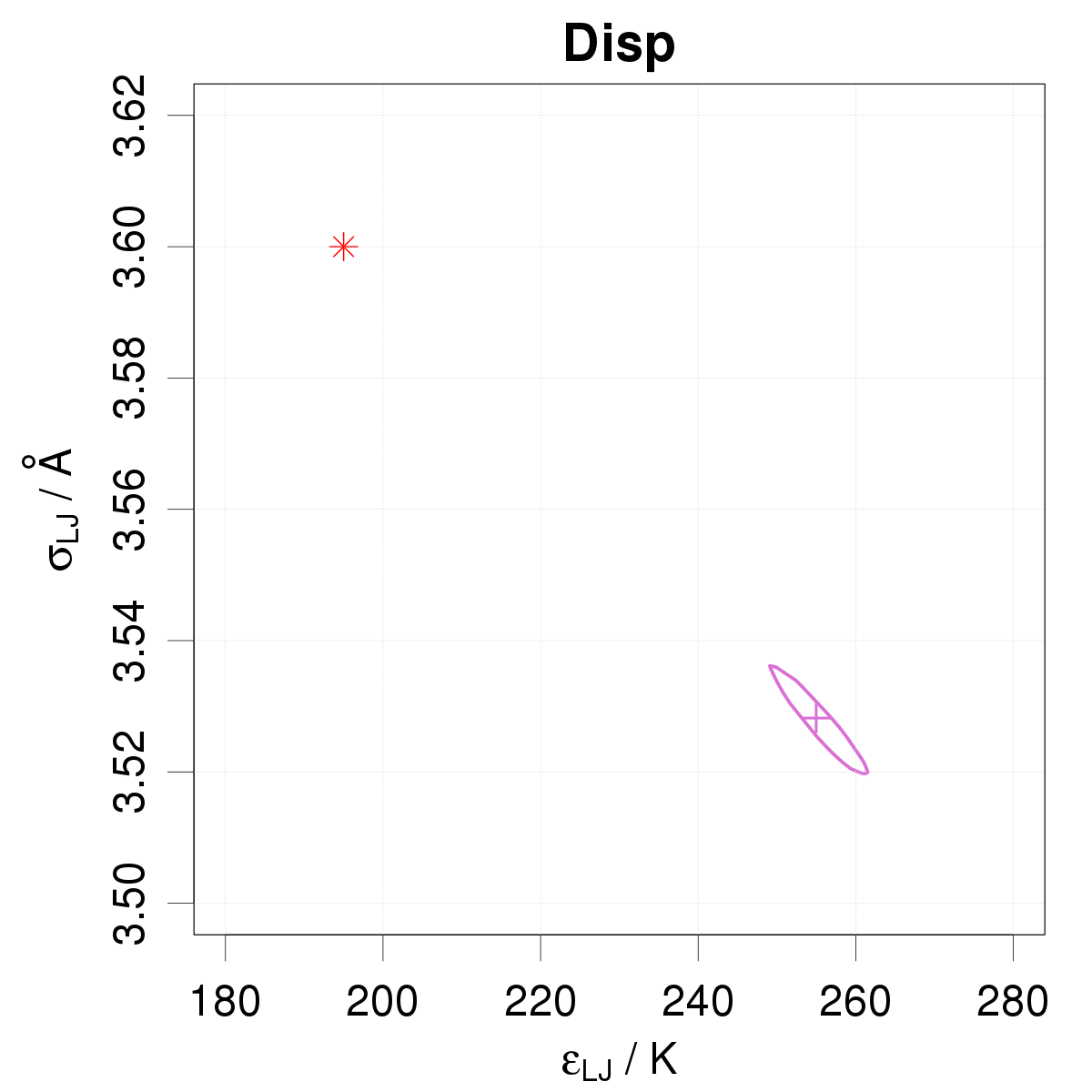} & \includegraphics[bb=0bp 0bp 1200bp 1200bp,clip,height=5cm]{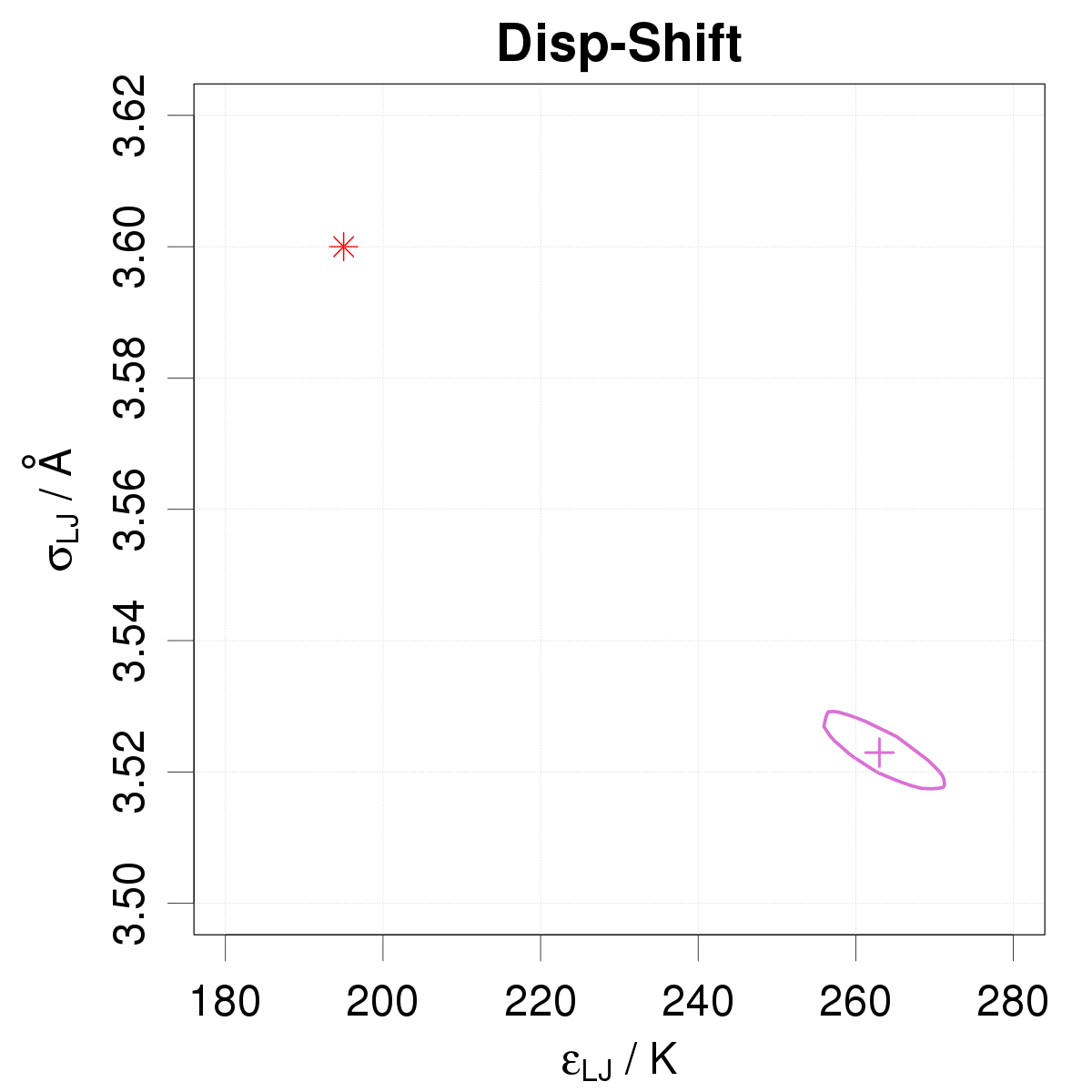} & \includegraphics[bb=0bp 0bp 1200bp 1200bp,clip,height=5cm]{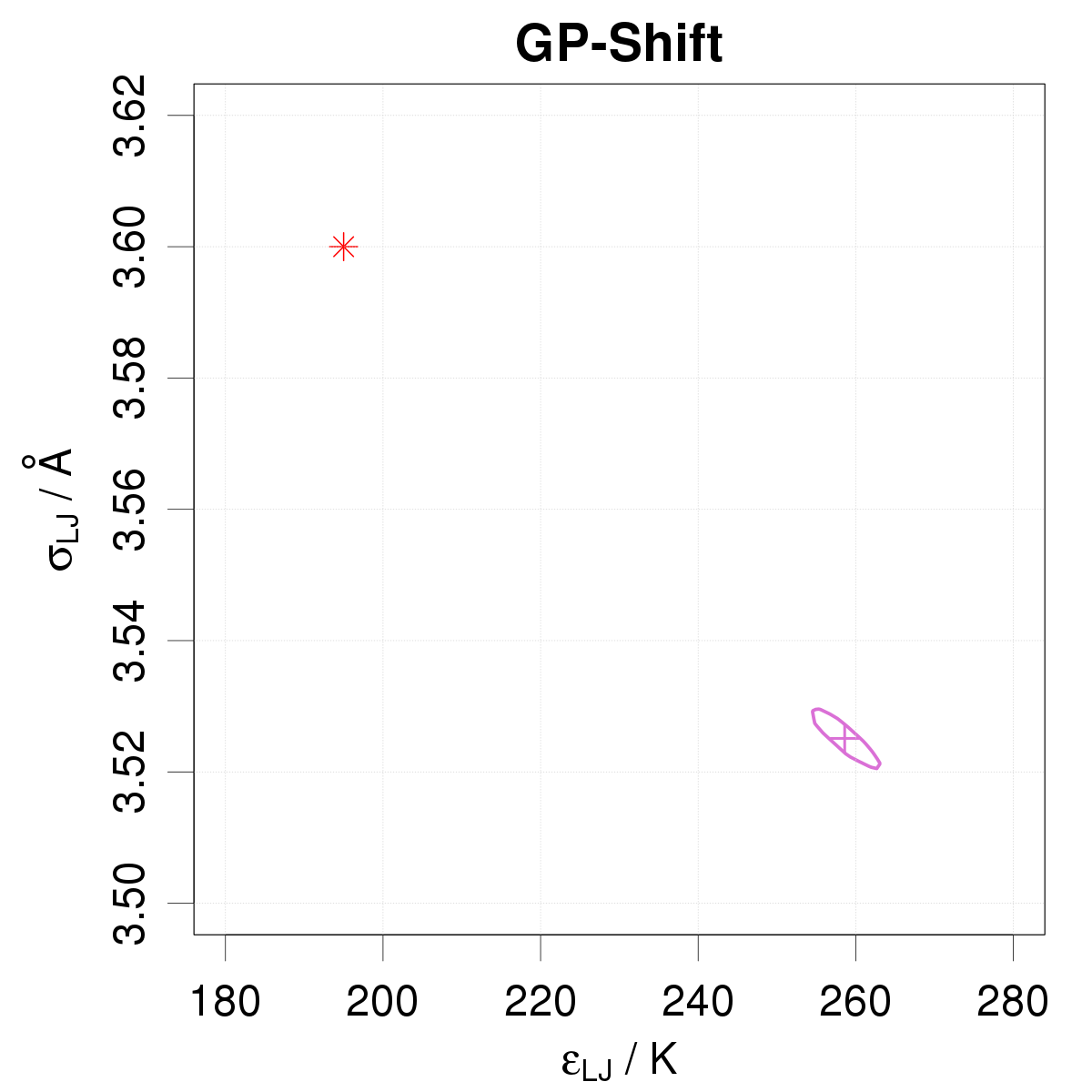}\tabularnewline
\includegraphics[bb=0bp 0bp 1200bp 1200bp,clip,height=5cm]{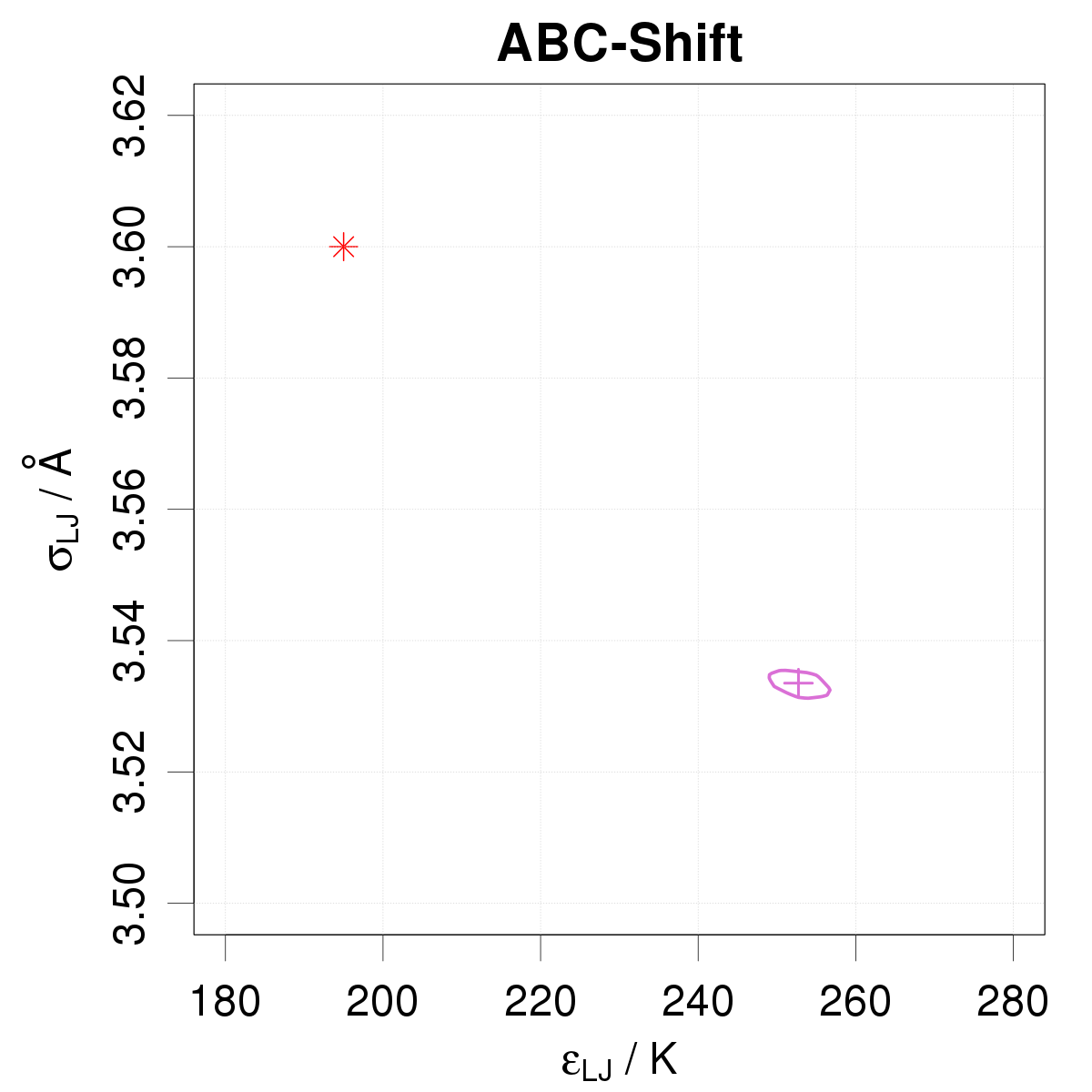} & \includegraphics[bb=0bp 0bp 1200bp 1200bp,clip,height=5cm]{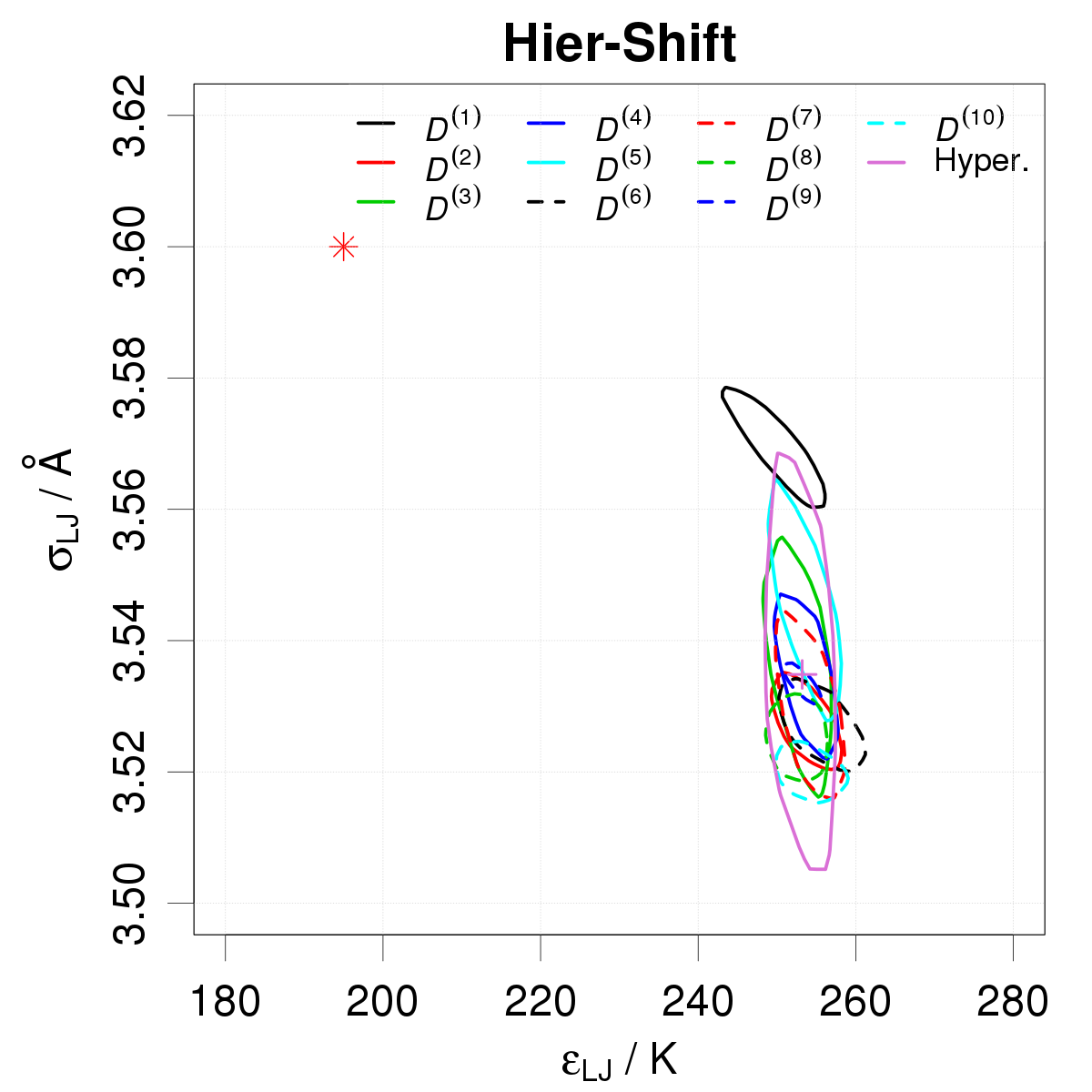} & \includegraphics[bb=0bp 0bp 1200bp 1200bp,clip,height=5cm]{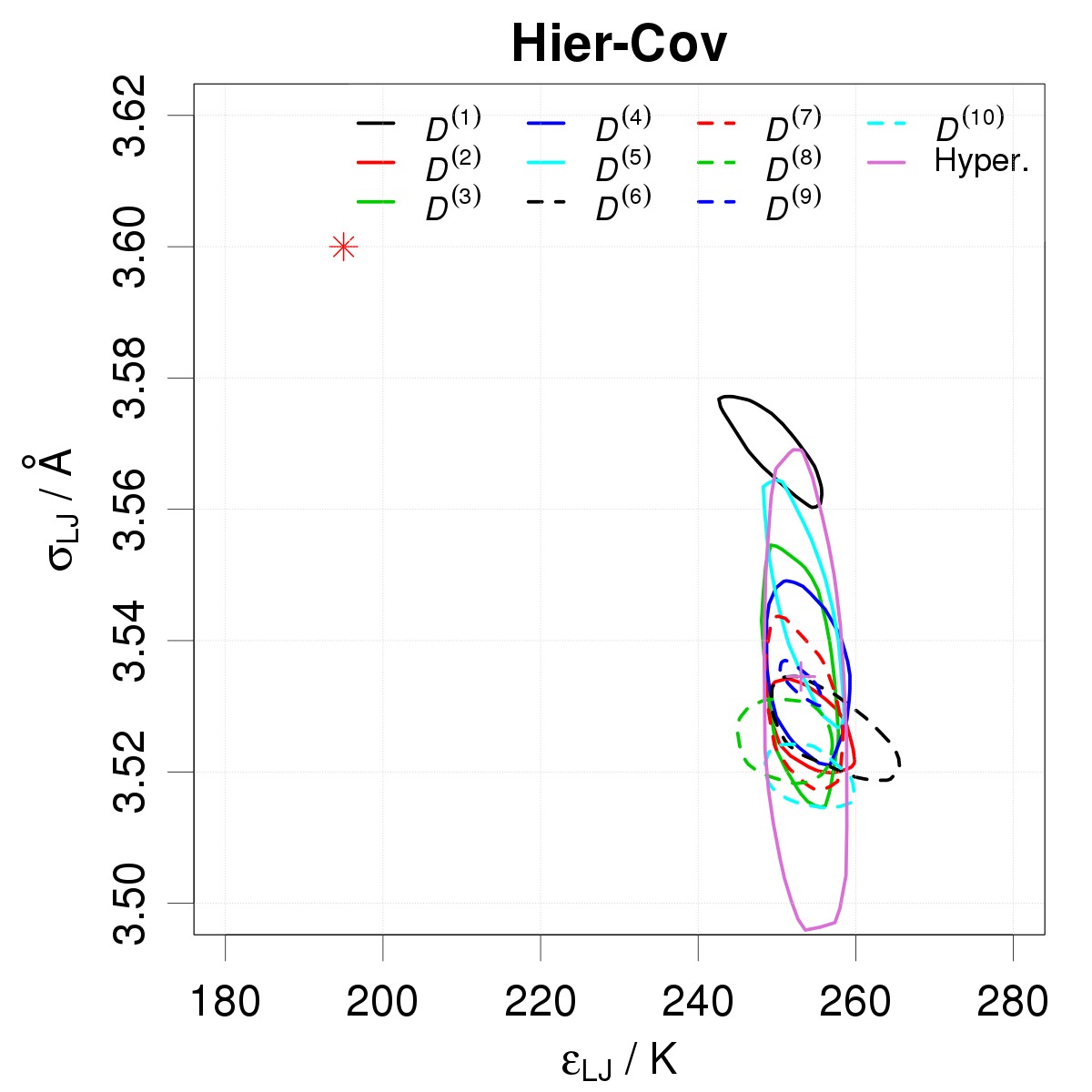}\tabularnewline
\includegraphics[bb=0bp 0bp 1200bp 1200bp,clip,height=5cm]{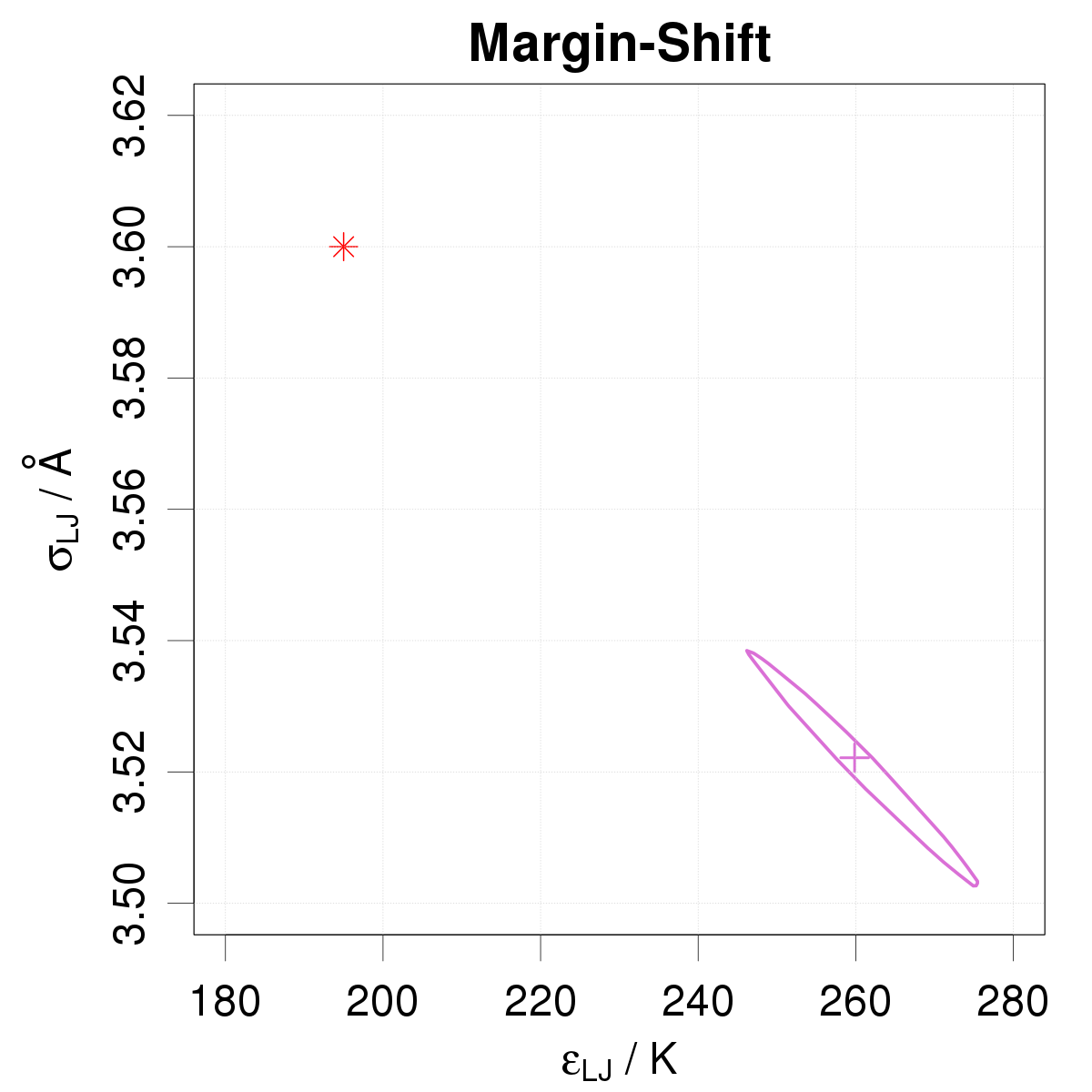} &  & \tabularnewline
\end{tabular}
\par\end{centering}
\caption{\label{fig:Sample-Synth3}Posterior samples of LJ parameters calibrated
on data SD-3: the elliptic contours represent 95~\% probability intervals;
the red cross depicts the reference value of the parameters. }
\end{figure}
\par\end{flushleft}

\noindent \begin{flushleft}
\begin{figure}
\begin{centering}
\includegraphics[bb=0bp 0bp 1800bp 1200bp,clip,width=1\textwidth]{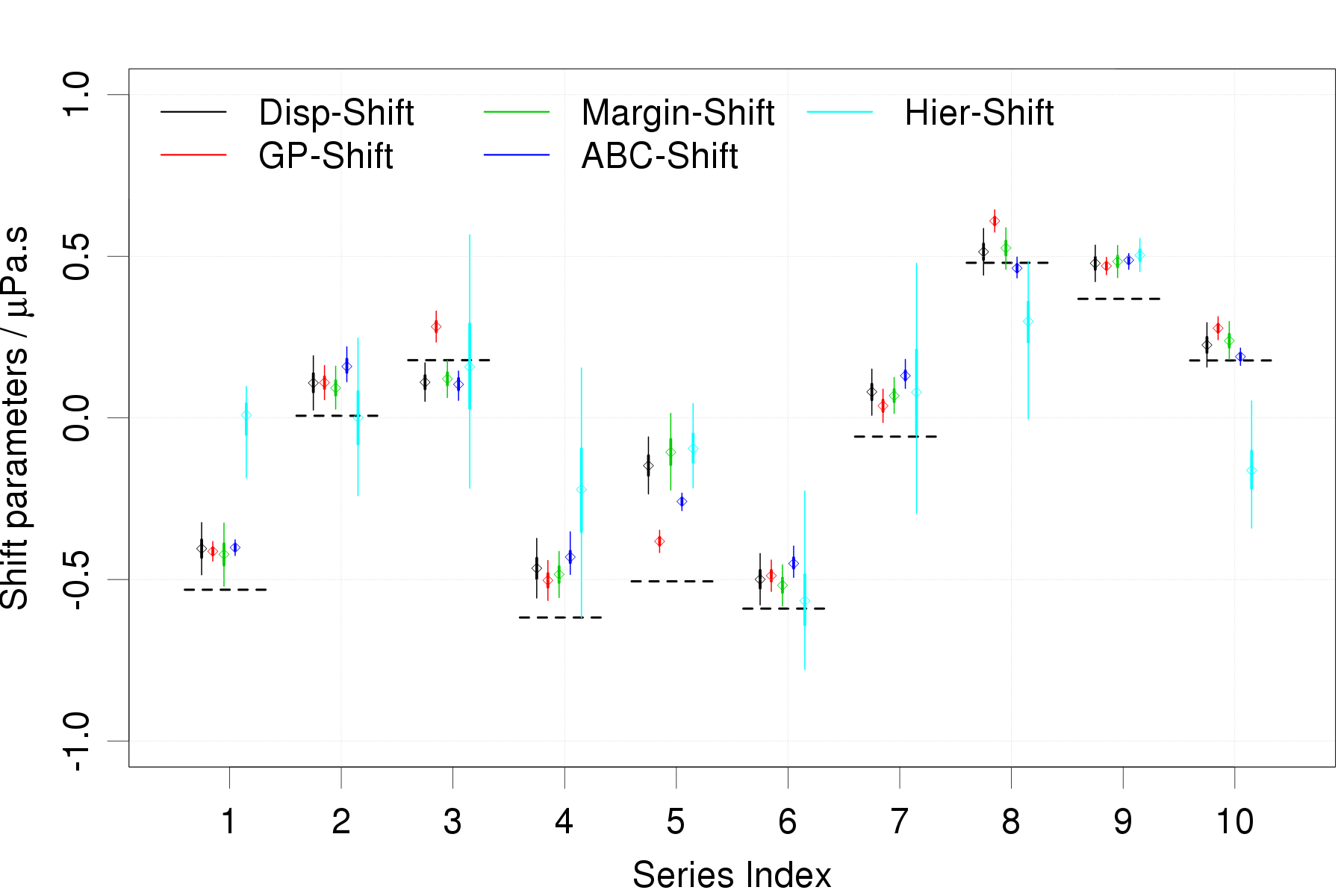}
\par\end{centering}
\caption{\label{fig:Shifts-Synth3}95\% confidence intervals for the shift
parameters recovered by five methods from the SD-3 dataset. The horizontal
bars represent the true values. }
\end{figure}
\par\end{flushleft}

\subsection{Krypton viscosity}

We revisit now an example used by Wu \emph{et al.} \cite{Wu2015}
to illustrate their hierarchical bayesian model, which aims to account
for discrepancy between series in a compound set of data originating
from several experimental setups and physical conditions. This example
is a real-life realization of the synthetic SD-3 set and we test the
same set of methods, except the Hier method defended by Wu \emph{et
al.} \cite{Wu2015} ($\mathcal{M}_{H1}$model) and its variants Hier-Shift
and Hier-Cov, which have been invalidated on the synthetic datasets. 

The dataset used here includes 50 points from 5 series of viscosity
measurements for Kr at temperatures covering a 120\,\textendash \,2000\,K
range. These experiments have been evaluated and selected by Bich
\cite{Bich1990} in his review of the viscosity of monoatomic gases,
where measurement uncertainties are also provided for all data. The
dataset and references are provided in Appendix~\ref{sec:Reference-data}.

The calibration/prediction results are shown in Figs.~\ref{fig:Predict-Exptl},
\ref{fig:Sample-Exptl}, and Table~\ref{tab:Fit-results-Exptl}.

Let us first consider the fit of the dataset by the Disp method, in
order to appraise the structure of the residuals (Fig.~\ref{fig:Predict-Exptl}).
The standard deviation of the residuals is $0.44$\,$\mu$Pa.s, and
they display a strong serial correlation. The maximal deviation of
the residuals is about $0.8$\,$\mu$Pa.s, whereas the dispersion
between data series in the $500$-$1000$\,K range is about 5 times
smaller. Series $D^{(3)}$ and $D^{(5)}$ are separated by more than
$1$\,$\mu$Pa.s between $1100$ and $1600$\,K. The data are clearly
inconsistent above $1000$\,K. Model inadequacy is revealed by the
marked tendencies of the residuals, within each series, and overall. 

Looking at the $RMSD$, the best results are achieved, as expected,
by the GP-Shift model. However, it is not able to resolve fully the
residual inconsistency between series $D^{(1)}$, $D^{(3)}$ and $D^{(5)}$
after correction by shift factors, hence a slightly large Birge ratio
($1.3$). With a Birge ratio of $2.1$, the ABC-Shift method achieves
a slightly poorer performance than the Disp-Shift and Margin-Shift
methods, as observed in the SD-3 case. It provides a rather too narrow
model prediction band, notably for the low-T series, $D^{(4)}$. 

The posterior pdfs for the parameters of ABC-Shift and Margin-Shift
are bimodal (Appendix~\ref{sec:Multimodality-in-posterior}). As
seen for SD-1, the modes correspond to extreme values of the hyperparameters
describing the covariance matrix of the LJ parameters. In the present
case both methods explore the modes corresponding to (1) $\rho\simeq-1$
and (2) very small values of $u_{\sigma}$. In case SD-1, the posterior
samples for these methods were located each on one of these modes.
The increase in parameters uncertainty due to the smaller reference
data set might be responsible for the multimodality. In fact, a trimodality,
involving the mode with very small values of $u_{\epsilon}$, was
formerly observed by Pernot \cite{Pernot2016} for the calibration
of Ar viscosity data with 41 points.

The shifts inferred by the relevant methods are shown in Fig.~\ref{fig:Shifts-Exptl}.
They are in global agreement, with a larger inter-method dispersion
for series $D^{(4)}$ and $D^{(5)}$.
\noindent \begin{flushleft}
\begin{figure}
\begin{centering}
\begin{tabular}{ccc}
\includegraphics[bb=0bp 0bp 1200bp 1200bp,clip,height=5cm]{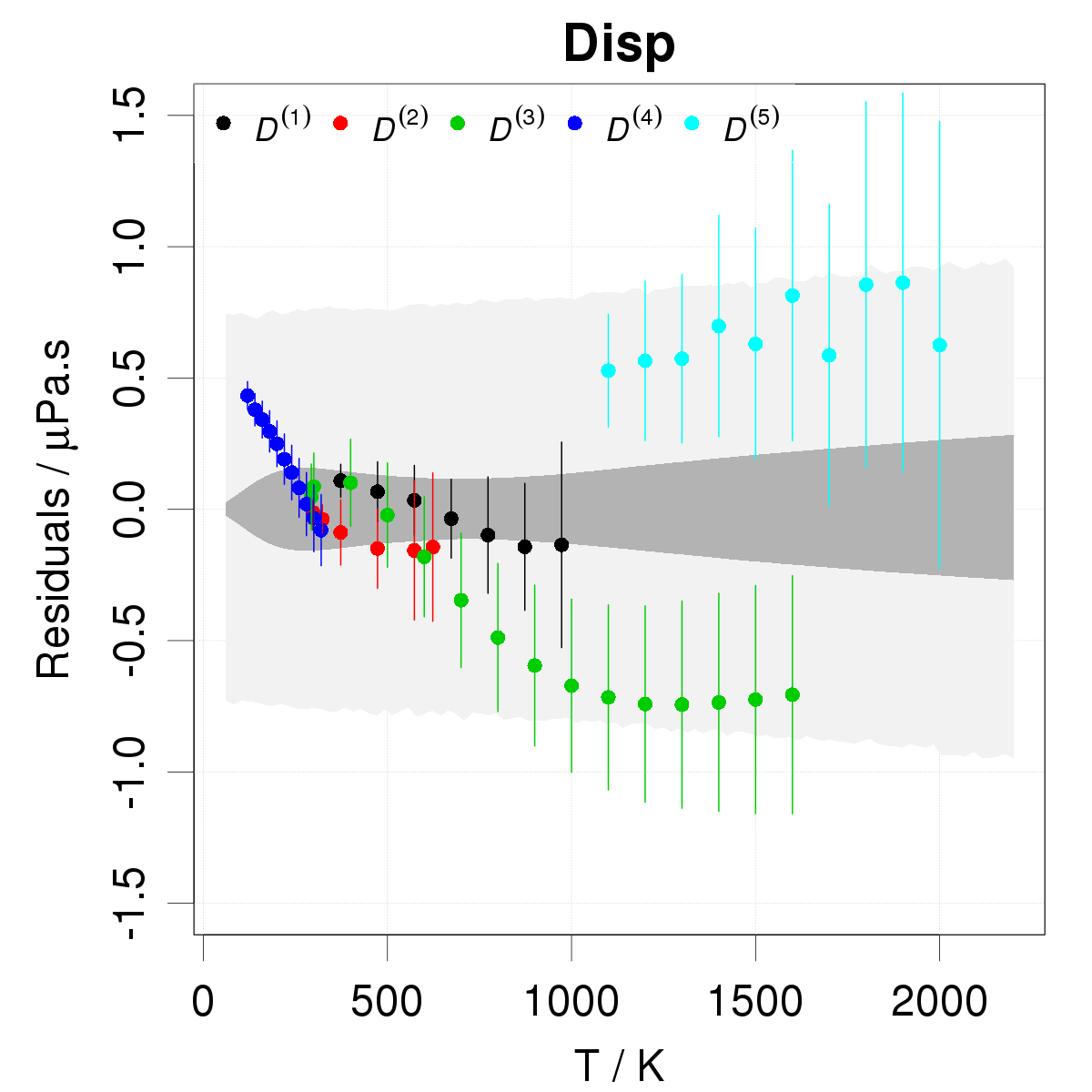} & \includegraphics[bb=0bp 0bp 1200bp 1200bp,clip,height=5cm]{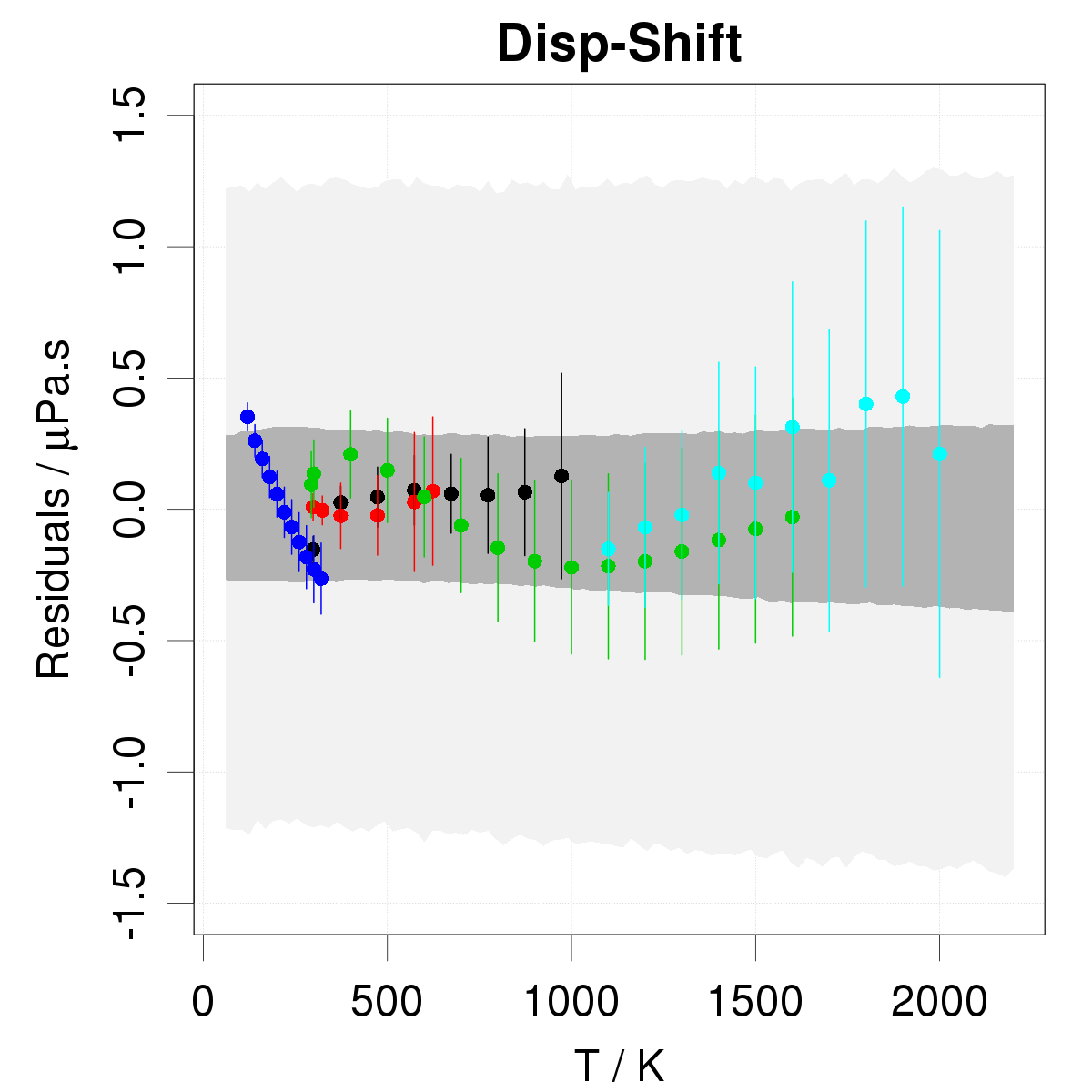} & \includegraphics[bb=0bp 0bp 1200bp 1200bp,clip,height=5cm]{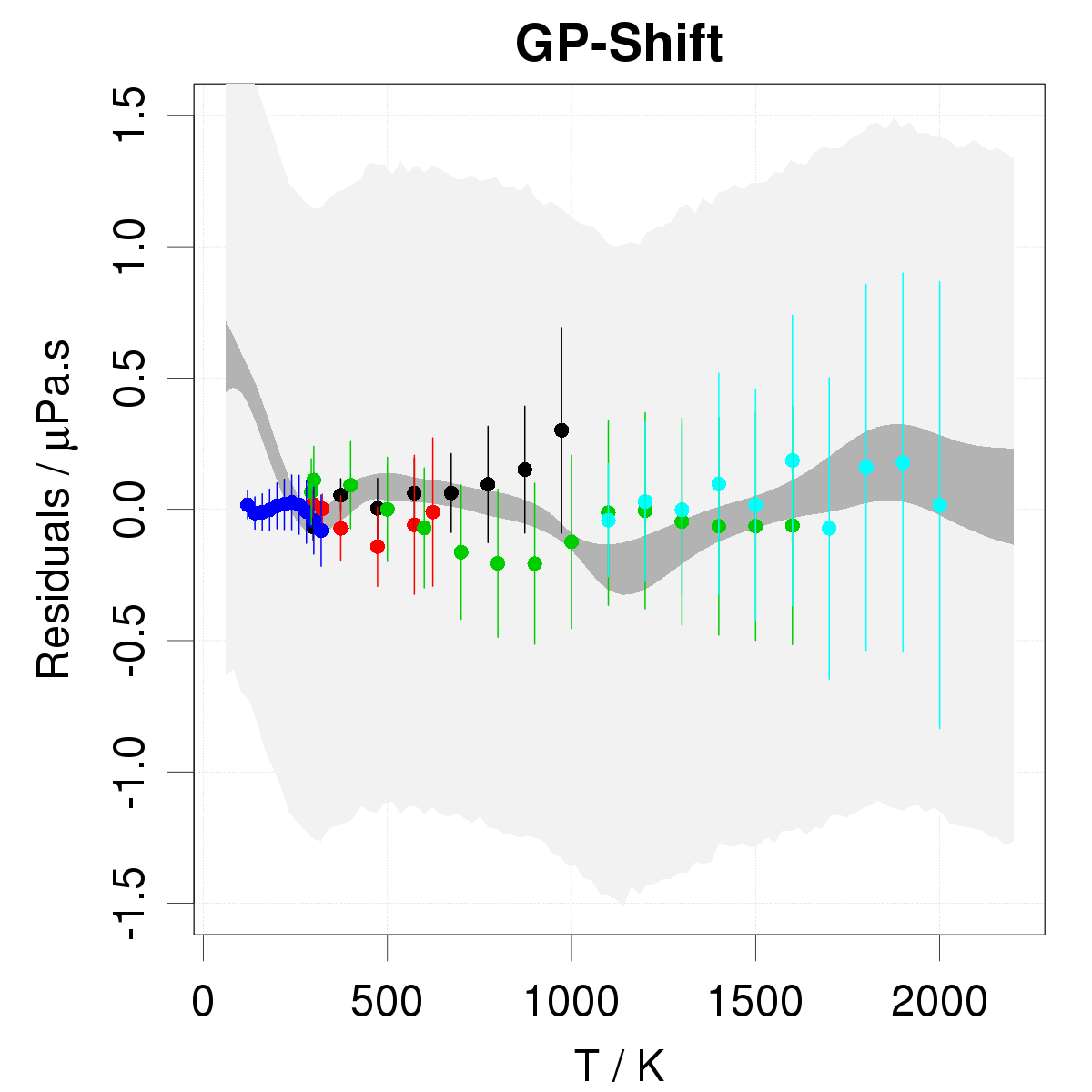}\tabularnewline
\includegraphics[bb=0bp 0bp 1200bp 1200bp,clip,height=5cm]{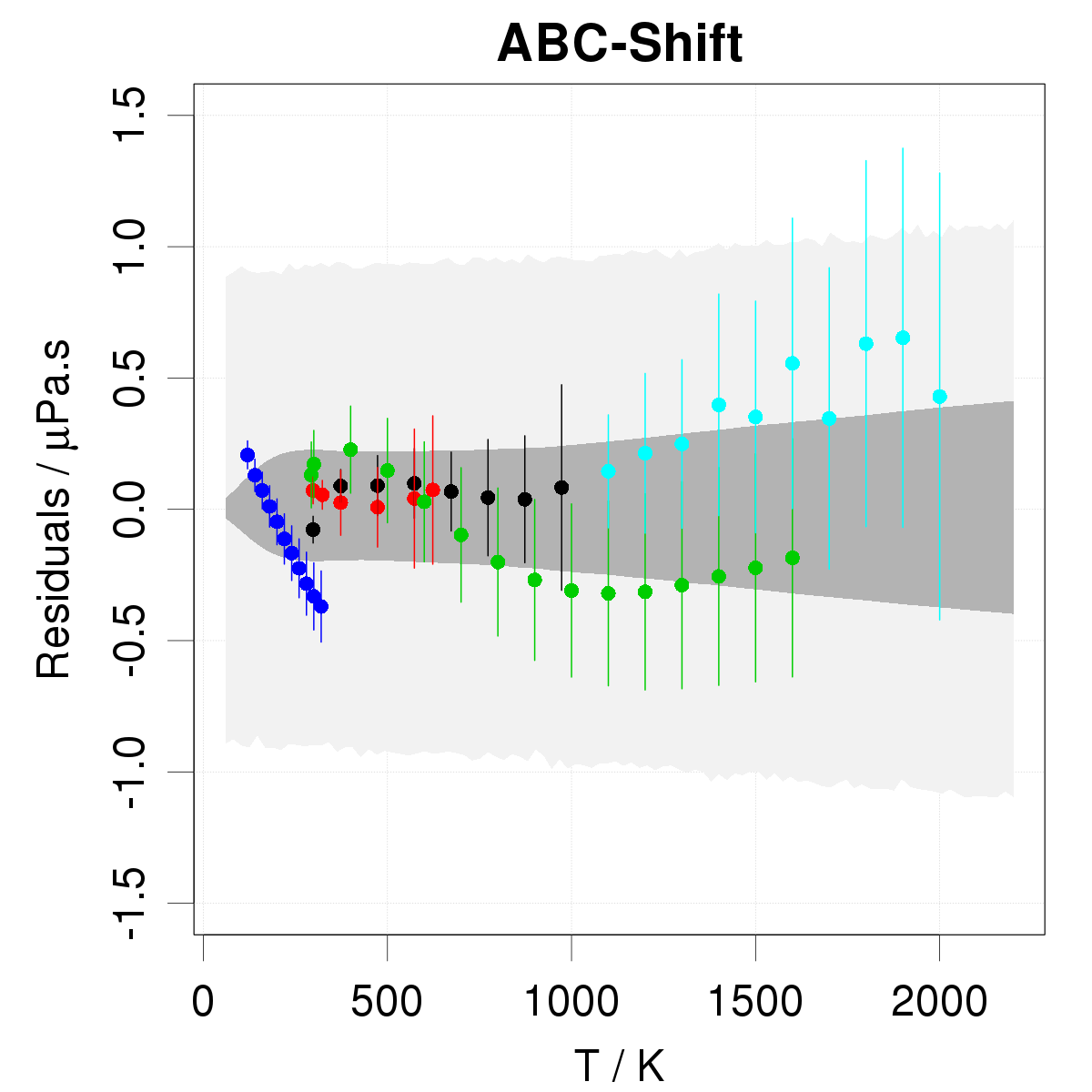} & \includegraphics[bb=0bp 0bp 1200bp 1200bp,clip,height=5cm]{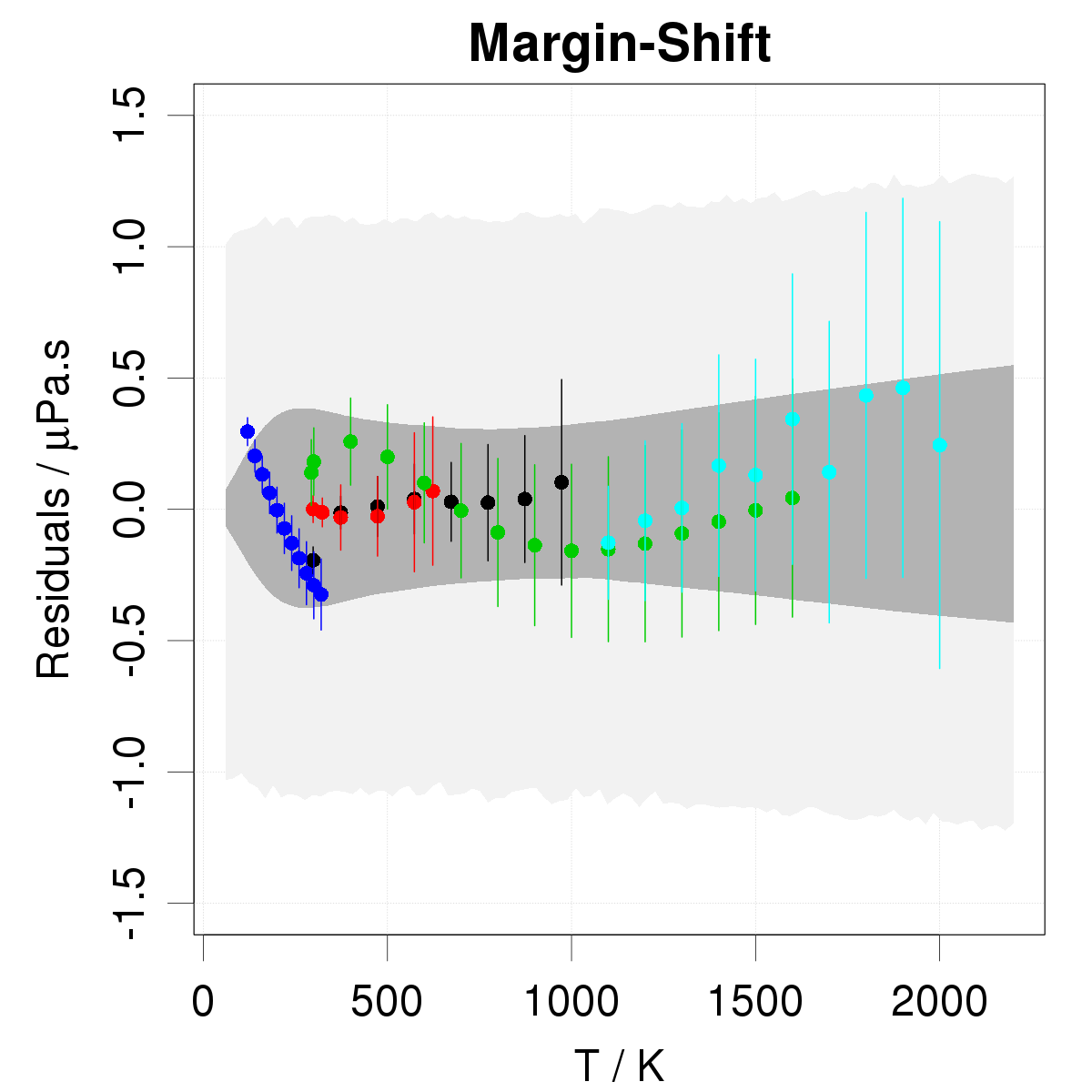} & \tabularnewline
\end{tabular}
\par\end{centering}
\caption{\label{fig:Predict-Exptl}Calibration residuals and posterior prediction
intervals for the Krypton set. For legibility, one has subtracted
the best fit to all curves. The dark gray area represents model prediction
95\,\% probability ($p_{M}$) and the light gray area experiment
prediction 95\,\% probability ($p_{e}$). }
\end{figure}
\par\end{flushleft}

\noindent \begin{flushleft}
\begin{table}
\noindent \begin{centering}
\begin{tabular}{rrccccccccccccccc}
\hline 
Model &  & $\epsilon_{LJ}$ &  & $\sigma_{LJ}$ &  & $\sigma$ &  & $\tau$  &  & $MR$ &  & $RMSD$ &  & $R_{B}$ &  & $u_{e|D}$\tabularnewline
 &  & (K) &  & ($\text{\AA}$) &  & ($\mu$Pa.s) &  & ($\mu$Pa.s) &  & ($\mu$Pa.s) &  & ($\mu$Pa.s) &  &  &  & ($\mu$Pa.s)\tabularnewline
\cline{1-1} \cline{3-3} \cline{5-5} \cline{7-7} \cline{9-9} \cline{11-11} \cline{13-13} \cline{15-15} \cline{17-17} 
\noalign{\vskip\doublerulesep}
Disp &  & 204(3)  &  & 3.535(5)  &  & 0.37(5)  &  & -  &  & 0.03  &  & 0.44  &  & 1.20  &  & 0.41 \tabularnewline
\noalign{\vskip\doublerulesep}
\noalign{\vskip\doublerulesep}
Disp-Shift &  & 192(2)  &  & 3.559(4) &  & 0.14(2) &  & 0.6(2) &  & 0.02  &  & 0.16  &  & 0.96 &  & 0.22\tabularnewline
\noalign{\vskip\doublerulesep}
GP-Shift &  & 192(2)  &  & 3.559(4) &  & - &  & 0.6(1) &  & 0.00  &  & 0.10  &  & 1.30 &  & 0.16\tabularnewline
\noalign{\vskip\doublerulesep}
Margin-Shift &  & 193(2)  &  & 3.557(5)  &  & -  &  & 0.5(2)  &  & 0.03  &  & 0.17  &  & 0.93  &  & 0.24 \tabularnewline
$u_{\epsilon},\,u_{\sigma}$ &  & 5(1) &  & 0.007(4)  &  &  &  &  &  &  &  &  &  &  &  & \tabularnewline
$\rho$ &  & \multicolumn{3}{c}{-0.8(4)} &  &  &  &  &  &  &  &  &  &  &  & \tabularnewline
\noalign{\vskip\doublerulesep}
ABC-Shift &  & 196.5(8)  &  & 3.550(2)  &  & -  &  & 0.3(1)  &  & 0.02  &  & 0.24  &  & 2.10  &  & 0.20 \tabularnewline
$u_{\epsilon},\,u_{\sigma}$ &  & 3(1)  &  & 0.004(4) &  &  &  &  &  &  &  &  &  &  &  & \tabularnewline
$\rho$ &  & \multicolumn{3}{c}{-0.5(6)} &  &  &  &  &  &  &  &  &  &  &  & \tabularnewline
\hline 
\noalign{\vskip\doublerulesep}
\end{tabular}
\par\end{centering}
\caption{\label{tab:Fit-results-Exptl}Summary of the main parameters for the
methods tested on the Krypton set.}
\end{table}
\par\end{flushleft}

\noindent \begin{flushleft}
\begin{figure}
\begin{centering}
\begin{tabular}{ccc}
\includegraphics[bb=0bp 0bp 1200bp 1200bp,clip,height=5cm]{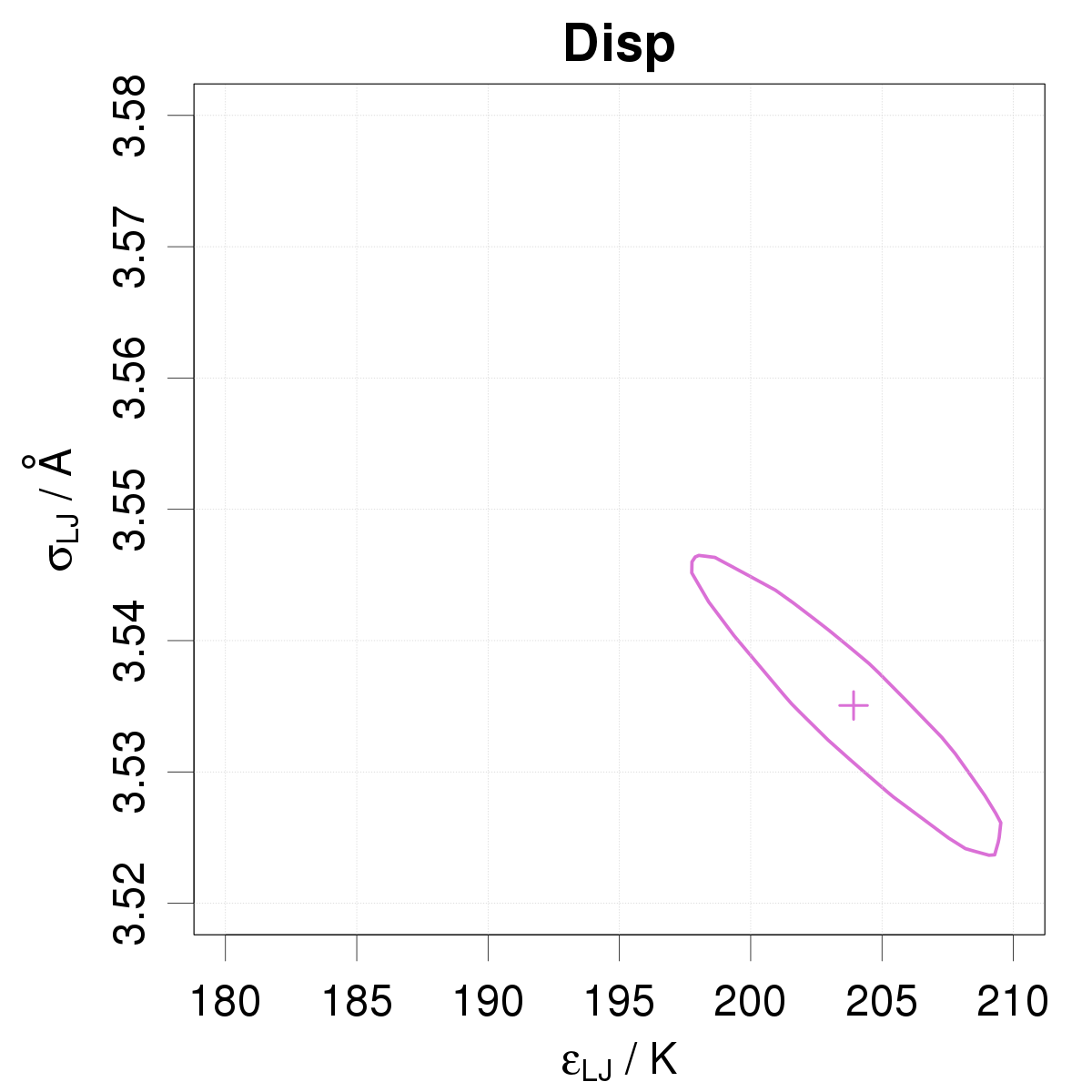} & \includegraphics[bb=0bp 0bp 1200bp 1200bp,clip,height=5cm]{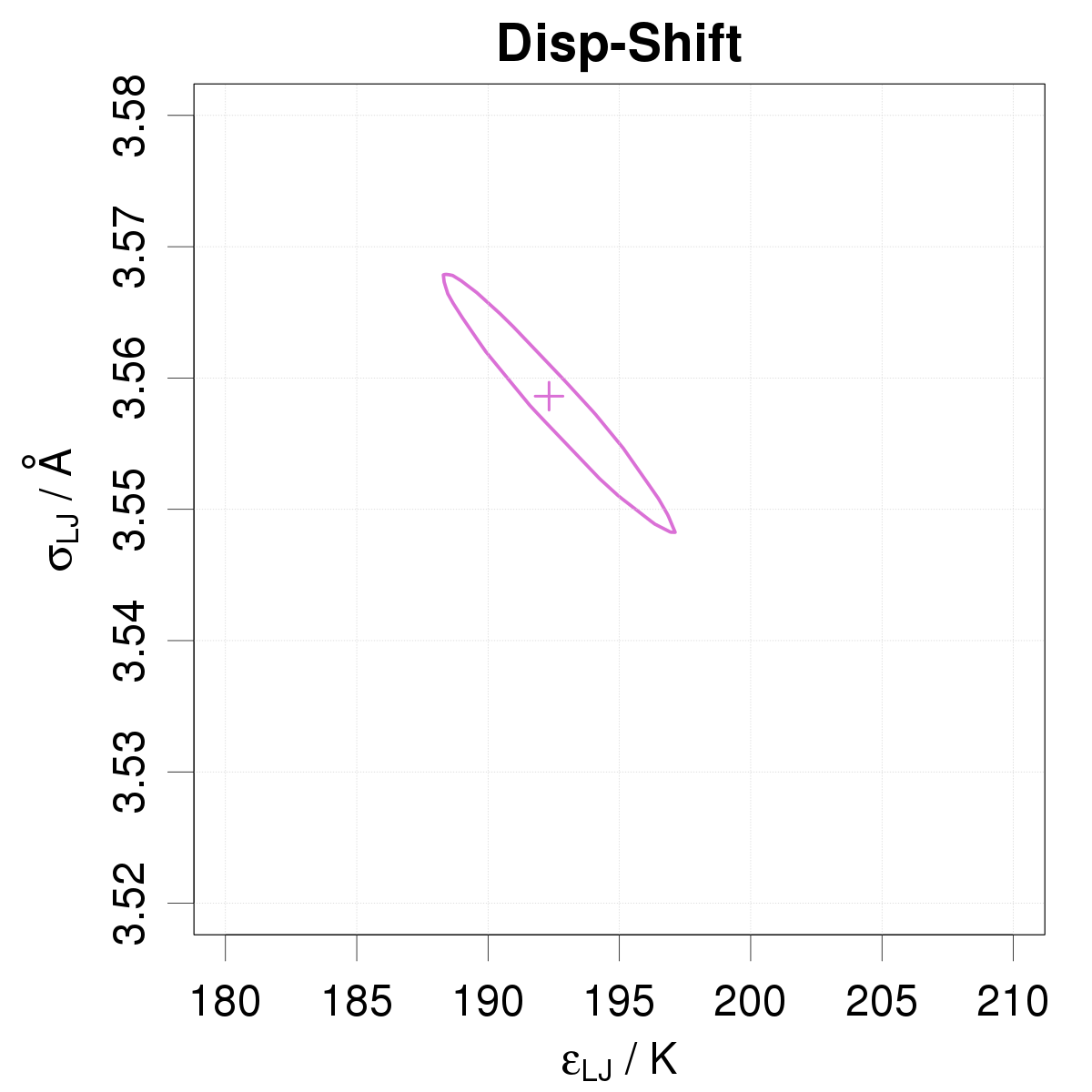} & \includegraphics[bb=0bp 0bp 1200bp 1200bp,clip,height=5cm]{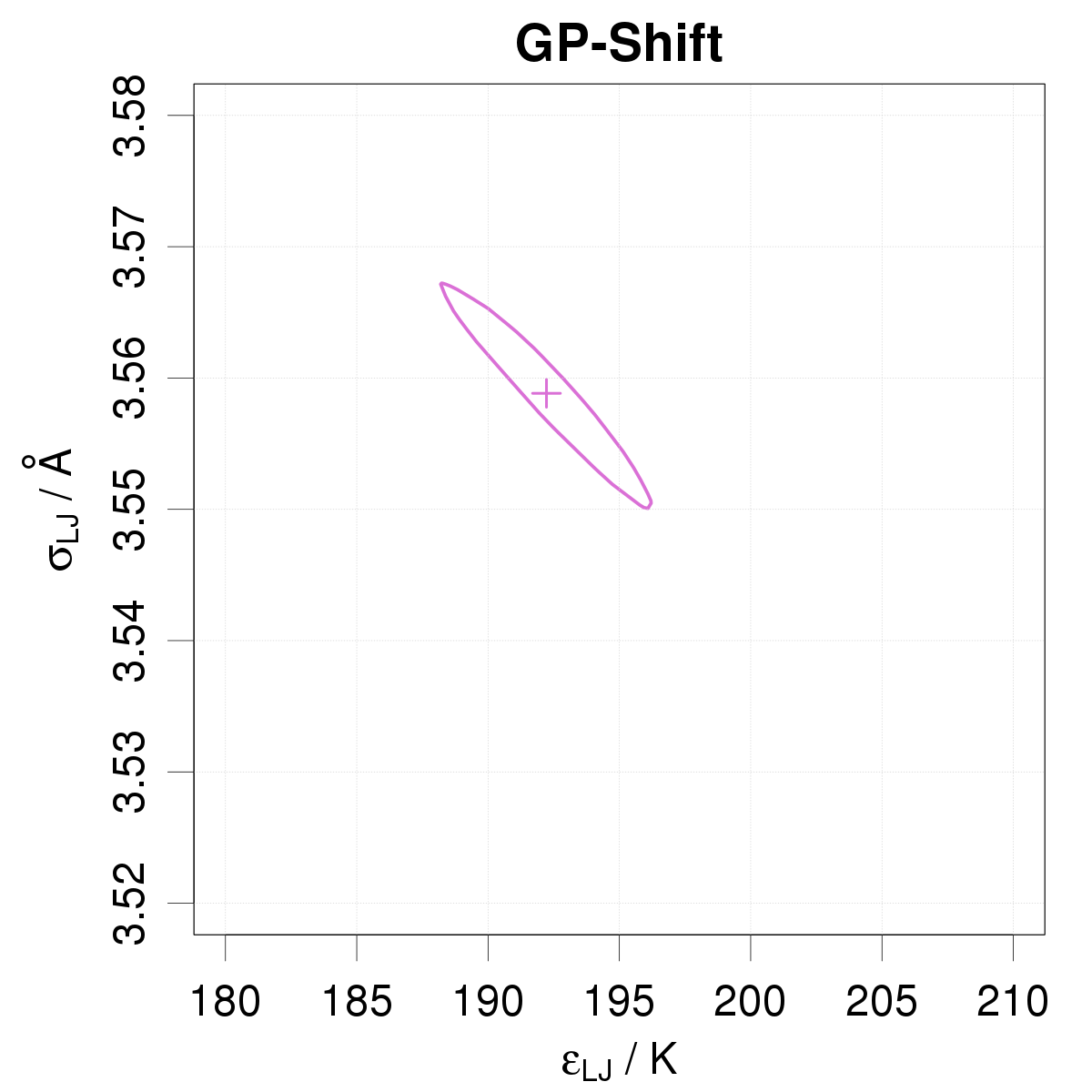}\tabularnewline
\includegraphics[bb=0bp 0bp 1200bp 1200bp,clip,height=5cm]{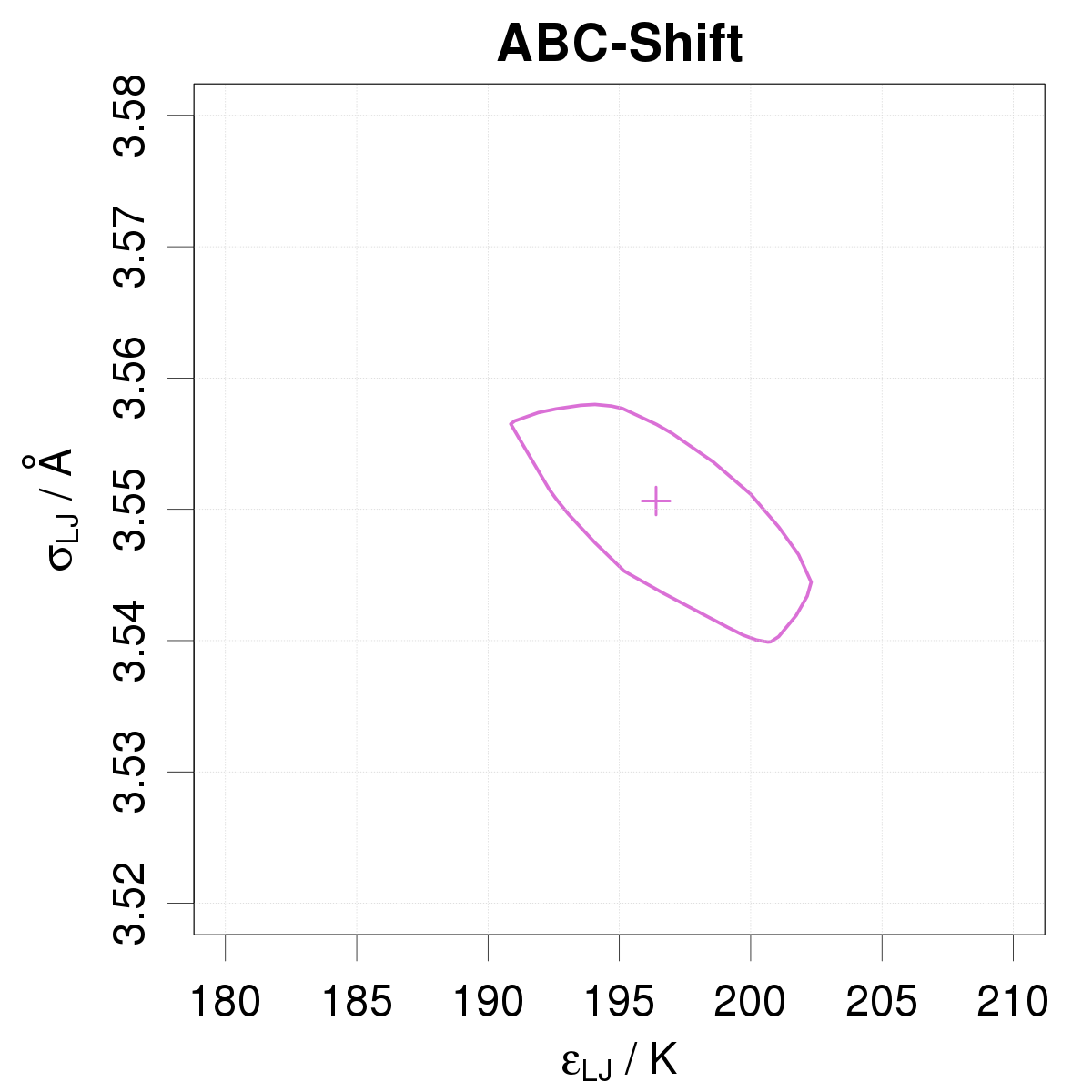} & \includegraphics[bb=0bp 0bp 1200bp 1200bp,clip,height=5cm]{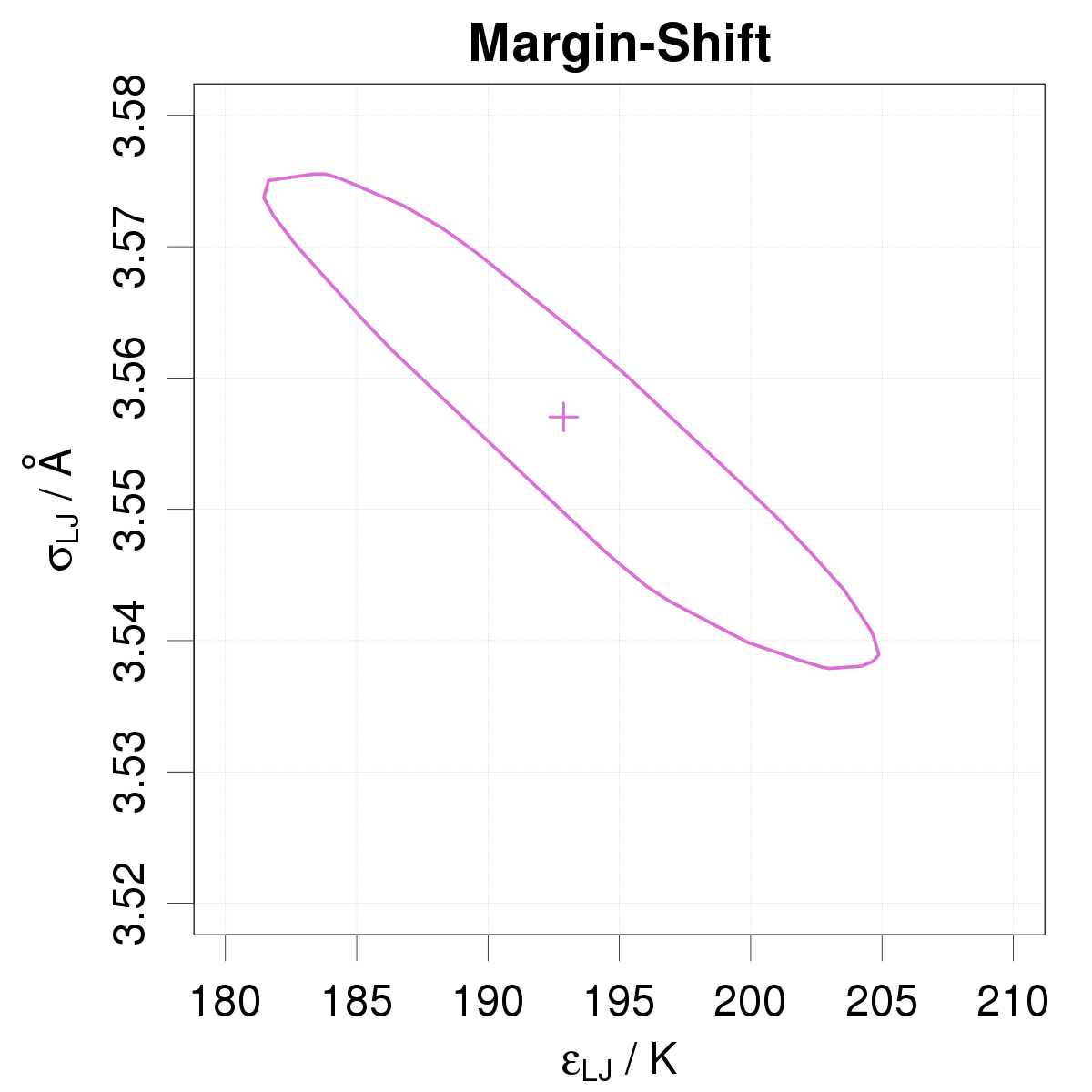} & \tabularnewline
\end{tabular}
\par\end{centering}
\caption{\label{fig:Sample-Exptl}Posterior samples of LJ parameters calibrated
on the Kr dataset: the elliptic contours represent 95~\% probability
intervals.}
\end{figure}
\par\end{flushleft}

\noindent \begin{flushleft}
\begin{figure}
\begin{centering}
\includegraphics[bb=0bp 0bp 1800bp 1200bp,clip,width=1\textwidth]{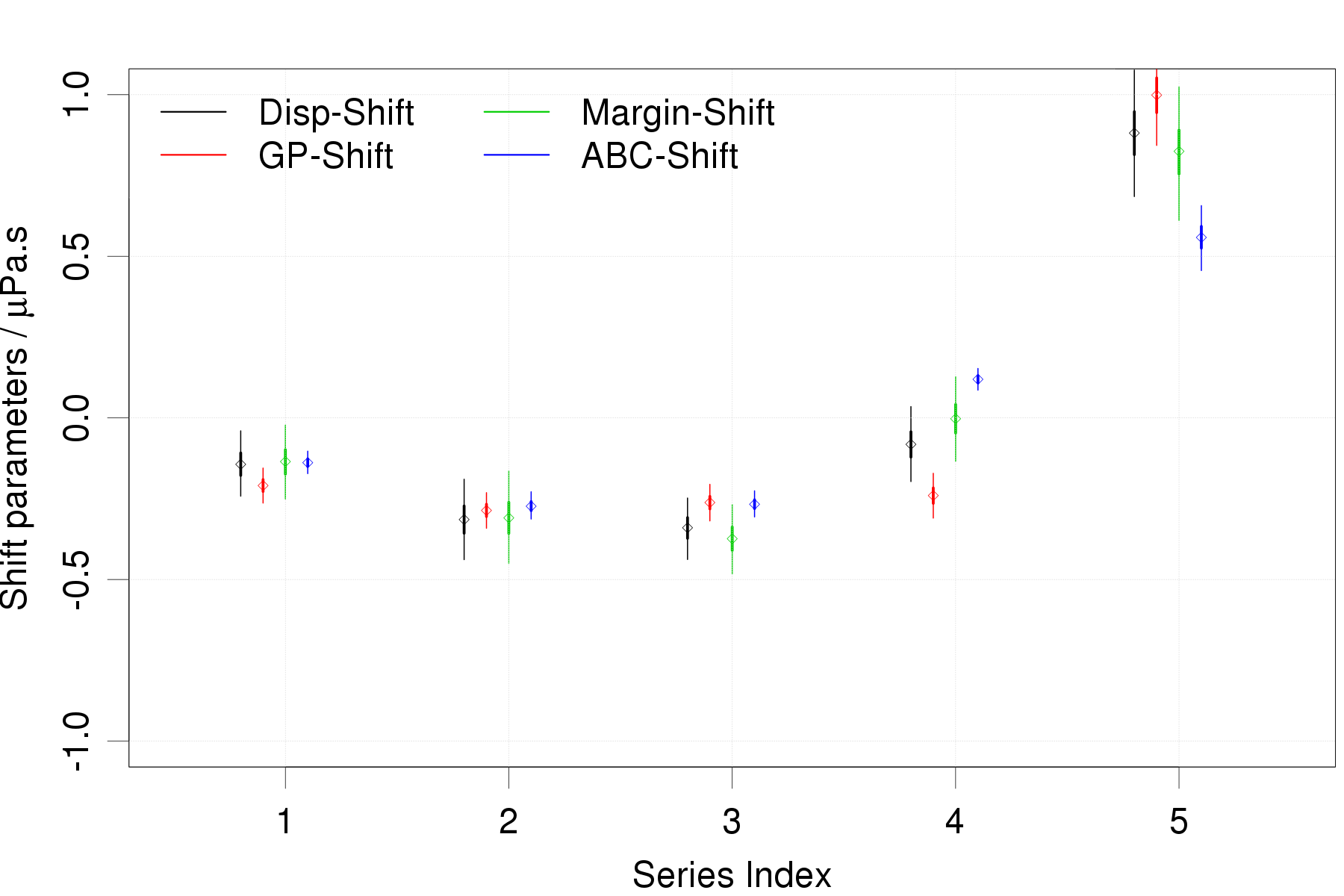}
\par\end{centering}
\caption{\label{fig:Shifts-Exptl}95\% confidence intervals for the shift parameters
recovered by four methods from the Kr dataset.}
\end{figure}
\par\end{flushleft}

\clearpage{}

\section{Discussion\label{sec:Discussion}}

A first observation resulting from the analysis of cases SD-1 and
SD-3 is that LJ parameters are very sensitive to model inadequacy,
and that none of the tested methods includes the true value of the
parameters in its posterior pdf/sample, even those correcting and/or
accounting for model discrepancy. Model improvement is the only way
to infer unbiased values of physical parameters. On the other hand,
effective values of the parameters can still be used to provide useful
predictions.

\subsection{Error sources discrimination}

Among the methods proposed in the literature to deal with calibration/prediction
of physico-chemical models, not all are able to deal with the combination
of inconsistent data and model errors. The numerical experiments in
the previous section have shown that it is essential to treat both
error sources by distinct/complementary statistical models. This is
well achieved by combination methods, such as Disp-Shift, GP-Shift,
Margin-Shift and ABC-Shift.

Wu \emph{et al.} \cite{Wu2015} have proposed a hierarchical model
on the LJ parameters to analyze ``cross-experiment uncertainty''
because it provides a higher prediction uncertainty than the standard
model. In our simulations, the idea of absorbing data inconsistency
in parameters uncertainty (model Hier) has been shown to fail systematically
by leading to (largely) overestimated prediction uncertainties. Even
when combining this approach with an explicit treatment of data inconsistency
(models Hier-Shift or Hier-Cov), the interactions between LJ parameters
and shift parameters lead to exaggerated prediction uncertainties.
Note that this is not a global rejection of hierarchical modeling,
which is perfectly sound and useful to infer, for instance, the shifts
of inconsistent datasets (model Shift).

\subsection{Posterior pdf multimodality in ABC \& Margin }

In PUI methods describing the ensemble of the LJ parameters by a multivariate
normal distribution $p(\boldsymbol{\vartheta}_{M}|\boldsymbol{\mu}_{\vartheta_{M}},\boldsymbol{V}_{\vartheta_{M}})$,
one observes typically a concentration of the hyperparameters on one
or several of the modes corresponding to extreme values of the covariance
matrix parameters, $u_{\sigma}$, $u_{\epsilon}$ and $\rho$ in the
present 2D case. The Margin method tends to favor an extreme negative
correlation coefficient ($\rho\simeq-1$), which is not the case for
the ABC method, leading to different parameter samples and prediction
bands. Besides the fact that it makes these methods difficult to calibrate
if one has not a good prior for the hyperparameters, the multimodality
might be problematic in the sense that it makes the prediction bands
very sensitive to the details of the calibration dataset. Except maybe
for very large calibration datasets, adding more calibration data
might shift the posterior pdf of the parameters onto one or the other
mode in an unpredictable way. As shown by Pernot \cite{Pernot2016},
the prediction bands corresponding to the three modes have different
geometries, and mode-shifting or the redistribution of probability
among modes might result in radical changes in the shape of prediction
bands. 

\subsection{VarInf-type methods}

The simple PUI methods based on data variance scaling (VarInf-type
methods) have been shown to provide poorly reliable prediction bands.
The scaling method based on the temperature analogy (Eq.~\ref{eq:VarInf_Rb})
has been found in case SD-1 to provide exaggerated prediction uncertainty.
A better mean prediction uncertainty was obtained with a prediction
variance based criterion (Eq.~\ref{eq:VarInf_MSE}). This approach
cannot be recommended on a general basis, and should be carefully
validated before being put into use.

\subsection{Calibration transferability}

A difficulty in calibration is the transferability of parameter uncertainty.
In presence of model inadequacy, parameter uncertainty represents
often a small contribution to MPU. In fact, unless there is an identification
problem, parameter uncertainty resulting from a model calibration
is expected to diminish when one increases the size of the calibration
dataset, which is not the case for model inadequacy. The main issue
is therefore the transferability of model inadequacy \cite{Campbell2006,Oliver2015}.
Additive corrections to the model are not transferable to other observables
than the ones for which the correction is designed. Methods such as
Disp and GP, are therefore not useful outside of their calibration
framework. As extensively discussed by Sargsyan \emph{et al.} \cite{Sargsyan2015}
an additional problem with these methods is the difficulty to constrain
their predictions by physical rules or boundary limits. This might
be notably problematic in sequential multiscale simulations if the
predictions at one level violate physical constraints required at
an upper level. 

If transferability or such constraints are not an issue, we have shown
that methods such as Disp and Disp-Shift are very interesting options,
notably because they provide correct model inadequacy coverage and
are easy to communicate and apply for further predictions \cite{Pernot2015}. 

The main approach that has been proposed in the literature to resolve
the transferability problem is to embed model inadequacy in parameter
uncertainty. Several methods to do this have been tested above: VarInf,
Margin, ABC and Hier (Section~\ref{subsec:Model-variance-adaptation}).
We have shown that they all suffer a priori from the problem of inadequate
prediction bands, but our simulations have shown that the ABC and
Margin method perform better than the other ones. Provided the reserves
on this type of method expressed above, and considering that the Margin
method has no or little room for improvement, the ABC method offers
probably the most promising framework. One could for instance envision
additional constraints on the prediction band shape in the ABC pseudo-likelihood
function $p_{reg}$ (Eq.~\ref{eq:ABC}) for a better prediction reliability
and stability.

\section{Conclusion}

We have reviewed and tested a series of standard and advanced statistical
models used in the computational chemistry literature to calibrate
physical models on experimental data. All models have been coded and
run in the same bayesian framework in the \texttt{stan} language.
Their prediction performances have been evaluated on synthetic and
experimental data. 

Focusing on the realistic scenario where measurement uncertainty,
experimental (residual) systematic errors and model inadequacy contribute
significantly to the error budget, we have shown that reliable calibration/prediction
methods depend on the capture and disambiguation of all the error
sources in the statistical analysis model. For instance, in order
to do meaningful predictions, variables of the statistical model have
to be unambiguously attributed to one of the error sources. 

In this scenario, the best performing methods are the GP-Shift, Margin-Shift
and ABC-Shift methods, both with advantages and drawbacks. The GP-Shift
method provides a good interpolator, but it is not usable for extrapolation
out of the calibration range nor to other observables. The Margin-Shift
and ABC-Shift methods do not present the drawbacks of the GP approach,
but their prediction bands are controlled by the model derivatives
and do not necessarily conform with the observed error distribution.
Moreover, we have shown that they are possibly unstable with regard
to the calibration dataset. 

The Hier(archical) methods proposed recently by Wu \emph{et al.} \cite{Wu2015}
have been shown to systematically overestimate prediction uncertainty.
On the one hand, the use of LJ parameters uncertainty to describe
experimental systematic errors conflicts with our definition of unambiguous
attribution of error sources. On the other hand, adaptation of LJ
parameters to the calibration scale of temperature results in a parameter
space too large to permit meaningful predictions. We cannot recommend
these methods in our study cases. 

There is at the moment no perfect and universal solution to the prediction
problem of inadequate physical models \cite{Kalyanaraman2016}. If
model inadequacy is the dominant source of uncertainty, and if the
residuals are not structured, description of model inadequacy by a
single stochastic variable is a robust and efficient alternative (Disp
method) \cite{Pernot2015}. We have also shown that adapting the ABC
method as proposed by Sargsyan \emph{et al.} \cite{Sargsyan2015}
to deal with random and systematic experimental errors offers a promising
framework for reliable predictions and calibration transferability.

\section*{Supporting Information}

The \texttt{R} scripts and \texttt{stan} codes necessary to reproduce
the tables and figures of the article are available on github: \url{https://github.com/ppernot/CalPred}.

\newpage{}

\newpage{}

\appendix
\appendixpage 
\addappheadtotoc

\section{Systematic errors, correlation and weighting\label{sec:Correlation-and-weighting}}

Let us consider two results of repeated measurements, $y_{1}$ and
$y_{2}$, with a common, unknown, measurement bias (systematic error).
The joint posterior pdf of the true value $\mu$ and the bias $s$
is
\begin{eqnarray}
p(\mu,s|y_{1},y_{2},u,\tau) & \propto & u^{-1}\exp\left(-\frac{1}{2u^{2}}\left((y_{1}-\mu-s)^{2}+(y_{2}-\mu-s)^{2}\right)\right)\nonumber \\
 &  & \times\tau^{-1}\exp\left(-\frac{1}{2}\frac{s^{2}}{\tau^{2}}\right)
\end{eqnarray}
where $u$ is the measurement uncertainty due to random effects, and
$s$ is a priori normally distributed with mean $0$ and standard
deviation $\tau$.

Integration on $s$ provides the marginal posterior for $\mu$
\begin{eqnarray}
p(\mu|y_{1},y_{2},u,\tau) & \propto & \int p(\mu,s|y_{1},y_{2},u,\tau)\,ds\\
 & \propto & \frac{1}{\sqrt{(u^{2}+\tau^{2})(1-\rho^{2})}}\\
 &  & \times\exp\left(-\frac{(y_{1}-\mu)^{2}-2\rho(y_{1}-\mu)(y_{2}-\mu)+(y_{2}-\mu)^{2}}{2(u^{2}+\tau^{2})(1-\rho^{2})}\right)\\
 & \propto & (\det\mathbf{V})^{-1/2}\exp\left(-\frac{1}{2}\boldsymbol{R}^{T}\boldsymbol{V}^{-1}\boldsymbol{R}\right)
\end{eqnarray}
where
\begin{equation}
\boldsymbol{R}=\left(\begin{array}{c}
y_{1}-\mu\\
y_{2}-\mu
\end{array}\right),
\end{equation}
\begin{equation}
\boldsymbol{V}^{-1}=\frac{1}{(u^{2}+\tau^{2})(1-\rho^{2})}\left(\begin{array}{cc}
1 & -\rho\\
-\rho & 1
\end{array}\right),
\end{equation}
with correlation coefficient
\begin{equation}
\rho=\tau^{2}/(u^{2}+\tau^{2}).
\end{equation}
The corresponding covariance matrix is thus
\begin{equation}
\boldsymbol{V}=(u^{2}+\tau^{2})\left(\begin{array}{cc}
1 & \rho\\
\rho & 1
\end{array}\right)
\end{equation}

More generally, for $N$ data points, one can design a covariance
matrix
\begin{equation}
\boldsymbol{V}=(u^{2}+\tau^{2})\left(\begin{array}{cccc}
1 & \rho & \cdots & \rho\\
\rho & 1 & \ddots & \vdots\\
\vdots & \ddots & \ddots & \rho\\
\rho & \cdots & \rho & 1
\end{array}\right),
\end{equation}
from which the estimate for $\mu$ will be

\begin{equation}
\mathrm{E}[\mu]=\overline{y}
\end{equation}

\begin{equation}
\mathrm{Var}[\mu]=\frac{1+(N-1)\rho}{N}\,(u^{2}+\tau^{2})
\end{equation}

\begin{itemize}
\item for $\rho=0$, one recovers the limit of independent measurements
and 
\begin{equation}
Var[\mu]=u^{2}/N,
\end{equation}
\item in the limit of very strong serial correlation ($\rho\simeq1)$, \emph{i.e.}
in cases where the random measurement errors are negligible before
the systematic error, one gets the maximal variance for $\mu$
\begin{equation}
Var[\mu]=(u^{2}+\tau^{2})\simeq\tau^{2},
\end{equation}
where the extreme correlation reduces the statistical weight of the
$N$ data points to a single one. 
\end{itemize}
Note that this maximal variance can also be achieved by using $\rho=0$
and data with modified uncertainties
\begin{equation}
\sigma'=\sigma_{s}\times\sqrt{N},
\end{equation}
as in the weighting scheme proposed by Turanyi \emph{et al.} \cite{Turanyi2012}. 

\section{Reference data\label{sec:Reference-data}}

Viscosity measurements of Kr at temperatures covering a 120\,\textendash \,2000\,K
range \cite{Kestin1972,Vogel1984,Dawe1970,Gough1976,Goldblatt1970},
as evaluated and selected by Bich \cite{Bich1990} in his review of
the viscosity of monoatomic gases. 

When compared with the uncertainties reported in the original studies,
the uncertainties provided by Bich are notably larger. As the author
does not explicit his estimation procedure, we assume that they include
systematic errors, unaccounted for in the reference studies. This
is not consistent with the analytic decomposition of errors presented
above, and we preferred to use the measurement uncertainties provided
in the original sources and estimate the systematic errors by adequate
calibration procedures:
\begin{itemize}
\item Kestin \emph{et al. }\cite{Kestin1972}: the nominal reproducibility
values in Table VII;
\item Vogel \cite{Vogel1984}: uncertainty between 0.1 and 0.3\,\%;
\item Dawe and Smith \cite{Dawe1970}, Gough \emph{et al.} \cite{Gough1976}
indicate a maximal error of 0.5\,\%. Assuming a factor 2, we used
0.25\,\%;
\item Goldblatt \emph{et al.} \cite{Goldblatt1970}: uncertainty between
0.15 and 0.4\,\%.
\end{itemize}
\begin{longtable}{cr@{\extracolsep{0pt}.}lr@{\extracolsep{0pt}.}lr@{\extracolsep{0pt}.}l}
\caption{Reference data.}
\tabularnewline
\hline
\hline 
Series  & \multicolumn{2}{c}{T / K} & \multicolumn{2}{c}{v / $\mu$Pas} & \multicolumn{2}{c}{u / $\mu$Pa s}\tabularnewline
\hline 
\endhead
1 \cite{Kestin1972}  & 298&15  & 25&389  & 0&025\tabularnewline
 & 373&15  & 31&219  & 0&031\tabularnewline
 & 473&15  & 38&055  & 0&057\tabularnewline
 & 573&15  & 44&280  & 0&066\tabularnewline
 & 673&15  & 49&995  & 0&075\tabularnewline
 & 773&15  & 55&336  & 0&110\tabularnewline
 & 873&15  & 60&375  & 0&120\tabularnewline
 & 973&15  & 65&195  & 0&200\tabularnewline
\hline 
2 \cite{Vogel1984} & 298&15  & 25&405  & 0&025\tabularnewline
 & 323&15  & 27&335  & 0&027\tabularnewline
 & 373&15  & 31&022  & 0&062\tabularnewline
 & 473&15  & 37&839  & 0&076\tabularnewline
 & 573&15  & 44&089  & 0&130\tabularnewline
 & 623&15  & 47&047  & 0&140\tabularnewline
\hline 
3 \cite{Dawe1970} & 293&20  & 25&070  & 0&063\tabularnewline
 & 300&00  & 25&650  & 0&064\tabularnewline
 & 400&00  & 33&130  & 0&083\tabularnewline
 & 500&00  & 39&700  & 0&099\tabularnewline
 & 600&00  & 45&660  & 0&110\tabularnewline
 & 700&00  & 51&170  & 0&130\tabularnewline
 & 800&00  & 56&340  & 0&140\tabularnewline
 & 900&00  & 61&240  & 0&150\tabularnewline
 & 1000&00  & 65&910  & 0&160\tabularnewline
 & 1100&00  & 70&390  & 0&180\tabularnewline
 & 1200&00  & 74&700  & 0&190\tabularnewline
 & 1300&00  & 78&870  & 0&200\tabularnewline
 & 1400&00  & 82&910  & 0&210\tabularnewline
 & 1500&00  & 86&830  & 0&220\tabularnewline
 & 1600&00  & 90&650  & 0&230\tabularnewline
\hline 
4 \cite{Gough1976} & 120&00  & 10&610  & 0&027\tabularnewline
 & 140&00  & 12&290  & 0&031\tabularnewline
 & 160&00  & 14&010  & 0&035\tabularnewline
 & 180&00  & 15&730  & 0&039\tabularnewline
 & 200&00  & 17&440  & 0&044\tabularnewline
 & 220&00  & 19&120  & 0&048\tabularnewline
 & 240&00  & 20&780  & 0&052\tabularnewline
 & 260&00  & 22&400  & 0&056\tabularnewline
 & 280&00  & 23&980  & 0&060\tabularnewline
 & 300&00  & 25&530  & 0&064\tabularnewline
 & 320&00  & 27&050  & 0&068\tabularnewline
\hline 
5 \cite{Goldblatt1970} & 1100&00  & 71&634  & 0&110\tabularnewline
 & 1200&00  & 76&007  & 0&150\tabularnewline
 & 1300&00  & 80&187  & 0&160\tabularnewline
 & 1400&00  & 84&343  & 0&210\tabularnewline
 & 1500&00  & 88&184  & 0&220\tabularnewline
 & 1600&00  & 92&170  & 0&280\tabularnewline
 & 1700&00  & 95&649  & 0&290\tabularnewline
 & 1800&00  & 99&539  & 0&350\tabularnewline
 & 1900&00  & 103&091  & 0&360\tabularnewline
 & 2000&00  & 106&328  & 0&430\tabularnewline
\hline 
\end{longtable}

\section{Multimodality in posterior pdf\label{sec:Multimodality-in-posterior}}

We present here two pairs-plots of the posterior pdf samples for a
subset of parameters of the ABC-Shift (Fig.~\ref{fig:S1}) and Margin-Shift
(Fig.~\ref{fig:S2}) methods applied to the Kr data. The bimodality
of the parameter distribution is particularly apparent on the histograms
of $u_{\epsilon}$ and $u_{\sigma}$. 
\noindent \begin{flushleft}
\begin{figure}
\begin{centering}
\includegraphics[clip,width=1\textwidth]{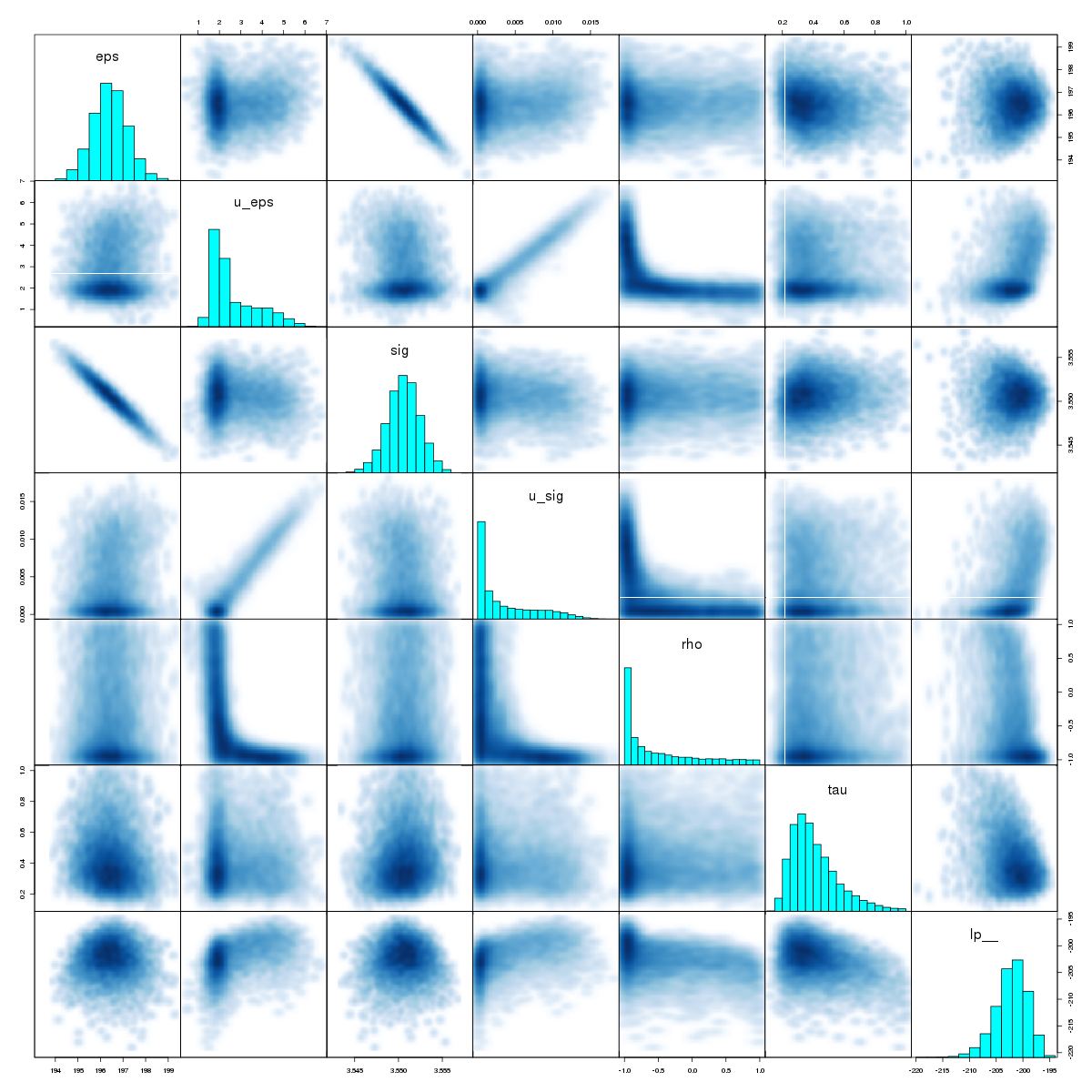}
\par\end{centering}
\caption{\label{fig:S1}Pairs plot of parameters and log-posterior for the
ABC-Shift method and Kr dataset. }
\end{figure}
\par\end{flushleft}

\noindent \begin{flushleft}
\begin{figure}
\begin{centering}
\includegraphics[bb=0bp 0bp 1200bp 1200bp,clip,width=1\textwidth]{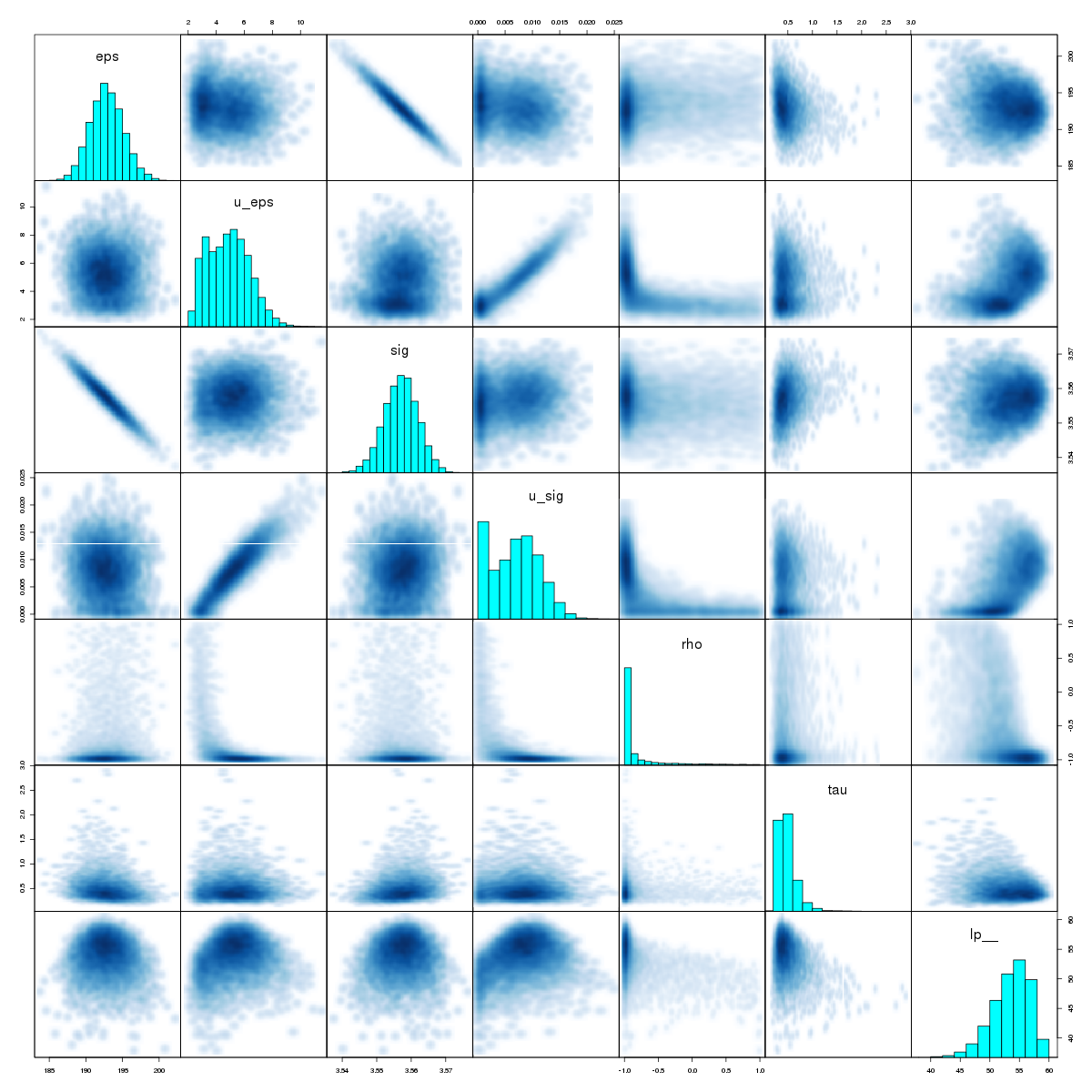}
\par\end{centering}
\caption{\label{fig:S2}Pairs plot of parameters and log-posterior for the
Margin-Shift method and Kr dataset. }
\end{figure}
\par\end{flushleft}
\end{document}